\documentclass[fleqn, usenatbib]{mnras}

\usepackage{newtxtext, newtxmath}
\usepackage[T1]{fontenc}
\usepackage{ae, aecompl}
\usepackage{amsmath}
\usepackage{amssymb}
\usepackage{scalerel}
\usepackage{natbib}
\bibpunct{(}{)}{;}{a}{}{;}
\usepackage{multirow}
\usepackage{graphicx}
\usepackage{gensymb}
\usepackage{threeparttable}
\usepackage{soul}
\usepackage{xcolor}
\usepackage{bm}
\hypersetup{colorlinks, linkcolor={blue}, citecolor={blue}, urlcolor={blue}}
\newcommand\iona[2]{#1$\;${\scshape{#2}}}

\newlength\myheight
\newlength\mydepth
\settototalheight\myheight{Xygp}
\newcommand*\inlinegraphics[1]{%
  \settototalheight\myheight{Xygp}%
  \settodepth\mydepth{Xygp}%
  \raisebox{-\mydepth}{\includegraphics[height=1.25\myheight]{#1}}%
}

\title[X-ray properties of broad-line DOGs]{X-ray properties of dust-obscured galaxies with broad optical/UV emission lines}

\author[F. Zou et al.]{
Fan Zou \inlinegraphics{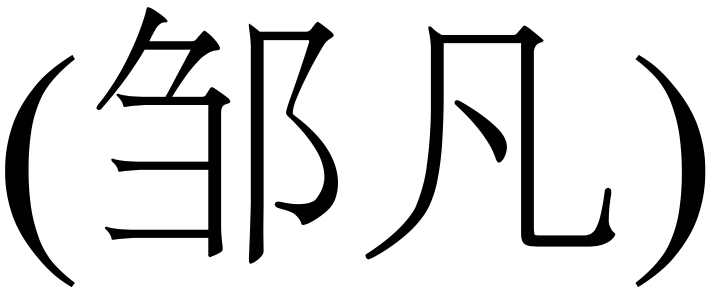},$^{1, 2}$\thanks{E-mail: fuz64@psu.edu} William N. Brandt,$^{1, 2, 3}$ Fabio Vito,$^{4, 5, 6}$ Chien-Ting Chen \inlinegraphics{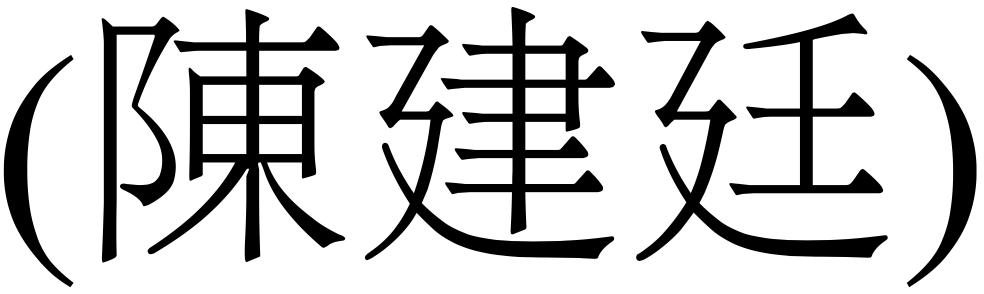},$^{7}$
\newauthor Gordon P. Garmire,$^{8}$ Daniel Stern,$^{9}$ and Ashraf Ayubinia$^{10, 11}$\\
$^{1}$Department of Astronomy and Astrophysics, 525 Davey Lab, The Pennsylvania State University, University Park, PA 16802, USA\\
$^{2}$Institute for Gravitation and the Cosmos, The Pennsylvania State University, University Park, PA 16802, USA\\
$^{3}$Department of Physics, 104 Davey Laboratory, The Pennsylvania State University, University Park, PA 16802, USA\\
$^{4}$Scuola Normale Superiore, Piazza dei Cavalieri 7, I-56126 Pisa, Italy\\
$^{5}$Instituto de Astrof{\'{\i}}sica and Centro de Astroingenier{\'{\i}}a, Facultad de F{\'{i}}sica, Pontificia Universidad Cat{\'{o}}lica de Chile, Casilla 306, Santiago 22, Chile\\
$^{6}$Chinese Academy of Sciences South America Center for Astronomy, National Astronomical Observatories, CAS, Beijing 100012, China\\
$^{7}$Marshall Space Flight Center, Huntsville, AL 35811, USA\\
$^{8}$Huntingdon Institute for X-ray Astronomy, LLC, 10677 Franks Road, Huntingdon, PA 16652, USA\\
$^{9}$Jet Propulsion Laboratory, California Institute of Technology, 4800 Oak Grove Drive, Pasadena, CA 91109, USA\\
$^{10}$CAS Key Laboratory for Research in Galaxies and Cosmology, Department of Astronomy, University of Science and Technology of China, Hefei 230026, China\\
$^{11}$School of Astronomy and Space Science, University of Science and Technology of China, Hefei 230026, China
}

\begin{document}
\label{firstpage}
\pagerange{\pageref{firstpage}--\pageref{lastpage}}
\maketitle

\begin{abstract}
Dust-obscured galaxies (DOGs) with extreme infrared luminosities may represent a key phase in the co-evolution of galaxies and supermassive black holes. We select 12 DOGs at $0.3\lesssim z\lesssim1.0$ with broad \iona{Mg}{ii} or H$\beta$ emission lines and investigate their \mbox{X-ray} properties utilizing snapshot observations ($\sim3~\mathrm{ks}$ per source) with \textit{Chandra}. By assuming that the broad lines are broadened due to virial motions of broad-line regions, we find that our sources generally have high Eddington ratios ($\lambda_\mathrm{Edd}$). Our sources generally have moderate intrinsic \mbox{X-ray} luminosities ($L_\mathrm{X}\lesssim10^{45}~\mathrm{erg~s^{-1}}$), which are similar to those of other DOGs, but are more obscured. They also present moderate outflows and intense starbursts. Based on these findings, we conclude that high-$\lambda_\mathrm{Edd}$ DOGs are closer to the peaks of both host-galaxy and black-hole growth compared to other DOGs, and that AGN feedback has not swept away their reservoirs of gas. However, we cannot fully rule out the possibility that the broad lines are broadened by outflows, at least for some sources. We investigate the relations among $L_\mathrm{X}$, AGN rest-frame $6~\mathrm{\mu m}$ monochromatic luminosity, and AGN bolometric luminosity, and find the relations are consistent with the expected ones.
\end{abstract}
\begin{keywords}
galaxies: active -- galaxies: evolution -- galaxies: nuclei -- X-rays: galaxies
\end{keywords}

\section{Introduction}
\label{sec: intro}
During the last few decades, astronomers have developed a co-evolution framework for supermassive black holes (SMBHs) and their host galaxies (e.g. \citealt{Sanders88, Hopkins06, Hopkins08, Alexander12}): mergers among gas-rich galaxies drive gas and dust down to the central SMBHs and trigger both strong accretion activity of the SMBHs and intense starbursts in the host galaxies. In the early stage of such evolution, the large amount of material would fuel the accretion to approach or even exceed the Eddington limit; meanwhile, the gas and dust causes severe obscuration. Then, radiation-driven outflows from near the central SMBHs sweep out the obscuring material, allowing the SMBHs to shine as unobscured quasars and may also suppress the star-formation (SF) activity in the host galaxies.\par
\bigbreak
Since the launch of the \textit{Spitzer Space Telescope}, studies of dusty galaxies have greatly improved. \citet{Dey08} found thousands of high-redshift dust-obscured galaxies (DOGs) with $f_\mathrm{24~\mu m}\ge0.3~\mathrm{mJy}$ and $(R-[24])_\mathrm{Vega}\ge14$. Generally, DOGs are thought to fit into the SMBH-galaxy co-evolution framework --- at least some DOGs are at the final stage of mergers when strong SMBH accretion and star-forming activities are obscured (e.g. \citealt{Narayanan10}). Their IR emission may be from central active galactic nuclei (AGNs), but may also be explained entirely by episodes of strong star formation (e.g. \citealt{Franceschini03, Bussmann09, Lanzuisi09, Teng10}). Such starbursting episodes are expected during or immediately after gas-rich mergers.\par
The \textit{Wide-field Infrared Survey Explorer} (\textit{WISE}; \citealt{Wright10}) also discovered a more extreme population of hyper-luminous infrared galaxies (HyLIRGs; $L_\mathrm{IR}>10^{13}~L_\odot$) with extreme MIR colors (e.g. \citealt{Eisenhardt12, Wu12, Tsai15, Assef16}). Their characteristic spectral energy distribution (SED) shape is due to their dust having temperatures much higher than those in DOGs (up to hundreds of $\mathrm{K}$ versus $30-40~\mathrm{K}$; e.g. \citealt{Pope08, Melbourne12, Wu12, Jones14, Tsai15}), and thus these hot dust-obscured galaxies are named ``Hot DOGs''. In the co-evolution framework, such extreme luminosities and high dust temperatures are thought to be powered by deeply buried, high-mass, and rapidly accreting SMBHs, caught during the peak of their post-merger accretion phases. Indeed, \mbox{X-ray} observations provide further evidence for this scenario. One of the advantages of \mbox{X-ray} observations is that \mbox{X-ray} emission can be used to identify AGNs directly, even when these are buried in large column densities of obscuring material, thanks to the high penetrating power of \mbox{X-rays} and the large contrast between AGN and stellar emission in the X-ray regime (e.g. \citealt{Brandt15}). It was found that these Hot DOGs indeed generally had high $L_\mathrm{X}$, which indicated strong AGN activity, as well as nearly Compton-thick obscuration in studies either focused on a few sources with high-quality observations \citep{Stern14, Assef16, Ricci17, Zappacosta18, Assef19} or statistically significant samples with shorter observations \citep{Vito18}. In particular, Hot DOGs as well as related objects (e.g. \citealt{Goulding18}) clearly occupy a separate region in the $N_\mathrm{H}-L_\mathrm{X}$ plane with higher column density, $N_\mathrm{H}$, than luminous optically type~1 and reddened quasars with similar luminosities. The \mbox{X-ray} spectra of most reddened type~1 quasars, thought to be transitioning from the heavily obscured phase to blue unobscured quasars, are affected by significantly lower obscuration levels (\citealt{Vito18} and references therein).\par
However, the \mbox{X-ray} obscuration of DOGs spans a wider $N_\mathrm{H}$ range than for Hot DOGs, i.e. from low-to-moderate to Compton-thick $N_\mathrm{H}$ \citep{Lanzuisi09, Corral16}. These results can be explained if the DOG population is heterogeneous, and the DOG IR emission can be produced through different physical processes. The population of the most X-ray obscured DOGs may be at similar evolutionary stages to Hot DOGs, when the central SMBHs are accreting rapidly (i.e. having high Eddington ratios $\lambda_\mathrm{Edd}$), and Hot DOGs just represent the extreme-luminosity tip of this population. Indeed, Hot DOGs have large SMBH masses ($\sim10^9M_\odot$; e.g. \citealt{Wu18}), and smaller SMBHs with high $\lambda_\mathrm{Edd}$ would not produce such high luminosities comparable to Hot DOGs' ($L_\mathrm{IR}\gtrsim10^{13}L_\odot$). For less X-ray obscured DOGs, they may have lower $\lambda_\mathrm{Edd}$, and their IR emission may be dominated by SF.\par
This explanation has not been well-tested yet. In this work, for the first time, we probe whether apparent high-$\lambda_\mathrm{Edd}$ DOGs are indeed more obscured. Under the co-evolution framework, we will examine whether high-$\lambda_\mathrm{Edd}$ DOGs are at the post-merger phases when the large reservoir of obscuring materials are fueling more intense SMBH and host-galaxy growth than other DOGs, and especially, whether these sources have entered the blow-out phase. The $\lambda_\mathrm{Edd}$ values are derived based on broad optical/UV lines by assuming that the broad lines are dominated by the virial motions of broad-line regions (BLRs), which is called the ``virial assumption'' hereafter. We selected 12 DOGs at $z\lesssim1$ with broad \iona{Mg}{ii} or H$\beta$ lines, and found that their $\lambda_\mathrm{Edd}$ values appear generally high ($\sim0.1-1$) under the virial assumption. We note that broad-line DOGs or Hot DOGs are not rare. Among 36 IR-bright DOGs in \citet{Toba17}, 17 (47\%) have such broad lines. The fraction of Hot DOGs with broad optical lines is around $20-30\%$ (Eisenhardt et al. in preparation). Though these fractions may be influenced by selection bias, it is almost certain that broad-line DOGs are not just a rare subsample among the whole DOG population. We proposed for snapshot observations with \textit{Chandra} ($\sim3~\mathrm{ks}$) for each of our sources. The snapshot strategy aims to cover the sample with economical but sensitive \mbox{X-ray} observations. These can provide the basic properties of our sources, and can also provide guidance for future long-exposure observations. Such a strategy has been proved to be efficient and has successfully revealed the basic X-ray properties of Hot DOGs (e.g. \citealt{Vito18}).\par
However, the virial assumption is still uncertain. A recent article about Hot DOGs, \citet{Jun20}, threw the origins of the broad lines into question. They argued that the broad emission lines in Hot DOGs might be explained by outflows within the narrow-line region because the widths of these lines are comparable to those of the outflowing [\iona{O}{iii}] lines. We also examine this issue for our DOGs (see Section~\ref{sec: origin_broad}), finding that the virial assumption is generally favored. Therefore, to ensure the fluency of the flow of our narrative, we still adopt the virial assumption throughout the whole paper unless noted and leave most of the related issues to Section~\ref{sec: origin_broad}.\par
This work presents a multi-wavelength study of our sources, with a focus on their basic \mbox{X-ray} properties. \citet{Corral16} have analyzed X-ray properties of DOGs in the \textit{Chandra} Deep Field-South and also presented their multi-wavelength properties (IR luminosity, star-formation rate, and stellar mass). Their sample can be regarded as a representative sample of the whole DOG population. Also, \citet{Toba20} recently presented a high-$\lambda_\mathrm{Edd}$ DOG ($\lambda_\mathrm{Edd}=0.7$) with X-ray coverage. We will compare our results with theirs in this work. This paper is organized as follows. In Section~\ref{sec: samp}, our sample and observations are described. In Section~\ref{sec: analyze_data}, we reduce our \mbox{X-ray} and other multiwavelength data. We present results in Section~\ref{sec: results}. We discuss the physical implications of our results and the validity of the virial assumption in Section~\ref{sec: discussion}. Finally, we summarize this work in Section~\ref{sec: sum_future}, Throughout this paper, we adopt a flat $\Lambda\mathrm{CDM}$ cosmology with $H_0=70~\mathrm{km~s^{-1}~Mpc^{-1}}$, $\Omega_\Lambda=0.73$, and $\Omega_M=0.27$.

\section{Sample selection and observations}
\label{sec: samp}
Our observed sample is drawn from the parent DOG sample in \citet{Toba17}, who selected 36 IR-bright DOGs ($f_\mathrm{22~\mu m}>3.8~\mathrm{mJy}$) at $0.05<z<1.02$ with extreme optical/IR colors ($i-[22]_\mathrm{AB}>7$) and clear [\iona{O}{iii}] lines in their SDSS spectra. We note that this selection criterion is slightly different from (but largely consistent with) the ``classic'' criterion ($f_\mathrm{24~\mu m}\ge0.3~\mathrm{mJy}$ and $R-[24]_\mathrm{vega}\ge14$; \citealt{Dey08}). This new criterion enables a more efficient way to select DOGs that have higher MIR fluxes and are more likely to have larger AGN contributions \citep{Toba15, Toba16}. \citet{Toba17} have also conducted detailed optical spectral analyses and obtained the [\iona{O}{iii}]-based outflow properties. Among them, we select 12 sources with broad \iona{Mg}{ii} or H$\beta$ lines (i.e. in SDSS quasar catalogs) at $0.3\lesssim z\lesssim1$. The criterion for a ``broad'' line is that its full width at half maximum (FWHM) $>1000~\mathrm{km~s^{-1}}$. Then we obtained \textit{Chandra} snapshot observations of the 12 DOGs in Cycle 20. \textit{Chandra} has high sensitivity and low background, and thus is well-suited for observing these possibly heavily \mbox{X-ray} obscured sources.\par
The spectroscopic redshifts of our sources are from \citet{Toba17} based on the stellar absorption lines or H$\beta$, and we have checked that the redshift measurements are not severely affected by outflows. The redshift values are consistent with the ones in \citet{Paris17}. Through visual inspection, we also found that the redshifts are consistent with several other lines that can provide robust source redshifts (e.g. the [\iona{O}{ii}] line; \citealt{Shen16}). We collect the foreground Galactic $N_\mathrm{H}$ values of our sources based on \citet{HI4PI16}. The basic properties of our sample and the \textit{Chandra} observations are summarized in Table~\ref{BasicPropTable}.\par

\begin{table*}
\caption{Basic properties of the targeted DOGs and \textit{Chandra} observations}
\label{BasicPropTable}
\centering
\begin{threeparttable}
\begin{tabular}{cccccccc}
\hline
\hline
SDSS Name & RA & Dec & Redshift & Galactic $N_\mathrm{H}$ & \textit{Chandra} ObsID & ObsMJD& $T_\mathrm{exp}$\\
 & (hms) & (dms) & & ($10^{20}~\mathrm{cm^{-2}}$) & & (d) & (ks)\\
(1) & (2) & (3) & (4) & (5) & (6) & (7) & (8)\\
\hline
J0756+4432 & 07:56:09.9 & +44:32:22.8 & 0.510 & 4.20 & 21142 & 58439 & 3.1\\
J0833+4508 & 08:33:38.5 & +45:08:33.5 & 0.925 & 2.82 & 21149 & 58502 & 3.7\\
J1010+3725 & 10:10:34.2 & +37:25:14.7 & 0.282 & 1.16 & 21151 & 58434 & 2.9\\
J1028+5011 & 10:28:01.5 & +50:11:02.5 & 0.776 & 1.06 & 21150 & 58548 & 4.4\\
J1042+2451 & 10:42:41.1 & +24:51:07.0 & 1.026 & 3.12 & 21143 & 58439 & 3.1\\
J1210+6105 & 12:10:56.9 & +61:05:51.5 & 0.926 & 1.78 & 21141 & 58477 & 3.1\\
J1235+4827 & 12:35:44.9 & +48:27:15.4 & 1.023 & 1.50 & 21140 & 58428 & 2.9\\
J1248+4242 & 12:48:36.1 & +42:42:59.3 & 0.682 & 1.81 & 21146 & 58429 & 5.3\\
J1324+4501 & 13:24:40.1 & +45:01:33.8 & 0.774 & 2.07 & 21144 & 58717 & 3.1\\
J1513+1451 & 15:13:54.4 & +14:51:25.2 & 0.882 & 2.10 & 21145 & 58479 & 3.1\\
J1525+1234 & 15:25:04.7 & +12:34:01.7 & 0.851 & 3.06 & 21148 & 58471 & 3.5\\
J1531+4533 & 15:31:05.1 & +45:33:03.4 & 0.871 & 1.28 & 21147 & 58434 & 3.3\\
\hline
\hline
\end{tabular}
\begin{tablenotes}
\item
\emph{Notes.} (2) and (3) SDSS J2000 coordinates. (4) Spectroscopic redshifts from \citet{Toba17}. (5) Galactic foreground column densities \citep{HI4PI16}. (6) -- (8) \textit{Chandra} observation IDs, modified Julian dates, and exposure times.
\end{tablenotes}
\end{threeparttable}
\end{table*}

We note that when initially designing this project, the ambiguity about the nature of broad lines had not been proposed by \citet{Jun20}, and thus we focused on the high-$\lambda_\mathrm{Edd}$ interpretation, and did not include all the DOGs with broad lines in \citet{Toba17}. Among the parent sample, 17 DOGs present broad lines, i.e. are included in the SDSS quasar catalogs (\citealt{Shen11, Kozlowski17}) or are identified as type~1 AGNs in \citet{Toba17}, but only 12 of them are included in our \textit{Chandra} program.

\section{Data reduction and analyses}
\label{sec: analyze_data}
In this section, we reduce the X-ray data (Section~\ref{sec: Xray}), analyze SEDs (Section~\ref{sec: SED}), and derive $M_\mathrm{BH}$ and $\lambda_\mathrm{Edd}$ through fitting the SDSS spectra (Section~\ref{sec: MBH}).
\subsection{X-ray data}
\label{sec: Xray}
\subsubsection{Data reduction}
Our \mbox{X-ray} data were reduced with \texttt{CIAO} 4.11 and \texttt{CALDB} 4.8.4.1. First, we reprocess all the observations with the \texttt{chandra\_repro} script, setting the option $check\_vf\_pha=\mathrm{yes}$ since our observations were taken in very faint mode. We then use the \texttt{fluximage} script to generate images, exposure maps, and point spread function (PSF) maps in the soft ($0.5-2~\mathrm{keV}$), hard ($2-7~\mathrm{keV}$), and full ($0.5-7~\mathrm{keV}$) bands. The maps are weighted by redshifted absorbed power law models, where the redshifts are set to be the source redshifts, the intrinsic $N_\mathrm{H}$ values are set to $5\times10^{23}~\mathrm{cm}^{-2}$, and the photon indices are set to $\Gamma=2.0$. The weighting parameters are based on prior estimations that our sources are highly obscured ($N_\mathrm{H}\gtrsim10^{23}~\mathrm{cm}^{-2}$; Section~\ref{sec: XSrcProp}). Their exact values are not important, and changing them does not materially influence the results.\par
\subsubsection{Source detection}
To assess the significance of detection, we compute the binomial no-source probability \citep{Broos07, Weisskopf07}:
\begin{equation}
\label{Pdetect}
P_B(X\ge S)=\sum_{X=S}^{N}\frac{N!}{X!(N-X)!}p^X(1-p)^{N-X},
\end{equation}
where $S$ is the number of counts in the source region, $N$ is the number of total counts in both the source region and the background region, $p=1/(1+\mathrm{BACKSCAL})$, and $\mathrm{BACKSCAL}$ is the ratio of areas between the background region and the source region. We adopt $P_B=0.01$ as the detection threshold, which corresponds to a significance of 99\%, as adopted in previous works (e.g. \citealt{Luo15, Vito19}). The expected number of false detections in our whole sample is only $\sim0.2$, and thus all the detections should be statistically reliable.\par
We extract source counts and background counts in a circular aperture with a $1.5''$ radius and an annulus with inner and outer radii of $10''$ and $40''$ centered at the source position, respectively (e.g. \citealt{Luo15}). The aperture radius is chosen to be small enough to prevent much contamination from the backgrounds and also to improve the significance of the detection, but also large enough to (visually) encircle nearly all the source photons. Adopting other radius choices (e.g. $2.0''$) would not significantly influence the results. The encircled-energy fraction of the source region is between $\sim0.87-0.97$, depending on the exact spectral shape and energy band. Using Eq.~\ref{Pdetect}, we detect one source (J1028+5011) in all three bands and five sources only in the hard and full bands (J1010+3725, J1235+4827, J1248+4242, J1324+4501, and J1525+1234). We note that those sources with only 2 counts (J1235+4827 and J1248+4242) are regarded as detected, largely owing to the very small background in a source-detection cell. Independently, we also run \texttt{wavdetect} with a much higher significance threshold ($10^{-6}$) both to detect the sources and obtain \mbox{X-ray} positions for the sources. We obtain the same results as those based on the binomial no-source probability in terms of detections besides the two-count sources, indicating the reliability of the detection of the sources with over 2 counts. For sources with counts $\le2$, we adopt their optical positions as the \mbox{X-ray} positions.\par
\begin{figure*}
\resizebox{\hsize}{!}{
\includegraphics{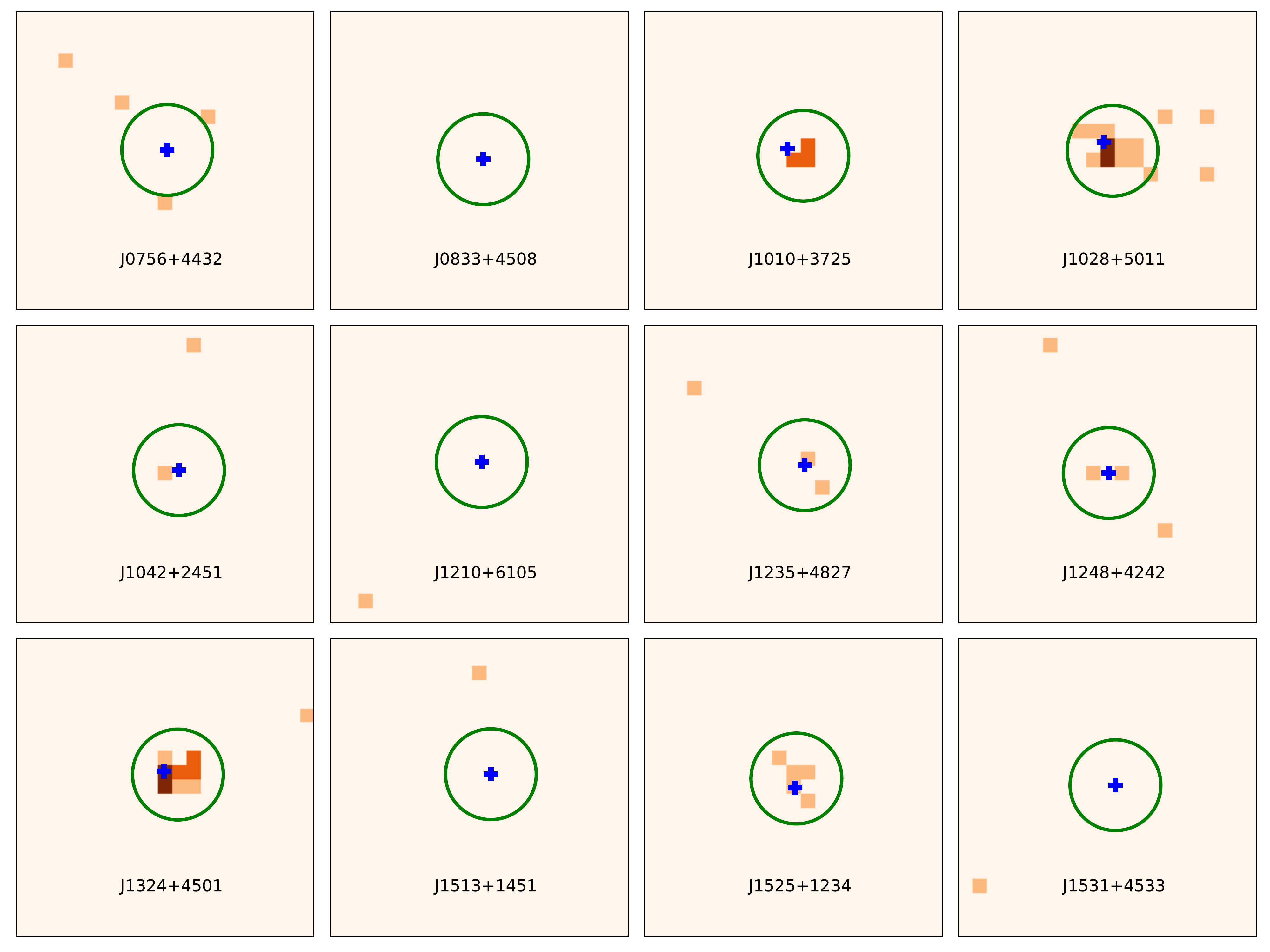}
}
\caption{Full-band ($0.5-7~\mathrm{keV}$) images of the 12 DOGs. All the images are on the same color scale (i.e. $0-3$ counts). The green circles are the source regions, i.e. $1.5''$-radius circles centered on the \mbox{X-ray} positions for sources with $>2$ counts or optical positions for other sources. The blue pluses are the optical positions.}
\label{SrcImgFig}
\end{figure*}

We show the images of full-band counts in Fig.~\ref{SrcImgFig}. We have checked if there are other noticeable sources close to the detected sources in the SDSS images, and found no neighboring sources within $\sim7''$. Thus, there should not be any contamination or source confusion.
\subsubsection{Source properties}
\label{sec: XSrcProp}
\begin{figure*}
\resizebox{\hsize}{!}{
\includegraphics{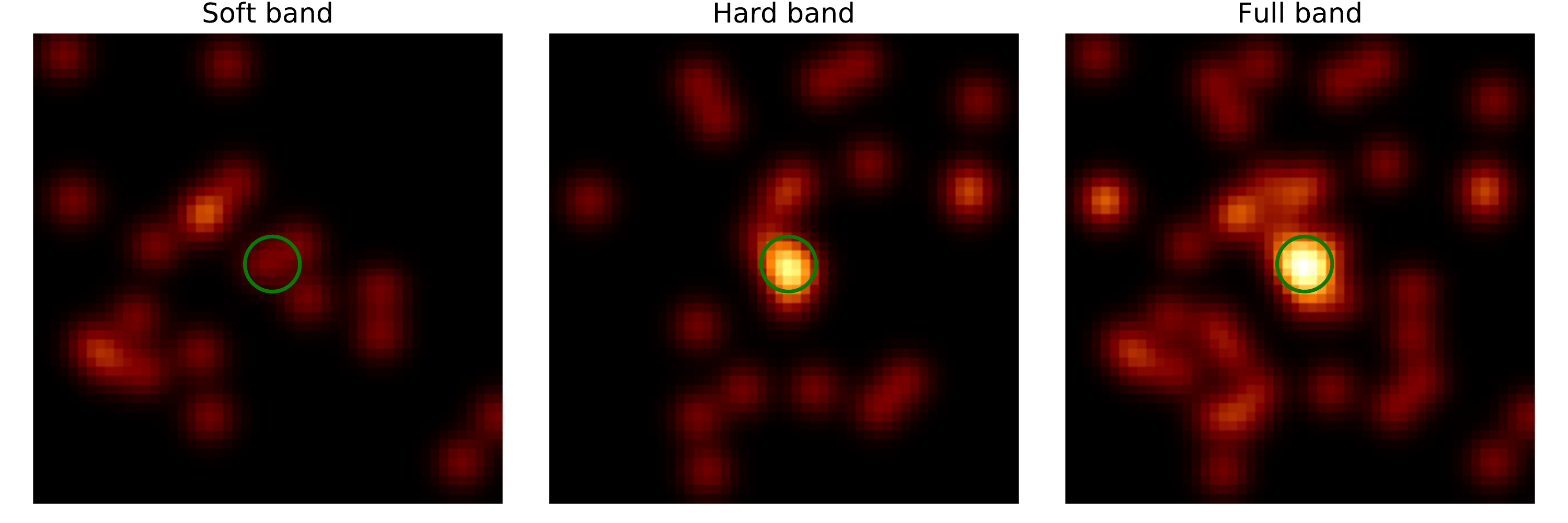}
}
\caption{The smoothed stacked images of the 8 sources with $\le2$ counts. The green circles are the source regions. The stacked signals are clearly seen in the hard band ($2-7~\mathrm{keV}$) and full band ($0.5-7~\mathrm{keV}$) but not in the soft band ($0.5-2~\mathrm{keV}$).}
\label{StackImgFig}
\end{figure*}

We then constrain the net source counts within the source apertures in the three bands based on the probability density function of net counts derived in \citet{Weisskopf07}. We also derive the $90\%$ confidence upper limits on the net counts for undetected sources. The results are shown in Table~\ref{XPropTable}. In addition, we stack the sources with $\le2$ counts, and obtain 1, 4, and 5 counts in the soft, hard, and full bands, respectively. The significances of the detections of the stacked signals are $P_B=0.10$ in the soft band and $P_B=4.6\times10^{-5}$ in the hard band. Therefore, the stacked signal is significantly detected in the hard band but not in the soft band, indicating likely heavy absorption. This can be seen in the stacked images displayed in Fig.~\ref{StackImgFig}.\par

\begin{table*}
\caption{X-ray counts and basic properties}
\label{XPropTable}
\centering
\begin{threeparttable}
\begin{tabular}{cccccc}
\hline
\hline
SDSS Name & Soft Band & Hard Band & Full Band & $N_\mathrm{H}$ & $\mathrm{log}L_\mathrm{X}$\\
 & (counts) & (counts) & (counts) & ($10^{22}~\mathrm{cm}^{-2}$) & ($\mathrm{erg~s^{-1}}$)\\
(1) & (2) & (3) & (4) & (5) & (6)\\
\hline
J0756+4432 & $<2.3$ & $<2.3$ & $<2.3$ & -- & $<43.7$\\
J0833+4508 & $<2.3$ & $<2.3$ & $<2.3$ & -- & $<44.1$\\
J1010+3725 & $<2.3$ & $6.0_{-1.6}^{+3.6}$ & $5.9_{-1.6}^{+3.6}$ & $9.3_{-4.6}^{+7.5}$ & $43.3_{-0.3}^{+0.3}$\\
J1028+5011 & $6.0_{-1.6}^{+3.6}$ & $8.0_{-1.9}^{+3.9}$ & $13.9_{-2.8}^{+4.8}$ & $4.1_{-2.0}^{+2.6}$ & $44.2_{-0.2}^{+0.2}$\\
J1042+2451 & $<3.9$ & $<2.3$ & $<3.9$ & -- & $<44.5$\\
J1210+6105 & $<2.3$ & $<2.3$ & $<2.3$ & -- & $<44.2$\\
J1235+4827 & $<2.3$ & $2.0_{-0.6}^{+2.6}$ & $2.0_{-0.6}^{+2.6}$ & $18.7_{-18.0}^{+60.3}$ & $44.2_{-0.7}^{+0.6}$\\
J1248+4242 & $<2.3$ & $2.0_{-0.6}^{+2.6}$ & $2.0_{-0.6}^{+2.6}$ & $67.2_{-51.1}^{+105.3}$ & $44.2_{-0.7}^{+0.8}$\\
J1324+4501 & $<2.3$ & $15.0_{-3.0}^{+5.0}$ & $15.0_{-3.0}^{+5.0}$ & $54.0_{-18.4}^{+24.4}$ & $45.2_{-0.2}^{+0.2}$\\
J1513+1451 & $<2.3$ & $<2.3$ & $<2.3$ & -- & $<44.1$\\
J1525+1234 & $<3.9$ & $4.0_{-1.2}^{+3.2}$ & $4.9_{-1.4}^{+3.4}$ & $17.5_{-9.8}^{+17.3}$ & $44.3_{-0.3}^{+0.3}$\\
J1531+4533 & $<2.3$ & $<2.3$ & $<2.3$ & -- & $<44.1$\\
\hline
\hline
\end{tabular}
\begin{tablenotes}
\item
\emph{Notes.} (2) -- (4) Net source counts in the soft, hard, and full bands, respectively. (5) Column density. (6) Intrinsic $2-10~\mathrm{keV}$ luminosity. The quoted uncertainty intervals are for a 68\% confidence level, while the upper limits are for a 90\% confidence level. For undetected sources, the $L_\mathrm{X}$ upper limits are given assuming $N_\mathrm{H}=2.2\times10^{23}~\mathrm{cm^{-2}}$.
\end{tablenotes}
\end{threeparttable}
\end{table*}

To constrain the $N_\mathrm{H}$ and intrinsic $2-10~\mathrm{keV}$ luminosity, $L_\mathrm{X}$, of the sources, we extract the spectra for the detected sources using the $\texttt{specextract}$ tool in $\texttt{CIAO}$, and the spectra are binned to have at least one count per bin. The binning is used to reduce the number of unnecessary free parameters in parameterizing the background, and can thus avoid possible biases.\footnote{https://giacomov.github.io/Bias-in-profile-poisson-likelihood/} We fit their spectra with a $\texttt{phabs*zphabs*clumin*zpowerlw}$ model in $\texttt{XSPEC}$ \citep{Arnaud96}. The $N_\mathrm{H}$ of the $\texttt{phabs}$ component is fixed to the corresponding Galactic foreground $N_\mathrm{H}$, while the $N_\mathrm{H}$ of $\texttt{zphabs}$ represents the sources' intrinsic $N_\mathrm{H}$. The limited counts make constraining the photon index of $\texttt{zpowerlw}$ impossible, and thus we fix its value at $\Gamma=2$. This value is approximately the expected value under the virial assumption, where the estimated $\lambda_\mathrm{Edd}$ values are high (e.g. \citealt{Shemmer08, Brightman13}). Fixing the value at $1.8$, the value for typical AGNs, does not change our results materially. The $\texttt{clumin}$ component is used to measure $L_\mathrm{X}$. In fitting, we adopt the modified Cash statistic \citep{Cash79, Kaastra17}, which is appropriate when analyzing low-count spectra. The best-fit $N_\mathrm{H}$ and $L_\mathrm{X}$ values for these detected sources are shown in Table~\ref{XPropTable}.\par
We then jointly fit the six spectra of the detected sources to assess the typical $N_\mathrm{H}$ for our sample. We link the $N_\mathrm{H}$ parameter among all the sources to retrieve a single, average value, and the $L_\mathrm{X}$ value of each source is fixed to its corresponding best-fit $L_\mathrm{X}$ in \mbox{Table~\ref{XPropTable}}. The fit returns $N_\mathrm{H}=2.2_{-0.3}^{+0.4}\times10^{23}~\mathrm{cm^{-2}}$. Most (5/6) of our undetected sources do not contain any counts within the source region, and these zero-count sources cannot offer any information about $N_\mathrm{H}$ without some prior knowledge about their $L_\mathrm{X}$. This zero-count case is different from the usual non-detection case. In the latter case, the observed signals may still include some source signals, which are however not significant enough compared to the background level. That is, there may be some hidden information about their spectral shapes (and $N_\mathrm{H}$) for typical undetected sources. In principle, if there are plenty of such undetected sources, we can use statistical methods (e.g. stacking) to estimate the average properties for these sources. However, our zero-count sources are already known not to have any source photons, and thus we are not able to extract any information about their \mbox{X-ray} spectral shapes and $N_\mathrm{H}$, unless we know their $L_\mathrm{X}$ a priori (which is not our case). Therefore, we simply assume that their $N_\mathrm{H}$ values are generally similar to those of our detected sources. Based on the constraints on their counts, we derive upper limits on their intrinsic $L_\mathrm{X}$, assuming $N_\mathrm{H}=2.2\times10^{23}~\mathrm{cm^{-2}}$, and the results are also shown in \mbox{Table~\ref{XPropTable}}. We note that their $L_\mathrm{X}$ upper limits are similar to the $L_\mathrm{X}$ values of our detected sources. This indicates that if we instead assume that the $L_\mathrm{X}$ of our undetected sources and detected sources are similar, we will obtain the $N_\mathrm{H}$ lower limits for these undetected sources to be around $2.2\times10^{23}~\mathrm{cm^{-2}}$. Given this constraint, we may adopt (reasonably) higher $N_\mathrm{H}$ values to derive the $L_\mathrm{X}$ upper limits for the undetected sources, but our analyses and conclusions would not change in such cases. For example, if we assume their $N_\mathrm{H}=10^{24}~\mathrm{cm^{-2}}$, the $L_\mathrm{X}$ upper limits would be $\sim0.7$ dex higher. If we put the $L_\mathrm{X}$ upper limits upward by 0.7 dex in Fig.~\ref{L6wLXFig}, Fig.~\ref{kbolwLbolFig}, and Fig.~\ref{outflow_XrayFig} (see Section~\ref{sec: results}), none of the conclusions in Section~\ref{sec: results} would change.\par
We note the default assumption here is that the X-ray emission is not severely contaminated by jets. We use radio power to examine this. Generally, an AGN can be regarded as a radio-loud AGN if it presents a radio excess compared with the IR-radio relation, in which case its X-ray emission may have a jet-linked contribution. Since the contributions of our sources to the total emission at IR wavelengths are much larger than those at optical (Section~\ref{sec: SED}), it is more reliable to use the IR-radio criterion instead of the conventional optical-radio criterion (e.g. \citealt{Hao14}) to discriminate radio-loud and radio-quiet sources. Thus, based on the $q_\mathrm{24, obs}$-method in \citet{Bonzini13}, we found that only one source (J1248+4242) is radio-loud, and it only marginally reaches the radio-loud threshold. Therefore, the jet contamination to the X-ray data is generally insignificant.

\subsection{SED fitting}
\label{sec: SED}
Here, we conduct SED fitting to decompose the total SEDs into AGN components and galaxy components and obtain the host-galaxy properties. In the optical-to-IR bands, we use similar photometric data as in \citet{Toba16}: SDSS DR12 optical photometry in five bands ($u$, $g$, $r$, $i$, and $z$; \citealt{Alam15}), \textit{WISE} NIR-to-MIR photometry in four bands ($3.4~\mathrm{\mu m}$, $4.6~\mathrm{\mu m}$, $12~\mathrm{\mu m}$, and $22~\mathrm{\mu m}$; \citealt{Cutri14}), and \textit{Akari} MIR-to-FIR photometry in six bands ($9~\mathrm{\mu m}$, $18~\mathrm{\mu m}$, $65~\mathrm{\mu m}$, $90~\mathrm{\mu m}$, $140~\mathrm{\mu m}$, and $160~\mathrm{\mu m}$; \citealt{Ishihara10, Yamamura10}). However, none of our sources is detected by \textit{Akari} \citep{Toba16}, and thus we adopt the $5~\sigma$ photometric upper limits: 0.05, 0.12, 2.4, 0.55, 1.4, and 6.2~Jy in each band, respectively \citep{Kawada07, Ishihara10}. We also include 2MASS photometry in three bands ($J$, $H$, and $Ks$) for J1010+3725 (other sources are not detected). The \mbox{X-ray} photometry (Section~\ref{sec: XSrcProp}) is also included in the SED fitting to provide better constraints on the AGN power.\par
We use \mbox{\texttt{X-CIGALE}} \citep{YangG19} to conduct the SED fitting. \mbox{\texttt{X-CIGALE}} was developed from the Code Investigating GALaxy Emission (\mbox{\texttt{CIGALE}}; e.g. \citealt{Boquien19}), which is well-known for its efficiency and accuracy (e.g. \citealt{Ciesla15}). It allows addition of the \mbox{X-ray} photometry into the total SEDs, which can effectively constrain the AGN power because \mbox{X-ray} emission is typically dominated by the AGN component. Additionally, \mbox{\texttt{X-CIGALE}} enables us to model the extinction in type~1 AGNs, which is consistent with our case. For the galaxy components, we adopt the same parameter settings as in \citet{YangG19}, which are also similar with those in other works (e.g. \citealt{Zou19}). More specifically, we adopt a delayed star-formation history. Stellar templates are from \citet{Bruzual03} assuming a Chabrier initial mass function \citep{Chabrier03}. The host-galaxy dust attenuation is assumed to follow a Calzetti law \citep{Calzetti00}, and the dust emission follows the models in \citet{Dale14}. The AGN models are based on a modern torus model \citep{Stalevski12, Stalevski16} with polar dust following the extinction law in the Small Magellanic Cloud \citep{Prevot84}. Since our sources present optical broad lines, the AGN systems should be relatively face-on, and thus we set the inclination angle to be $0\degree-50\degree$ with a step of $10\degree$. However, they are also heavily obscured. Based on the typical $N_\mathrm{H}$ measured from the \mbox{X-ray} data (Section~\ref{sec: NHLX}), the $E(B-V)$ is expected to be $\sim1-10$ \citep{Bohlin78, Maiolino01}. Similarly, Hot DOGs also show such heavy obscuration, and \citet{Assef15} found that the mean $E(B-V)$ can reach up to 6 for Hot DOGs. To account for the obscuration, we set the $E(B-V)$ of the polar dust to be 0, 0.05, 0.5, 2, 4, 6, 8, 10, 15, 20, 25, 35, and 50. In the \mbox{X-ray} module, we set the \mbox{X-ray} power-law photon index to be $\Gamma=2$, as adopted in Section~\ref{sec: XSrcProp}. The input \mbox{X-ray} fluxes are absorption-corrected, and their units are converted to mJy following Eq. 1 in \citet{YangG19}. Other settings are the same as in \citet{YangG19}.\par
\begin{figure*}
\resizebox{\hsize}{!}{
\includegraphics{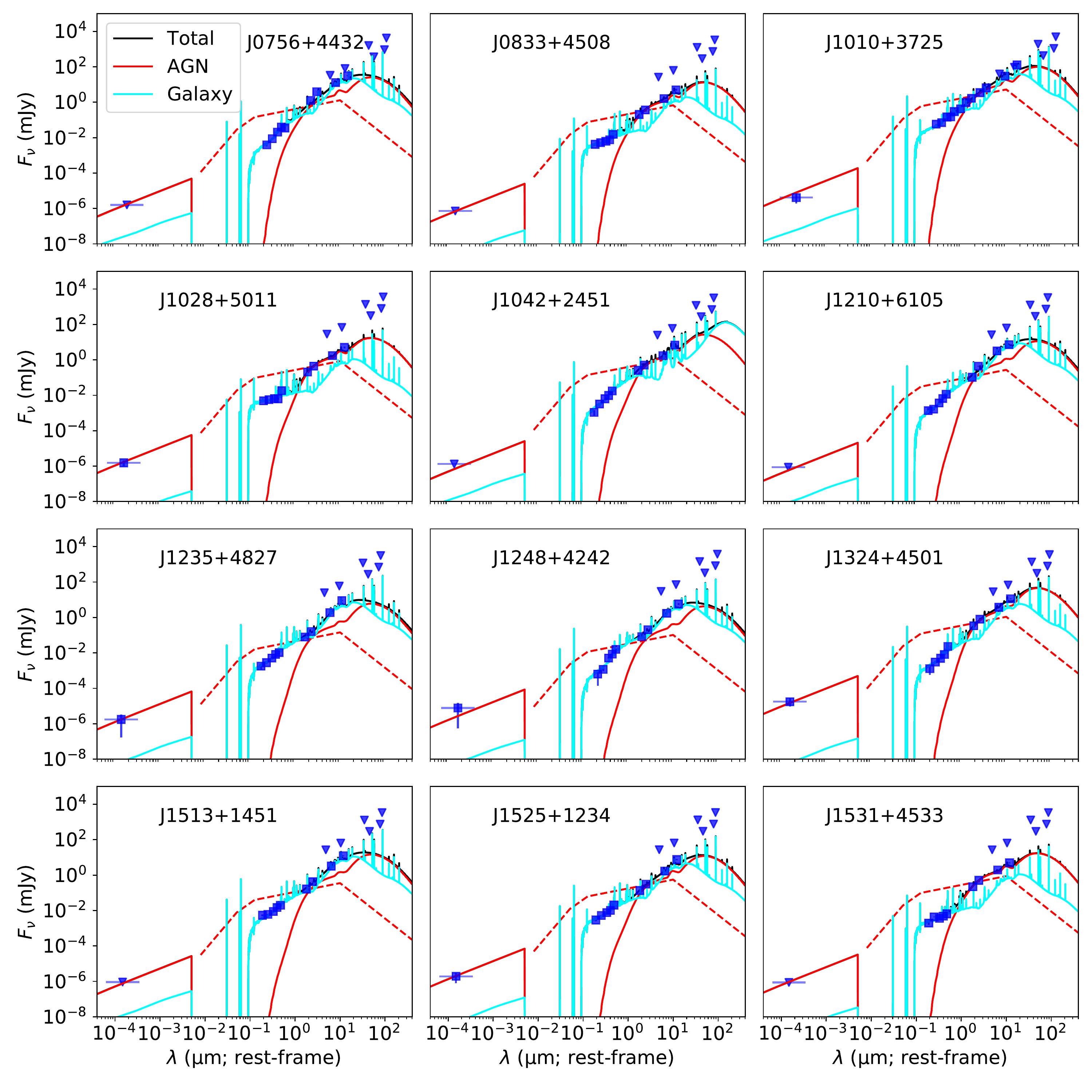}
}
\caption{Decomposition of the SEDs. The blue square data points are the measured photometry, and the triangle points are upper limits. The total SED is decomposed into an AGN component (red solid line) and a galaxy component (cyan solid line). The red dashed line is the intrinsic AGN disk SED, which is only modeled at $\lambda\ge0.008~\mathrm{\mu m}$ \citep{YangG19}.}
\label{SEDFig}
\end{figure*}

Fig.~\ref{SEDFig} displays the best-fit SEDs for our sample, and these SEDs are well-characterized across a wide range of wavelength. Generally, the \mbox{X-ray} emission is dominated by the AGN component, while the optical emission is dominated by the galaxy component because the AGN continuum is heavily obscured in the optical bands. This is also the case for Hot DOGs (e.g. \citealt{Assef15}). The intrinsic AGN disk SED \citep{YangG19} is shown in Fig.~\ref{SEDFig}, and is generally higher than the observed total SED in the optical band. The dominant component of the SEDs in the IR band is undetermined and depends on the exact power of each component, indicating the origins of the IR emission may be either starburst- or AGN-dominated \citep{Narayanan10}. We note that even though the AGN continuum is obscured in the UV-to-optical bands, these sources still present broad lines. This notable phenomenon is also seen in some Hot DOGs (e.g. \citealt{Ricci17, Wu18, Jun20}). One possible cause is that some emission from BLRs ``leaks'' out due to reflection or partial coverage, especially when the intrinsic AGN power is much stronger than the galaxy emission \citep{Assef16}. The other likely potential cause is that the lines are broadened by outflows \citep{Jun20}. \citet{Assef16} also considered the possibility that the emission is from a secondary unobscured AGN, but disfavored it because of the lack of soft X-ray photons. Their argument also works in our case. This issue about the broad lines will be discussed in Section~\ref{sec: origin_broad} in more detail.\par

The SED fitting procedure returns the global star-formation rates (SFRs) and stellar masses ($M_\star$) of the host galaxies, and the AGN bolometric luminosity, $L_\mathrm{bol}$, across the whole wavelength range. $L_\mathrm{bol}$ includes disk emission, dust emission, and \mbox{X-ray} emission, and the \mbox{X-ray} emission is over 1~dex lower than the disk and dust contribution. The contribution of radio emission is negligible because most of our sources are radio-quiet sources. This also indicates that the $\gamma$-ray contribution should be negligible because most $\gamma$-ray emitting AGNs are blazars with strong radio emission (e.g. \citealt{Dermer16}). Indeed, none of our sources is detected by \textit{Fermi}. \mbox{\texttt{X-CIGALE}} also outputs the fractional contributions of AGNs in the IR band ($f_\mathrm{AGN}$), but these are physically-motivated values (i.e. all the re-emitted emission is regarded as the IR emission) without any specific integration ranges in wavelength. To obtain more observationally relevant values, we calculate the fractional AGN contribution to the total SED between $8-1000~\mathrm{\mu m}$, $f_\mathrm{IR}$, and its error is assumed to be scaled from the error of the output $f_\mathrm{AGN}$, i.e. $\mathrm{Err}\{f_\mathrm{IR}\}=\mathrm{Err}\{f_\mathrm{AGN}\}f_\mathrm{IR}/f_\mathrm{AGN}$. We also derive the rest-frame observed $L_\mathrm{6~\mu m}$ based on the decomposed AGN component. The error of $L_\mathrm{6~\mu m}$ is also estimated by scaling down the error of the AGN dust luminosity, which is an output parameter of \mbox{\texttt{X-CIGALE}}. The results are shown in Table~\ref{SEDPropTable}.\par
We have compared our $L_\mathrm{bol}$ with those in \citet{Toba17}. The $L_\mathrm{bol}$ values in \citet{Toba17} do not include \mbox{X-ray} emission, and thus we subtract the \mbox{X-ray} contributions for the comparison. The mean offset is $-0.24~\mathrm{dex}$ (ours are smaller). Even though the offset is small, it may not just be by chance. The AGN templates used in \citet{Toba17} are mainly type~2 templates from \citet{Silva04}, i.e. the obscuration is mainly from the torus. However, our sources are observed to be type~1 with significant expected polar obscuration. The type of the adopted AGN template, as represented by the viewing angle from the disk axis, may lead to the difference. Indeed, if we adopt type~2 templates (viewing angle = $90\degree$) in \mbox{\texttt{X-CIGALE}}, the offset is significantly smaller ($0.03~\mathrm{dex}$). Since the heavily obscured type~1 templates are more physically reasonable, we adopt the corresponding results here. This comparison indicates that there may be underlying systematic uncertainties (e.g. due to template choices). Though it is hard to quantify the uncertainties, our conclusions are not materially affected when varying the AGN templates used, but we note that the $L_\mathrm{bol}$ values derived with type~2 templates would return higher $\lambda_\mathrm{Edd}$ in Section~\ref{sec: MBH}.

\begin{table*}
\caption{SED fitting results}
\label{SEDPropTable}
\centering
\begin{threeparttable}
\begin{tabular}{ccccccc}
\hline
\hline
SDSS Name & $\mathrm{log}M_\star$ & logSFR & $f_\mathrm{IR}$ & $L_\mathrm{bol}$ & $L_\mathrm{6~\mu m}$\\
 & ($M_\odot$) & ($M_\odot~\mathrm{yr}^{-1}$) & & ($10^{45}~\mathrm{erg~s^{-1}}$) & ($10^{45}~\mathrm{erg~s^{-1}}$)\\
(1) & (2) & (3) & (4) & (5) & (6)\\
\hline
J0756+4432 & $10.59\pm0.02$ & $2.26\pm0.02$ & $(41.9\pm0.1)\%$ & $5.66\pm0.27$ & $1.07\pm0.05$\\
J0833+4508 & $10.26\pm0.10$ & $1.92\pm0.12$ & $(81.1\pm6.6)\%$ & $12.28\pm0.94$ & $2.41\pm0.20$\\
J1010+3725 & $10.24\pm0.05$ & $1.95\pm0.04$ & $(61.8\pm5.0)\%$ & $5.71\pm0.50$ & $1.04\pm0.12$\\
J1028+5011 & $9.92\pm0.08$ & $1.61\pm0.07$ & $(90.7\pm3.0)\%$ & $11.46\pm2.02$ & $1.77\pm0.40$\\
J1042+2451 & $11.21\pm0.13$ & $2.66\pm0.22$ & $(44.6\pm9.4)\%$ & $26.33\pm3.64$ & $3.92\pm0.60$\\
J1210+6105 & $10.77\pm0.03$ & $2.47\pm0.03$ & $(42.5\pm6.1)\%$ & $7.63\pm1.34$ & $1.26\pm0.24$\\
J1235+4827 & $10.82\pm0.05$ & $2.53\pm0.04$ & $(32.2\pm11.3)\%$ & $5.01\pm1.65$ & $0.82\pm0.31$\\
J1248+4242 & $10.52\pm0.11$ & $1.91\pm0.05$ & $(32.2\pm8.6)\%$ & $1.42\pm0.39$ & $0.21\pm0.06$\\
J1324+4501 & $10.50\pm0.13$ & $2.15\pm0.11$ & $(81.7\pm7.6)\%$ & $20.72\pm2.65$ & $2.95\pm0.38$\\
J1513+1451 & $10.82\pm0.08$ & $2.53\pm0.08$ & $(42.5\pm10.6)\%$ & $8.35\pm1.92$ & $1.44\pm0.36$\\
J1525+1234 & $10.44\pm0.06$ & $2.12\pm0.05$ & $(61.1\pm5.4)\%$ & $8.76\pm1.32$ & $1.12\pm0.19$\\
J1531+4533 & $10.00\pm0.17$ & $1.70\pm0.16$ & $(90.7\pm15.2)\%$ & $12.70\pm2.69$ & $2.69\pm0.66$\\
\hline
\hline
\end{tabular}
\begin{tablenotes}
\item
\emph{Notes.} (2) Host stellar mass. (3) Host SFR. (4) the fractional AGN contribution to the IR emission between $8-1000~\mathrm{\mu m}$. (5) AGN bolometric luminosity. (6) Observed AGN monochromatic luminosity at rest-frame $6~\mathrm{\mu m}$. The intervals are for a 68\% confidence level.
\end{tablenotes}
\end{threeparttable}
\end{table*}

\subsection{Virial $M_\mathrm{BH}$ and $\lambda_\mathrm{Edd}$}
\label{sec: MBH}
\subsubsection{Estimation of $M_\mathrm{BH}$ and $\lambda_\mathrm{Edd}$}
\label{sec: estimate_MBH}
$M_\mathrm{BH}$ values for AGNs can be estimated based on their single-epoch spectra by assuming virial equilibrium in the broad-line region (e.g. \citealt{Shen13}). Generally, the $M_\mathrm{BH}$ measurements follow:
\begin{align}
\label{Eq_calcMBH}
\mathrm{log}\frac{M_\mathrm{BH}}{M_\odot}=a+b~\mathrm{log}\frac{L_\mathrm{AGN}(\lambda)}{\mathrm{10^{44}~erg~s^{-1}}}+2~\mathrm{log\frac{FWHM}{km~s^{-1}}},
\end{align}
where $L_\mathrm{AGN}(\lambda)$ is the intrinsic AGN monochromatic luminosity at wavelength $=\lambda$, and $(a, b)$ are constants needing calibration. Here, we use the calibrations adopted in \citet{Shen11} to estimate $M_\mathrm{BH}$: $\lambda=5100$ \AA\ and $(a, b)=(0.910, 0.50)$ for H$\beta$; $\lambda=3000$ \AA\ and $(a, b)=(0.740, 0.62)$ for \iona{Mg}{ii}. Among the $M_\mathrm{BH}$ measurements for our 12 sources, 11 of them are based on \iona{Mg}{ii}, and one (J1010+3725; whose \iona{Mg}{ii} region is not covered by SDSS) is based on H$\beta$.\par
We do not directly use the $M_\mathrm{BH}$ in the SDSS quasar catalogs \citep{Shen11, Kozlowski17} because when measuring $L_\mathrm{AGN}(\lambda)$, they assume that the AGNs are unobscured and outshine their host galaxies. However, our sources are generally heavily obscured AGNs that appear much fainter than the host galaxies in the optical bands, though intrinsically they would outshine the host galaxies (see Section~\ref{sec: SED}). Also, because of their faintness, we cannot decompose the AGN components from their SDSS spectra. Hence, similar to \citet{Wu18}, we rely on SED fitting to measure $L_\mathrm{AGN}(\lambda)$ (see Section~\ref{sec: SED} for more details about SED fitting). \mbox{\texttt{X-CIGALE}} outputs the intrinsic AGN $L_\nu$ at $2500$ \AA\ at viewing angle $=30^\circ$, and then we derive $L_\mathrm{AGN}(\lambda)$ based on the intrinsic SED adopted in \mbox{\texttt{X-CIGALE}}: $L_\mathrm{AGN}(\lambda)=21.1\cos{\theta}(1+2\cos{\theta})\lambda^{-0.5}\nu L_\nu(2500)$, where $\theta$ is the best-fit viewing angle and $\lambda$ is in \AA. The angular dependence is due to the fact that the disk emission is anisotropic, as adopted in \mbox{\texttt{X-CIGALE}}. The $L_\mathrm{AGN}(\lambda)$ values are tabulated in Table~\ref{MBHTable}.\par
Though \citet{Paris17} have cataloged \iona{Mg}{ii} FWHM values for our sources, they did not provide FWHM errors, and they used different algorithms from those in \citet{Shen11} to derive FWHMs. Hence, we decide to refit the optical spectra consistently following a similar method to \citet{Shen11} and \citet{Timlin20}. We use \mbox{\texttt{PyQSOFit}} \citep{Guo18}, a \texttt{Python} software package dedicated to fitting quasar spectra, to conduct the fitting. The lines are fitted locally. For \iona{Mg}{ii}, a pseudo-continuum (i.e. power-law + \iona{Fe}{ii} emission) is used to fit the spectra at rest-frame [2200, 2700] \AA\ and [2900, 3090] \AA\ first. The \iona{Fe}{ii} emission templates are implemented in \mbox{\texttt{PyQSOFit}}. Then we fit the \iona{Mg}{ii} line with three Gaussian lines after subtracting the pseudo-continuum within rest-frame [2700, 2900] \AA\ except for J0833+4508 (see below). We note that this line model is slightly different from that in \citet{Shen11}, who used one narrow Gaussian line ($\mathrm{FWHM<1200~km~s^{-1}}$) and three broad Gaussian lines ($\mathrm{FWHM>1200~km~s^{-1}}$) for the fitting, and their FWHMs were only referred to the broad components. We adopt a different model because the narrow \iona{Mg}{ii} cores are often not explicitly distinguished from the broad component because of limited spectral quality and relatively small line widths. Hence, as suggested in the literature (e.g. \citealt{Bahk19, Le20}), we do not specifically subtract a narrow core. For J0833+4508, \iona{Mg}{ii} clearly presents double peaks, and thus we add another two narrow Gaussian lines to fit two peaks of its \iona{Mg}{ii} and further add two additional narrow Gaussian lines to fit two peaks of its [\iona{Fe}{iv}] doublet (2829.3~\AA\ and 2835.7~\AA; which is rarely seen in typical AGN spectra). We note that the SDSS spectrum of J0833+4508 presents many emission lines, including several strong high-ionization lines (e.g. [\iona{Ne}{v}], [\iona{Fe}{v}], and [\iona{Fe}{vii}]). Therefore, this source may be a Coronal-Line Forest AGN (e.g. \citealt{Rose11, Rose15}), whose torus together with the line of sight may provide a specific geometry that enables us to view the inner wall of the torus \citep{Glidden16}. The detailed analyses of these additional lines, however, are beyond the scope of this work. For H$\beta$ lines, as shown in \citet{Toba17}, all of our sources except J1010+3725 and J1028+5011 only present narrow H$\beta$ lines, and thus we only add broad components for J1010+3725 and J1028+5011. We fit H$\beta$ together with [\iona{O}{iii}]. Their pseudo-continuum window is rest-frame [4435, 4700] \AA\ and [5100, 5535] \AA, and their line-fitting window is rest-frame [4700, 5100] \AA. Each peak of the [\iona{O}{iii}] doublet (i.e. [\iona{O}{iii}]~$\lambda4959$ and [\iona{O}{iii}]~$\lambda5007$) is fitted with one narrow Gaussian core and one broad Gaussian wing representing outflows. A single narrow Gaussian line is used to model H$\beta$ for sources other than J1010+3725 and J1028+5011. For J1010+3725 and J1028+5011, we adopt the model in \citet{Shen11} and use one narrow Gaussian line plus three broad Gaussian lines to fit their H$\beta$ lines. The FWHM values are displayed in Table~\ref{MBHTable}. We compare our FWHM values with those in \citet{Shen11} and \citet{Paris17} and present the best-fit \iona{Mg}{ii} and H$\beta$-[\iona{O}{iii}] spectra in Appendix~\ref{append: optspec}.\par
Again, we note that only two of our sources (J1010+3725 and J1028+5011) have broad H$\beta$ lines (i.e. $\mathrm{FWHM>1000~km~s^{-1}}$), while all the sources show broad \iona{Mg}{ii} lines (except J1010+3725 whose spectrum does not cover the \iona{Mg}{ii} line). Outflows seem to be a straightforward way to explain this phenomenon --- the \iona{Mg}{ii} lines are broadened by outflows, and are thus less affected by absorption, while the virially broadened H$\beta$ lines are obscured. However, there are still several other possible explanations. One possibility is due to ``leaking'' emission, as mentioned in Section~\ref{sec: SED}. The broad \iona{Mg}{ii} lines may be from the leaked UV emission from BLRs, and such leaked emission is much weaker in the optical \citep{Assef16}, which can explain the lack of broad H$\beta$ lines. The other is that our sources may be similar to the observational class of AGNs in \citet{Roig14}, where the \iona{Mg}{ii} line is much stronger than the Balmer lines, though the underlying physical interpretations for this observational class are also unclear.\par

With $L_\mathrm{AGN}(\lambda)$ and FWHM measurements for all sources, we obtain the $M_\mathrm{BH}$ using Eq.~\ref{Eq_calcMBH}, and then the Eddington ratio is
\begin{align}
\lambda_\mathrm{Edd}=\frac{L_\mathrm{bol}/(\mathrm{erg~s^{-1}})}{1.26\times10^{38}M_\mathrm{BH}/M_\odot}.
\end{align}
Similar to \citet{Wu18}, the errors are propagated from $L_\mathrm{bol}$, $L_\mathrm{AGN}(\lambda)$, and FWHM, and do not include possible systematic errors. We warn that the systematic errors may be large, and will discuss this further in Section~\ref{sec: reliab_MBH}. The $M_\mathrm{BH}$ and $\lambda_\mathrm{Edd}$ values are displayed in Table~\ref{MBHTable}. Generally, $\lambda_\mathrm{Edd}$ is between 0.1 and 1.0. The median $\lambda_\mathrm{Edd}$ for our sample is 0.24. As displayed in the table, one (and only one) source, J1028+5011, has both \iona{Mg}{ii}-based $M_\mathrm{BH}$ and H$\beta$-based $M_\mathrm{BH}$ measurements. These two measurements are consistent within $1-2\sigma$. We decide to use its \iona{Mg}{ii}-based $M_\mathrm{BH}$ in this work because its broad H$\beta$ emission is not significant (see Fig.~\ref{HbOIIISpecs}) and has more potential to be confounded by the continuum.\par

\begin{table*}
\caption{$\lambda_\mathrm{Edd}$ measurements and the corresponding underlying parameters}
\label{MBHTable}
\centering
\begin{threeparttable}
\begin{tabular}{cccccc}
\hline
\hline
SDSS Name & line & $L_\mathrm{AGN}(\lambda)$ & FWHM & $\mathrm{log}M_\mathrm{BH}$ & $\lambda_\mathrm{Edd}$\\
 & & ($\mathrm{10^{45}~erg~s^{-1}}$) & ($\mathrm{km~s^{-1}}$) & ($M_\odot$) & \\
(1) & (2) & (3) & (4) & (5) & (6)\\
\hline
J0756+4432 & \iona{Mg}{ii} & $2.30\pm0.11$ & $2269\pm162$ & $8.30\pm0.06$ & $0.23\pm0.04$\\
J0833+4508 & \iona{Mg}{ii} & $5.18\pm0.38$ & $3136\pm5613$ & $8.80\pm1.55$ & $0.16\pm0.56$\\
J1010+3725 & H$\beta$ & $1.83\pm0.15$ & $4144\pm280$ & $8.78\pm0.06$ & $0.07\pm0.01$\\
\multirow{2}{*}{J1028+5011} & \iona{Mg}{ii} & $5.45\pm1.22$ & $4267\pm693$ & $9.08\pm0.15$ & $0.08\pm0.03$\\
& H$\beta$ & $4.18\pm0.94$ & $6030\pm804$ & $9.28\pm0.13$ & $0.05\pm0.02$\\
J1042+2451 & \iona{Mg}{ii} & $12.43\pm1.86$ & $2408\pm387$ & $8.80\pm0.15$ & $0.33\pm0.12$\\
J1210+6105 & \iona{Mg}{ii} & $1.74\pm0.37$ & $1549\pm586$ & $7.89\pm0.33$ & $0.78\pm0.62$\\
J1235+4827 & \iona{Mg}{ii} & $1.04\pm0.41$ & $2602\pm190$ & $8.20\pm0.12$ & $0.25\pm0.11$\\
J1248+4242 & \iona{Mg}{ii} & $0.27\pm0.09$ & $5326\pm4812$ & $8.46\pm0.79$ & $0.04\pm0.07$\\
J1324+4501 & \iona{Mg}{ii} & $4.33\pm0.67$ & $1819\pm837$ & $8.27\pm0.40$ & $0.88\pm0.82$\\
J1513+1451 & \iona{Mg}{ii} & $1.93\pm0.47$ & $2038\pm246$ & $8.15\pm0.12$ & $0.47\pm0.17$\\
J1525+1234 & \iona{Mg}{ii} & $3.78\pm0.77$ & $2085\pm533$ & $8.36\pm0.23$ & $0.31\pm0.17$\\
J1531+4533 & \iona{Mg}{ii} & $5.84\pm0.29$ & $2604\pm218$ & $8.67\pm0.07$ & $0.22\pm0.06$\\
\hline
\hline
\end{tabular}
\begin{tablenotes}
\item
\emph{Notes.} (2) The emission line used to estimate $M_\mathrm{BH}$. (3) Intrinsic AGN monochromatic luminosity at 3000 \AA\ (for \iona{Mg}{ii}) or 5100 \AA\ (for H$\beta$). (4) FWHM of the emission line. (5) Virial black-hole mass. (6) Eddington ratio. The intervals are for a 68\% confidence level.
\end{tablenotes}
\end{threeparttable}
\end{table*}

Considering the measured $\lambda_\mathrm{Edd}$ values, we conclude that they are relatively higher than for typical DOGs, even though there are few direct $\lambda_\mathrm{Edd}$ measurements for other DOG samples in the literature. For individual sources, \citet{Melbourne11} measured $M_\mathrm{BH}$ for four DOGs and obtained the lower limits to be $(1-9)\times10^8~M_\odot$. Based on the luminosity measurements in \citet{Melbourne12}, the inferred upper limit on $\lambda_\mathrm{Edd}$ is $\sim0.2$ (see also \citealt{Wu18}), which is smaller than our median $\lambda_\mathrm{Edd}$. For other sample studies, we can roughly estimate $M_\mathrm{BH}$ based on their stellar masses, $M_\star$, using empirical $M_\mathrm{BH}-M_\star$ relations (e.g. \citealt{Reines15, Sun15, Shankar16}). As an example, we apply this estimation to \citet{Corral16}. In \citet{Corral16}, some DOGs even do not present AGN activity; for those with AGN activity, we obtain the median $\lambda_\mathrm{Edd}$ to be only $\sim0.02$. Therefore, our $\lambda_\mathrm{Edd}$ values appear among the highest in typical DOG samples. Compared with typical quasars, the $\lambda_\mathrm{Edd}$ values of our sources are also higher. We select a luminosity-matched quasar sample from the SDSS DR12 quasar catalog \citep{Paris17, Kozlowski17}, and its median $\lambda_\mathrm{Edd}$ is 0.08, smaller than our median $\lambda_\mathrm{Edd}$.\par
Nevertheless, our $\lambda_\mathrm{Edd}$ values do not appear to be at the topmost level. They are generally smaller than those of Hot DOGs \citep{Wu18, Jun20} and high-redshift quasars (e.g. \citealt{Pons19}). However, given the large statistical uncertainties of the measurements as well as many uncertain factors to be discussed in Section~\ref{sec: reliab_MBH} that may systematically influence the $\lambda_\mathrm{Edd}$ measurements, it is hard to state conclusively that our sources are indeed less extreme.

\subsubsection{The reliability of the $M_\mathrm{BH}$ and $\lambda_\mathrm{Edd}$ measurements}
\label{sec: reliab_MBH}
Here, we discuss the reliability of the $M_\mathrm{BH}$ and $\lambda_\mathrm{Edd}$ measurements.\par
The most important source of uncertainty is the validity of our virial assumption about the origins of the broad lines. Section~\ref{sec: origin_broad} will discuss this issue in detail. Generally, the evidence in Section~\ref{sec: origin_broad} supports that the virial assumption is not severely problematic for our whole sample, but it is hard to quantitatively assess the reliability of the assumption for individual sources.\par
Additionally, the $M_\mathrm{BH}$ values for our sample are based on the single-epoch virial mass method whose uncertainty is large even for unobscured quasars ($\sim0.5~\mathrm{dex}$; e.g. \citealt{Shen13}), and 11 measurements are based on \iona{Mg}{ii} while one is based on H$\beta$. \iona{Mg}{ii} is generally thought to be a suitable line for measuring $M_\mathrm{BH}$ (e.g. \citealt{Wang19}), though some works show that the $M_\mathrm{BH}$ measurements from broad \iona{Mg}{ii} lines may be biased (e.g. \citealt{YangQ19}).\par
The measurements of $L_\mathrm{AGN}(\lambda)$ based on SED fitting may also be uncertain. In the optical band, the contribution of the AGN is generally $\sim2-3~\mathrm{dex}$ fainter than the host galaxy. Hence, the constraint on AGN optical luminosity is mainly from the IR and \mbox{X-ray} photometry. However, unlike Hot DOGs, DOGs do not necessarily dominate over their host galaxies in the IR band (Section~\ref{sec: SED}), and this further makes it difficult to measure $L_\mathrm{AGN}(\lambda)$ accurately.\par
As an independent examination, we calculate the $M_\mathrm{BH}/M_\star$ ratios for our sources, and the median value is 0.007, higher than the typical value in the local universe ($\sim0.002$; e.g. \citealt{Kormendy13}). The $M_\mathrm{BH}/M_\star$ ratio ($\sim0.14$) of J1028+5011 is even comparable to the highest value (1/8) known so far \citep{Trakhtenbrot15}. It is still unclear whether the (Hot) DOG population really has a similar $M_\mathrm{BH}/M_\star$ ratio to normal quasars. For instance, \citet{Narayanan10} predicted that the ratio should be lower than the normal ratio, \citet{Wu18} showed that the ratio is comparable, while \citet{Matsuoka18} and \citet{Fan19} showed that the ratio is much higher. As discussed above, there are also many other factors that may significantly affect the $M_\mathrm{BH}$ measurements, and thus we can hardly know whether these high ratios are real. To say the least, if we assume that the ratios should be similar to or lower than those of normal quasars due to the delayed growth of SMBHs compared to their host galaxies, i.e. the $M_\mathrm{BH}$ values are overestimated, our conclusion about the high-$\lambda_\mathrm{Edd}$ nature is actually enhanced.\par
Therefore, both the $M_\mathrm{BH}$ and $\lambda_\mathrm{Edd}$ measurements are uncertain, and it is difficult to quantitatively assess the possible systematic uncertainties. Thus the exact values and individual measurements should not be over-interpreted. Despite that, it is still likely that our sources have high $\lambda_\mathrm{Edd}$ values generally given several indirect pieces of evidence shown later favoring the idea: they are consistent with the expected $\lambda_\mathrm{Edd}-k_\mathrm{bol}$ relation (Section~\ref{sec: luminrelation}); they present moderate outflows (Section~\ref{sec: lEddNH}); and they are also consistent with the predictions under the co-evolution framework (Section~\ref{sec: lEdd_insights}).

\section{Results}
\label{sec: results}
\subsection{The $N_\mathrm{H}-L_\mathrm{X}$ plane}
\label{sec: NHLX}
\begin{figure}
\resizebox{\hsize}{!}{
\includegraphics{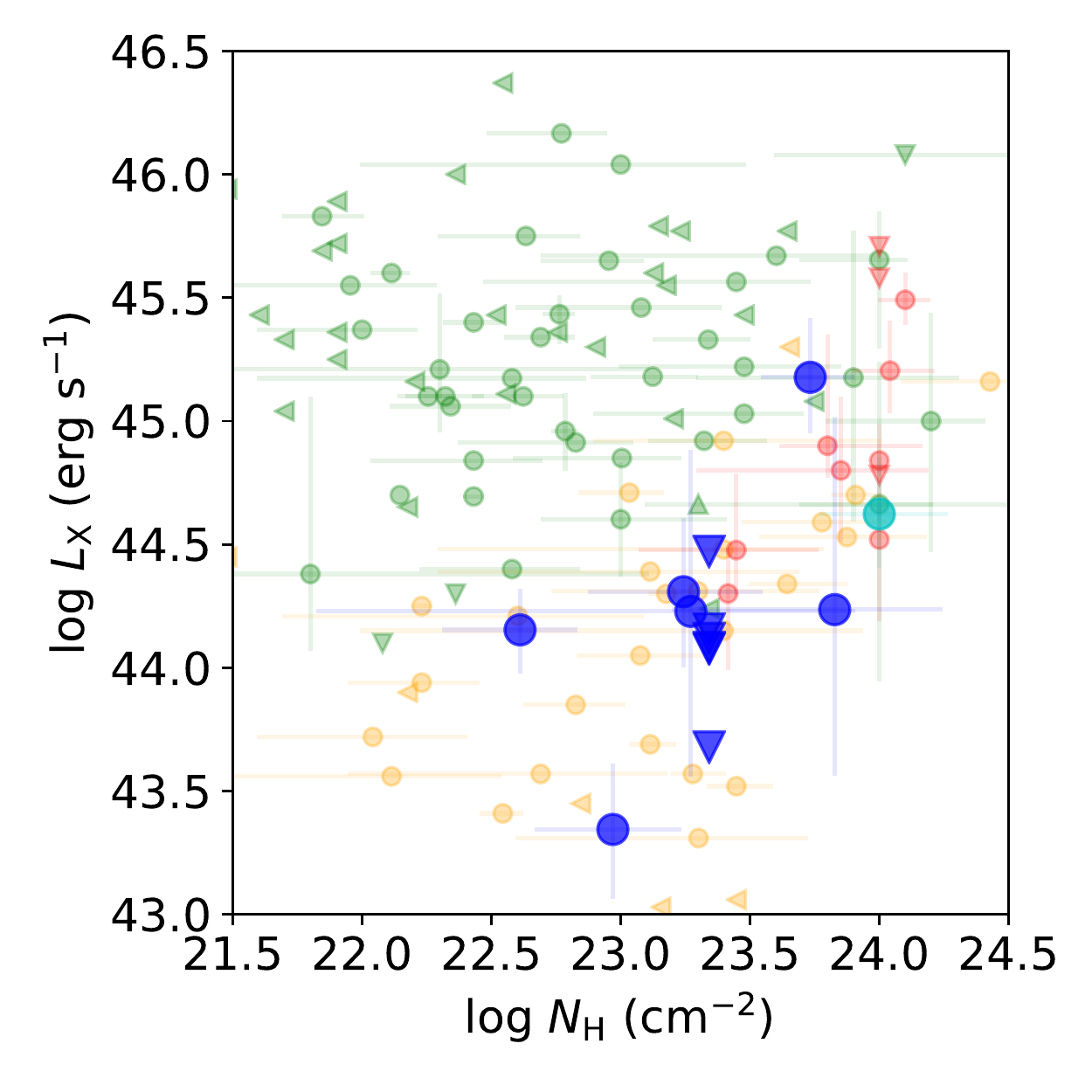}
}
\caption{Comparison between our sample and previous studies in the $N_\mathrm{H}-L_\mathrm{X}$ plane. The blue large circular points are our detected sources with 68\% confidence error bars, and the blue downward triangles at $N_\mathrm{H}=2.2\times10^{23}~\mathrm{cm^{-2}}$ present 90\% $L_\mathrm{X}$ upper limits for the undetected sources. The cyan large point is the high-$\lambda_\mathrm{Edd}$ DOG in \citet{Toba20}. The green, orange, and red points are reddened type~1 quasars, DOGs, and Hot DOGs, respectively. Our sample lies in the regime of moderate luminosity and high obscuration at the edge of the region populated by all the DOGs.}
\label{NH_LX_map_fig}
\end{figure}

We display our sources in the $N_\mathrm{H}-L_\mathrm{X}$ plane in Fig.~\ref{NH_LX_map_fig}. As a comparison, we also plot other kinds of AGNs collected from the literature: the high-$\lambda_\mathrm{Edd}$ DOG in \citet{Toba20}, reddened type~1 quasars \citep{Urrutia05, Martocchia17, Mountrichas17, Goulding18, Lansbury19}, DOGs \citep{Lanzuisi09, Corral16}, and Hot DOGs \citep{Stern14, Assef16, Ricci17, Vito18, Zappacosta18}. The figure shows that our sample and the DOG in \citet{Toba20} lie in a region with high $N_\mathrm{H}$ and moderate $L_\mathrm{X}$, and thus are located near the right edge of the region populated by DOGs, indicating that the physical processes in our DOGs are different from those in less-obscured DOGs. This property will be further discussed in Section~\ref{sec: lEdd_insights}. We also note that if we set the $N_\mathrm{H}$ values of the undetected sources to be as large as $10^{24}~\mathrm{cm^{-2}}$, their $L_\mathrm{X}$ would be $\sim0.7$ dex higher, and thus they will populate the region around the point of \citet{Toba20} in Fig.~\ref{NH_LX_map_fig}, and our conclusion is still unchanged.

\subsection{The relations among $L_\mathrm{X}$, $L_\mathrm{6~\mu m}$, and $L_\mathrm{bol}$}
\label{sec: luminrelation}
\begin{figure}
\resizebox{\hsize}{!}{
\includegraphics{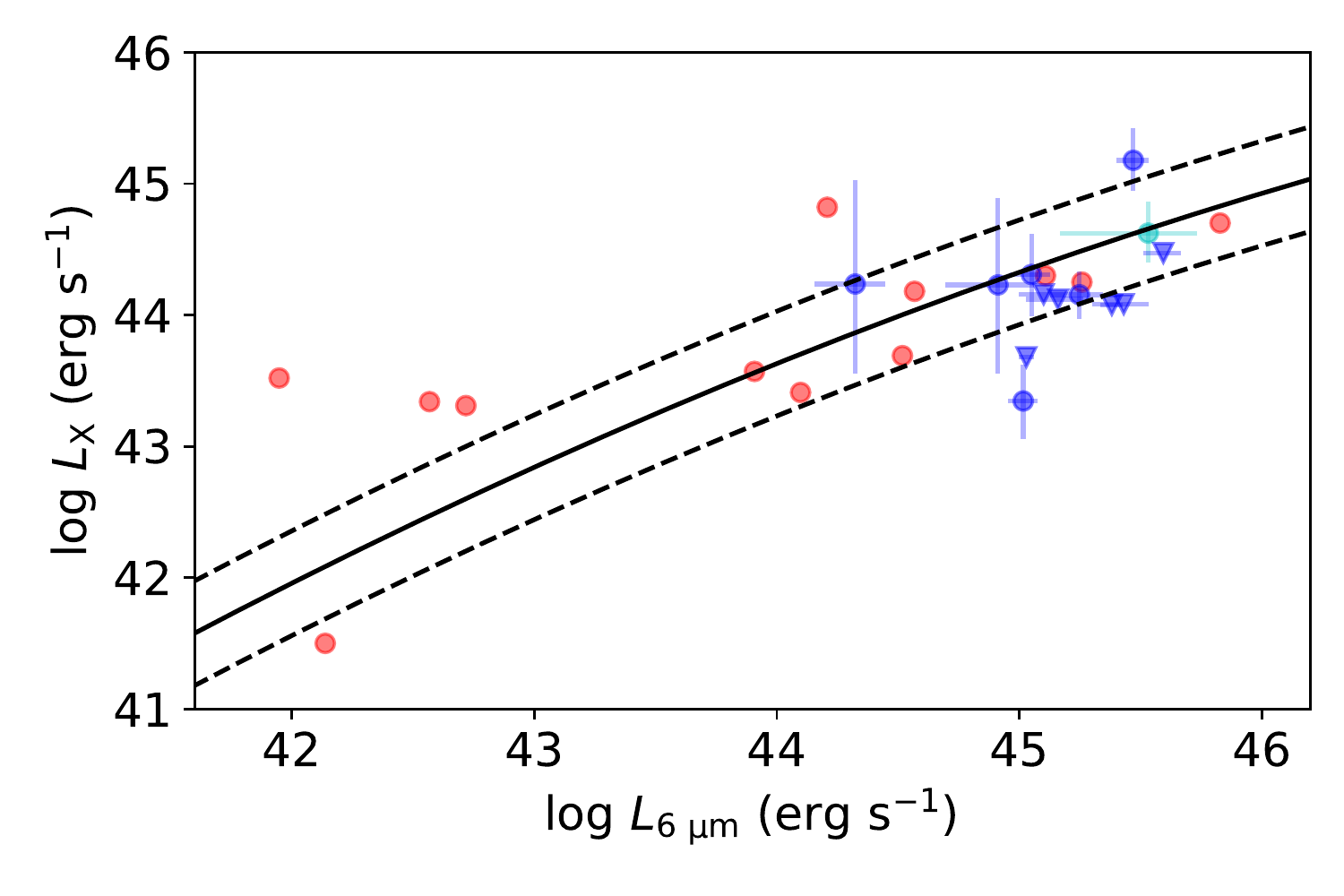}
}
\caption{The relation between $L_\mathrm{X}$ and $L_\mathrm{6~\mu m}$. The blue circular points are our \mbox{X-ray} detected sources, while the blue triangles indicate upper limits of $L_\mathrm{X}$. As a comparison, the sample from \citet{Corral16} is shown as the red points, the high-$\lambda_\mathrm{Edd}$ DOG in \citet{Toba20} is shown as the cyan point, and the relation in \citet{Stern15} as well as its $1~\sigma$ uncertainty are displayed as the black lines. All the samples are generally consistent with the relation.}
\label{L6wLXFig}
\end{figure}

Many works have shown that $L_\mathrm{X}$ is tightly correlated with the MIR luminosity (usually characterized by $L_\mathrm{6~\mu m}$) in AGNs over a wide range of MIR luminosity (between $\sim10^{42}-10^{47}~\mathrm{erg~s^{-1}}$; e.g. \citealt{Lutz04, Fiore09, Gandhi09, Lanzuisi09, Stern15, Chen17}). Generally, the relation is almost linear, but it flattens at high luminosities \citep{Stern15, Chen17}. The reason for using $L_\mathrm{6~\mu m}$ instead of the integral luminosity across the whole IR band is that the former is more representative of the hot-dust emission around AGNs. We display such a relation in Fig.~\ref{L6wLXFig}, in which we also show the relation in \citet{Stern15} with the $1~\sigma$ uncertainty and the DOG sample from \citet{Corral16}. We subtract 0.02 to convert $\mathrm{log}L_\mathrm{12~\mu m}$ in \citet{Corral16} to $\mathrm{log}L_\mathrm{6~\mu m}$. The conversion factor is derived from the AGN template in \citet{Assef10}, and is small. The figure shows that both of the DOG samples are consistent with the relation in \citet{Stern15} within $\lesssim2~\sigma$. The consistency is expected because both $L_\mathrm{X}$ and $L_\mathrm{6~\mu m}$ are from the same AGN component, and the absorbed AGN emission will be re-emitted in the IR bands. It also serves as independent evidence that the decomposition of the SEDs in Section~\ref{sec: SED} is generally reliable. We note that it is important to use the $L_\mathrm{6~\mu m}$ purely contributed by AGN emission, instead of the total $\mathrm{6~\mu m}$ emission, because contamination from the host galaxies may be large in DOGs \citep{Corral16}. Indeed, the median contribution of the host galaxies at $6~\mathrm{\mu m}$ in our sample is 54\%.\par

\begin{figure*}
\resizebox{\hsize}{!}{
\includegraphics{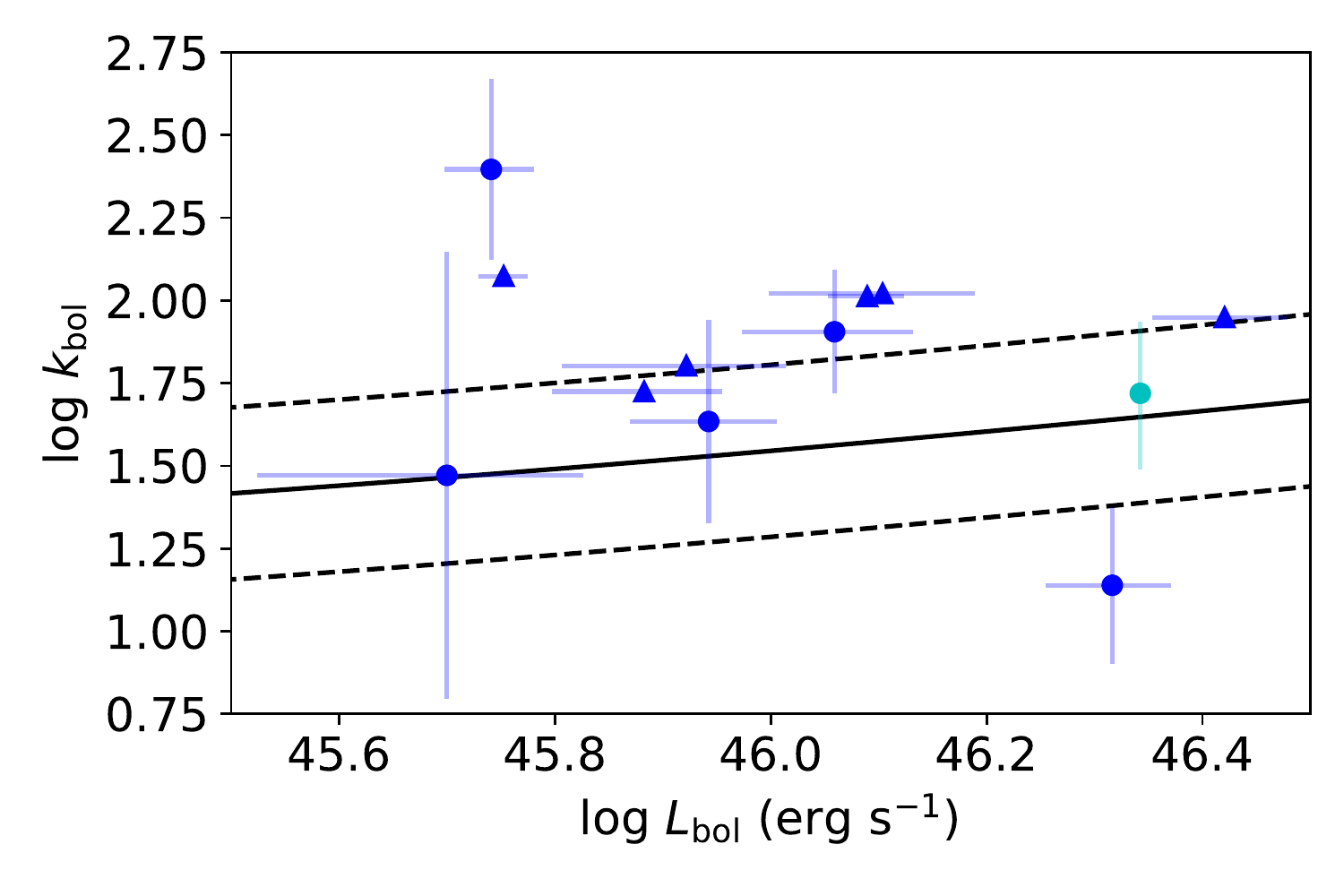}
\includegraphics{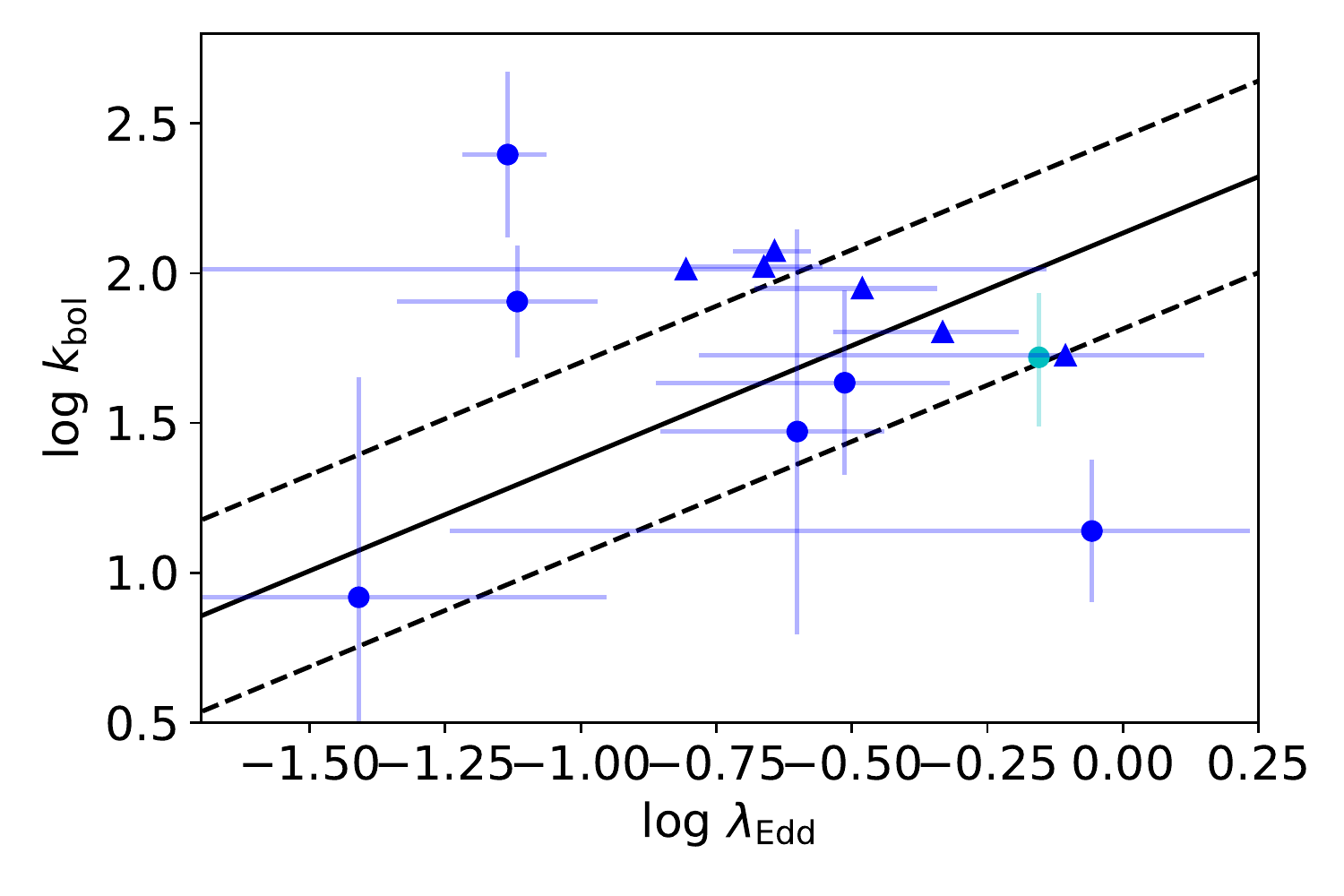}
}
\caption{The bolometric correction $k_\mathrm{bol}$ versus $L_\mathrm{bol}$ (left) and $\lambda_\mathrm{Edd}$ (right). The blue triangles indicate lower limits of $k_\mathrm{bol}$. The cyan point is the high-$\lambda_\mathrm{Edd}$ DOG in \citet{Toba20}. The black lines are the relations in \citet{Lusso12} with $1~\sigma$ uncertainties. These sources are generally consistent with the standard relations.}
\label{kbolwLbolFig}
\end{figure*}

Previous works also found that the \mbox{X-ray} bolometric correction, $k_\mathrm{bol}$, defined as $L_\mathrm{bol}/L_\mathrm{X}$, depends on both $L_\mathrm{bol}$ (e.g. \citealt{Hopkins07, Marconi04}) and $\lambda_\mathrm{Edd}$ (e.g. \citealt{Vasudevan09, Lusso10, Lusso12}). We show the relations for our sample in Fig.~\ref{kbolwLbolFig}, in which we also display the empirical relations in \citet{Lusso12} as a comparison. Our sample is generally consistent with both of the two relations within $\lesssim2~\sigma$. Especially, the $\lambda_\mathrm{Edd}-k_\mathrm{bol}$ relation indicates that the measured $\lambda_\mathrm{Edd}$ values are generally good. We also note that our sample seems to deviate more from the expected $L_\mathrm{bol}-k_\mathrm{bol}$ relation than from the $\lambda_\mathrm{Edd}-k_\mathrm{bol}$ relation. This may be at least partly because the $\lambda_\mathrm{Edd}$ values of the sample objects used in \citet{Lusso12} are generally lower than ours, and sources with lower $\lambda_\mathrm{Edd}$ tend to have lower $k_\mathrm{bol}$ because the coronal power may become stronger relative to the disk power when $\lambda_\mathrm{Edd}$ decreases (e.g. \citealt{Cao09}).

\subsection{Outflow}
\label{sec: lEddNH}
\citet{Toba17} showed that our sources have outflows, as manifested in their [\iona{O}{iii}] profiles. We analyze this property from an X-ray point of view here.
\subsubsection{The $\lambda_\mathrm{Edd}-N_\mathrm{H}$ plane}
\label{sec: lEddnH_plane}
The presence of outflows is expected for sources with effective Eddington ratios above 1 (e.g. \citealt{Ishibashi18}). Basically, the obscuring material in AGNs has to be massive enough that the gravity from the central SMBH can resist the radiation pressure; otherwise, the material will be blown out and form outflows. Therefore, there may be an outflow (or ``forbidden'') region in which long-lived obscuring clouds cannot survive in the $\lambda_\mathrm{Edd}-N_\mathrm{H}$ plane (e.g. \citealt{Fabian08, Fabian09, Kakkad16, Ishibashi18}), and thus the outflow region tends to have relatively lower $N_\mathrm{H}$ than the allowed region. Hence, the location of a source in the plane may serve as an indicator for the existence of outflows. Also, the absorption cross-section of dust integrated over all wavelengths is much larger than that of gas, and thus radiation trapping by dust can reshape the forbidden region significantly.\par
\begin{figure}
\resizebox{\hsize}{!}{
\includegraphics{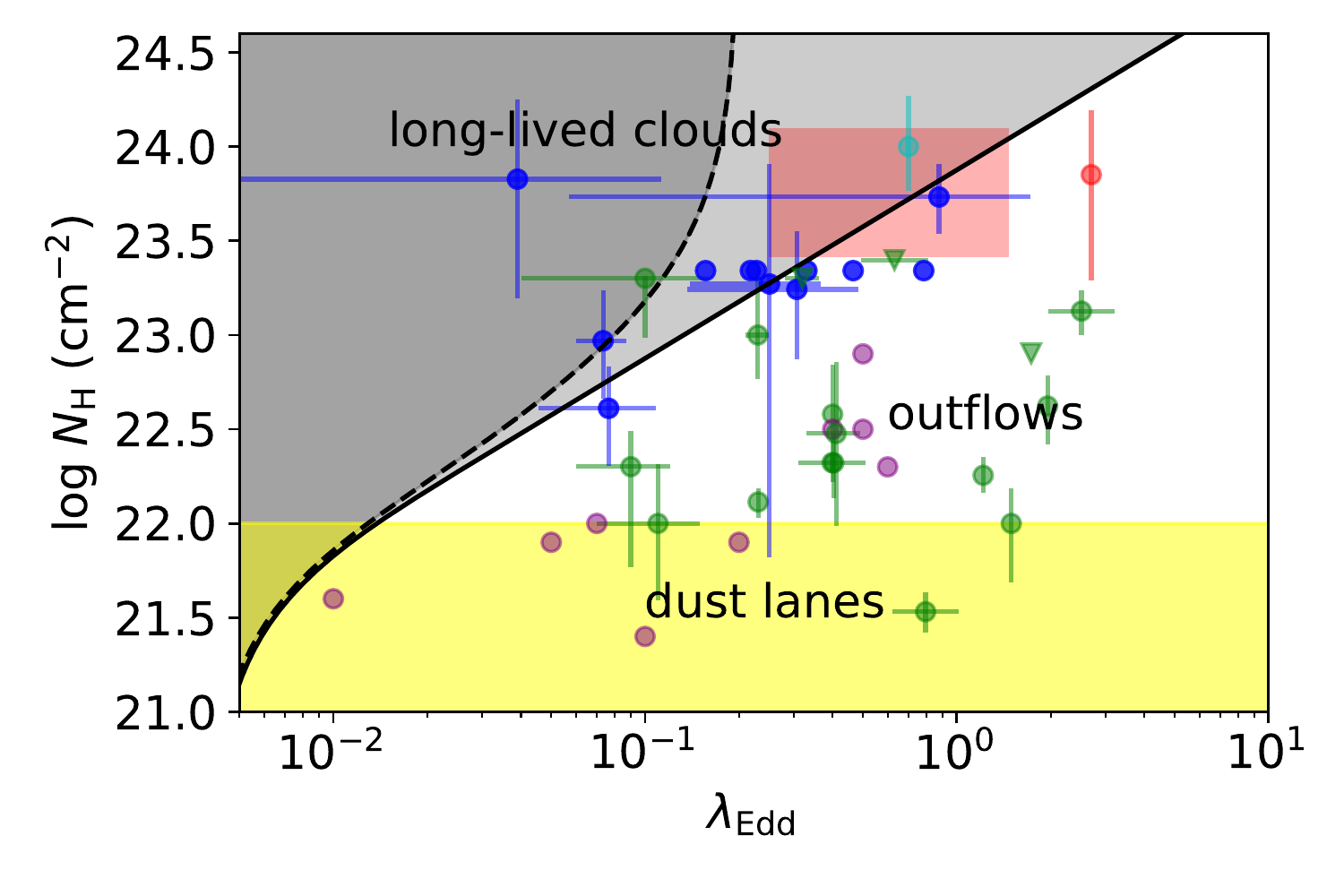}
}
\caption{Our sample in the $\lambda_\mathrm{Edd}-N_\mathrm{H}$ plane. The blue points are our sources. The cyan point is the high-$\lambda_\mathrm{Edd}$ DOG in \citet{Toba20} The purple points are the sources with strong outflows. The green points are reddened quasars. The red point is a Hot DOG \citep{Ricci17}, and the red shaded region is the expected region for Hot DOGs \citep{Vito18, Wu18}. The grey shaded region is the allowed region for long-lived clouds, and the boundaries between the allowed region and the forbidden region are the black solid line at the single-scattering limit and the dashed line at the radiation-trapping limit \citep{Ishibashi18}. The yellow region is the plausible one for obscuration caused by the host galaxy. Both our high-$\lambda_\mathrm{Edd}$ DOGs and Hot DOGs are closer to the boundary between the allowed region and the forbidden region than those with strong outflows.}
\label{lambdaEdd_NH_fig}
\end{figure}

We display our sample in the $\lambda_\mathrm{Edd}-N_\mathrm{H}$ plane in Fig.~\ref{lambdaEdd_NH_fig} together with the boundaries at the full radiation-trapping limit and the single-scattering limit \citep{Ishibashi18}, in which we assume $N_\mathrm{H}=2.2\times10^{23}~\mathrm{cm^{-2}}$ for the six sources without $N_\mathrm{H}$ measurements. We also display the high-$\lambda_\mathrm{Edd}$ DOG in \citet{Toba20}, sources with strong outflows \citep{Brusa15, Kakkad16}, reddened quasars \citep{LaMassa16, Glikman17, LaMassa17, Lansbury19}, and Hot DOGs \citep{Ricci17, Vito18, Wu18}. We warn that \citet{Jun20} pointed out that the $\lambda_\mathrm{Edd}$ measurements of Hot DOGs might not be reliable due to the same ambiguity in this paper about the origins of the broad lines. Neglecting this issue, the figure shows that our sources and Hot DOGs are generally just beside the boundary between the allowed region and the outflow region. The high-$\lambda_\mathrm{Edd}$ DOG in \citet{Toba17} has slightly larger $\lambda_\mathrm{Edd}$ and $N_\mathrm{H}$ values than our sources. Though it is within the allowed region at the single-scattering limit, it is in the outflow region at the radiation-trapping limit. Hence, it can still be regarded as at the near boundary of the allowed region. In contrast, the sample with strong outflows and reddened quasars which are also expected to present strong outflows \citep{Lansbury19} lie much further away from the boundary. Therefore, we would expect our sources to present only moderate outflows. Indeed, this is verified through the [\iona{O}{iii}] width. Though the definition of the [\iona{O}{iii}] width varies among the literature, simple comparisons can still be made in a basic manner. As shown in \citet{Brusa15} and \citet{Lansbury19} (see also \citealt{Temple19}), their sources typically have [\iona{O}{iii}] widths larger than $\sim1000~\mathrm{km~s^{-1}}$, but our sources generally do not broaden the [\iona{O}{iii}] lines by over $\sim1000~\mathrm{km~s^{-1}}$ (\citealt{Toba17}; see also Section~\ref{sec: outflow_strength}).\par

On the other hand, we can consider the relation among $N_\mathrm{H}$, $\lambda_\mathrm{Edd}$, and the presence of outflows in a different way. If we believe that $N_\mathrm{H}$ is high and our sources have outflows, then based on Fig.~\ref{lambdaEdd_NH_fig}, the $\lambda_\mathrm{Edd}$ values should be $\gtrsim0.1$ to ensure our sources enter the outflow region, supporting our basic assumption that our sources are high-$\lambda_\mathrm{Edd}$ DOGs.

\subsubsection{Outflow strength}
\label{sec: outflow_strength}
Here, we use the following quantity defined in \mbox{\citet{Toba17}} (see their Eq. 7) to indicate the outflow strength:
\begin{align}
\sigma_0=\sqrt{v_{\rm [O III]}^2+\sigma_{\rm [O III]}^2},
\end{align}
where $v_{\rm [O III]}$ and $\sigma_{\rm [O III]}$ are the velocity offset and dispersion of [\iona{O}{iii}], respectively. $\sigma_0$ is purely derived from the SDSS spectra, and roughly equals one half of the intrinsic bulk velocity of the outflows. We do not use other physical quantities (e.g. outflow energy injection rate) as indicators of the outflow strength because estimating them requires many strong assumptions (e.g. \citealt{Fiore17, Toba17}).\par
\begin{figure}
\resizebox{\hsize}{!}{
\includegraphics{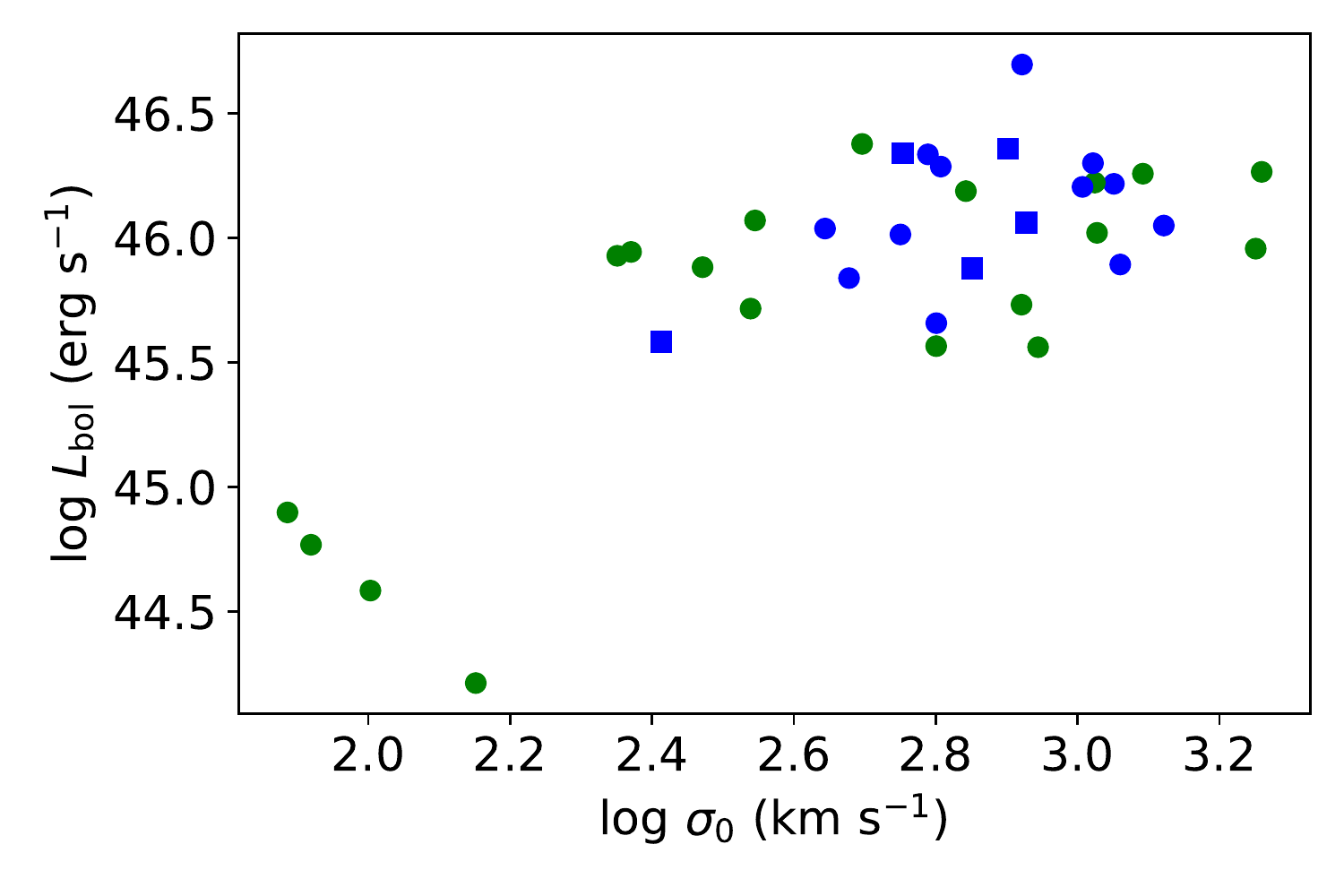}
}
\caption{$L_\mathrm{bol}$ versus $\sigma_0$ for our parent sample \citep{Toba17}. The blue circular points are our sources, the blue squares are other broad-line sources not included in our sample, and the green points are the remaining sources in \citet{Toba17}. Broad-line sources do not populate a significantly different region from the parent sample.}
\label{outfwlowLbolFig}
\end{figure}

To compare our broad-line DOGs with DOGs without broad lines, we display $L_\mathrm{bol}$ against $\sigma_0$ for our parent sample \citep{Toba17} in Fig.~\ref{outfwlowLbolFig}. Note that the $L_\mathrm{bol}$ values in the figure are from \citet{Toba17} instead of ours to prevent the influence of the systematic offset between the two $L_\mathrm{bol}$ measurements. The figure indicates that broad-line DOGs are not significantly different from the rest of the parent sample, indicating that the presence of broad \iona{Mg}{ii} or H$\beta$ lines is independent from the outflow strength.

\begin{figure}
\resizebox{\hsize}{!}{
\includegraphics{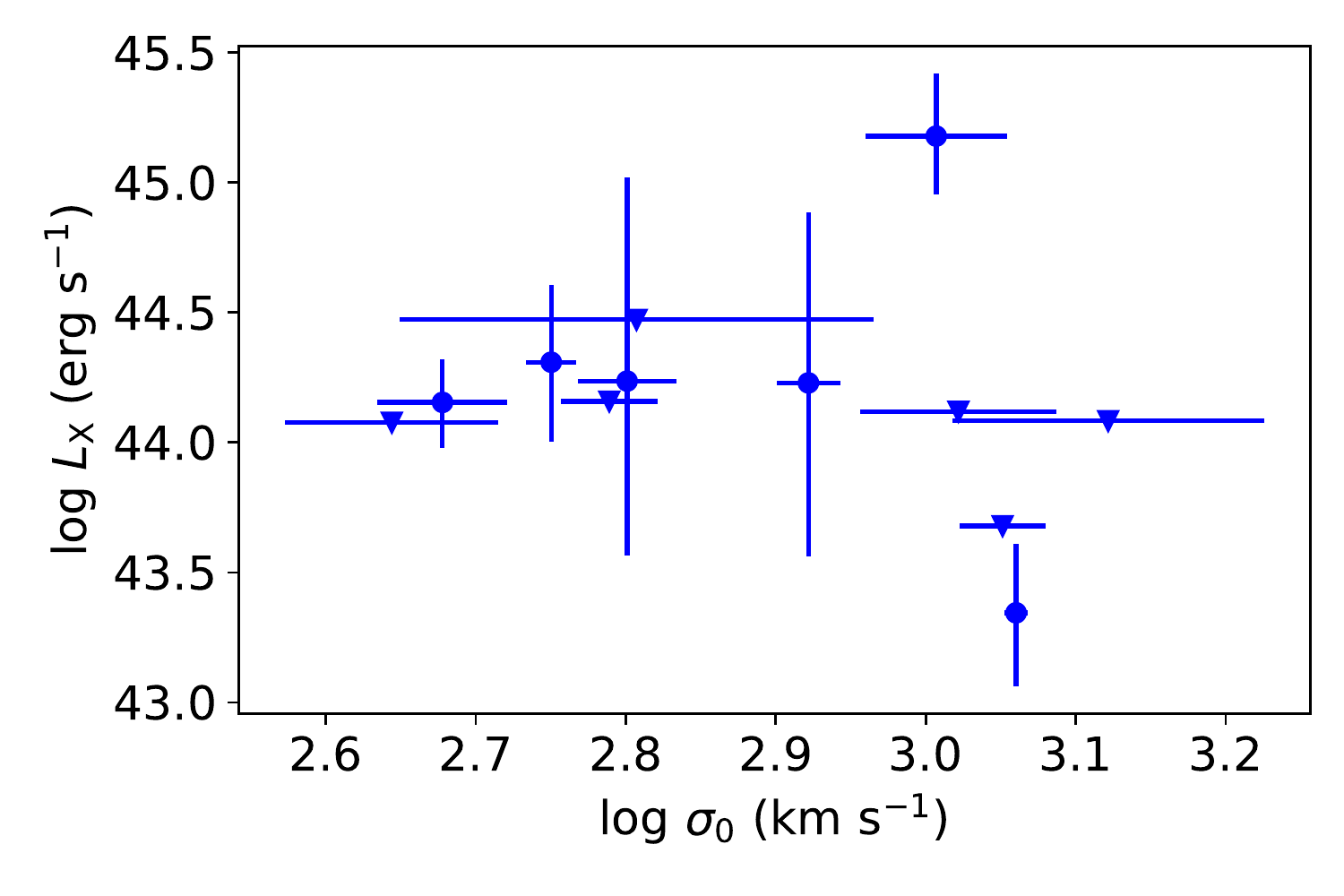}
}
\caption{$L_\mathrm{X}$ versus $\sigma_0$ for our sources. There is no apparent relation between the two variables.}
\label{outflow_XrayFig}
\end{figure}

\citet{Toba17} found that $\sigma_0$ is correlated with several AGN properties for their sample (i.e. our parent sample). Here we examine the relation between $\sigma_0$ and $L_\mathrm{X}$. We plot $L_\mathrm{X}$ versus $\sigma_0$ in \mbox{Fig.~\ref{outflow_XrayFig}}. There is not a significant association between the two variables, and the $p$-value of the generalized Kendall's $\tau$ test for this censored dataset is 0.56, confirming the non-correlation. However, this analysis may be highly limited by the sample size and, maybe more importantly, the $L_\mathrm{X}$ range of our data. Some previous works (e.g. \citealt{Harrison16, Perna17, Chen19}) did find a positive relationship between the outflow strength and $L_\mathrm{X}$, but their data generally cover a much wider $L_\mathrm{X}$ range. Also, our sources are generally consistent with their relations at our $L_\mathrm{X}$ values. For instance, \citet{Perna17} showed that the [\iona{O}{iii}] outflow width would be $\sim800~\mathrm{km~s^{-1}}$ at $L_\mathrm{X}\sim10^{44}~\mathrm{erg~s^{-1}}$ for typical quasars, and these values are consistent with our results. A more quantitative comparison would require much additional work to bridge the difference among the detailed definitions of the outflow width in the literature, but this is beyond the scope of this work.

\subsection{Host-galaxy properties}
\label{sec: host_MS}
In the $M_\star$-SFR plane, star-forming galaxies primarily occupy a nearly linear region called the main sequence (MS; e.g. \citealt{Whitaker12, Speagle14}), which evolves with redshift. The upward outliers from the MS are called starburst galaxies (e.g. \citealt{Rodighiero11}) such as some ULIRGs and submilimeter galaxies, and may be triggered by major mergers.\par
\begin{figure}
\resizebox{\hsize}{!}{
\includegraphics{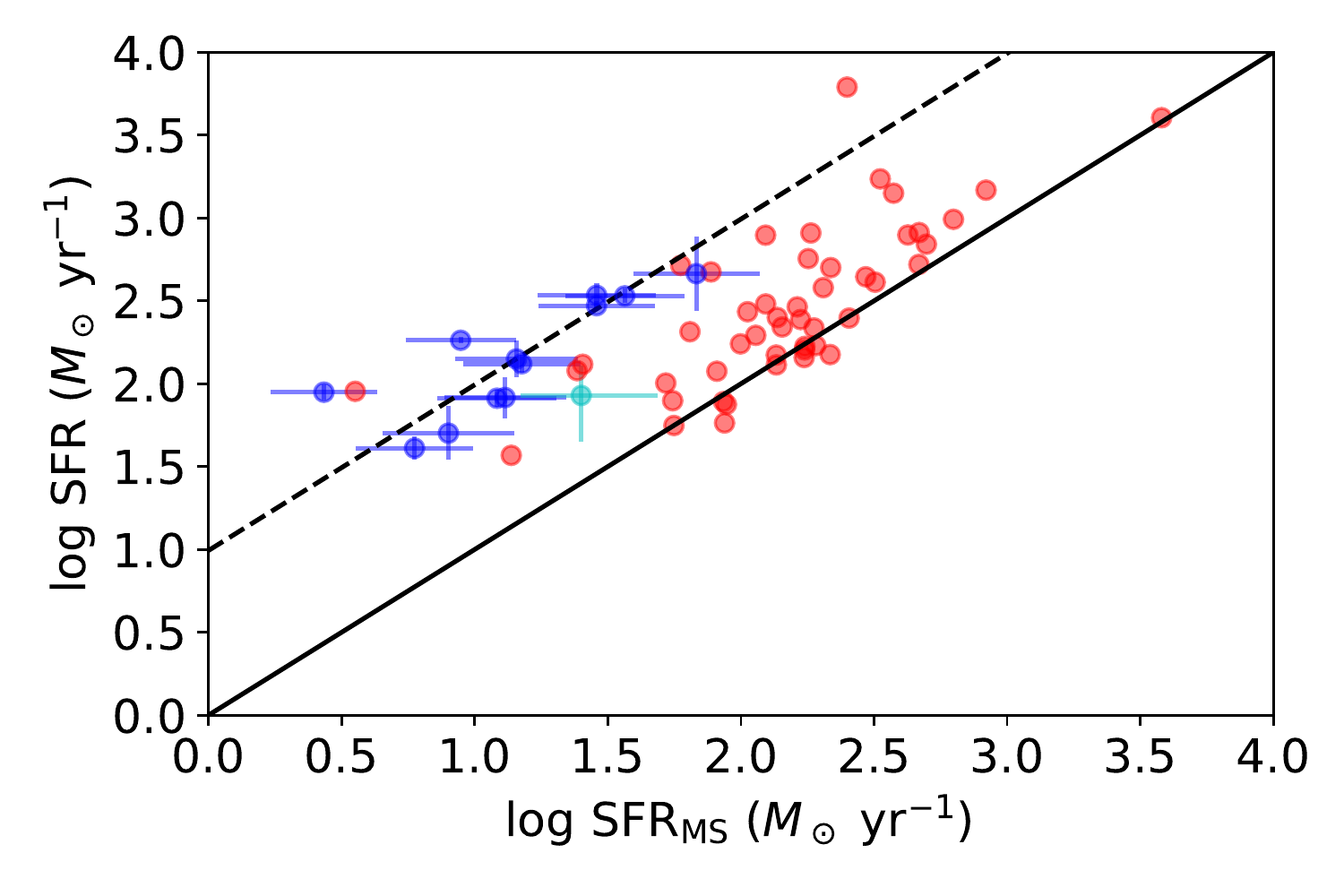}
}
\caption{Our SFRs versus the MS-predicted SFRs based on \citet{Speagle14}. The blue points are our sources, in which the error bars also include the dispersion of the MS. The cyan point is the high-$\lambda_\mathrm{Edd}$ DOG in \citet{Toba20}. The red points are the DOG sample from \citet{Corral16}. The black solid line is a one-to-one relation, and the black dashed line indicates the mean offset of our sample from the MS. High-$\lambda_\mathrm{Edd}$ DOGs are undergoing intenser starbursts then those in \mbox{\citet{Corral16}}.}
\label{MstarwSFRFig}
\end{figure}

In Fig.~\ref{MstarwSFRFig}, we compare our derived SFRs with the MS-predicted SFRs \citep{Speagle14}, in which the uncertainty of the MS ($\sim0.2~\mathrm{dex}$) is included in the error bars. We also display the DOGs from \citet{Corral16} and \citet{Toba20} in the figure. As the figure shows, our sources are about 10 times above the MS, and thus are undergoing intense starbursts. Though the sources in \citet{Corral16} are also above the MS, they are less extreme. The DOG in \citet{Toba20} is also further away from the MS line than those in \citet{Corral16}. This indicates that high-$\lambda_\mathrm{Edd}$ DOGs are closer to the peak of galaxy growth. We will further discuss this result in Section~\ref{sec: lEdd_insights}.

\section{Discussion}
\label{sec: discussion}
\subsection{Physical insights into high-$\lambda_\mathrm{Edd}$ DOGs}
\label{sec: lEdd_insights}
In Fig.~\ref{NH_LX_map_fig}, we compare the location of our sample in the $N_\mathrm{H}-L_\mathrm{X}$ plane with reddened type~1 quasars, Hot DOGs, and other DOGs. The Hot DOGs show higher \mbox{X-ray} obscuration than the reddened quasars \citep{Vito18}. This is likely because Hot DOGs are at the peak of SMBH accretion when the feedback has not yet swept away the reservoir of surrounding gas, while reddened quasars are generally at the post-merger phase and the feedback is already ongoing. However, reddened quasars still show different levels of \mbox{X-ray} obscuration because their evolutionary stages are heterogeneous, and some of them may be just after the peak of SMBH accretion and thus are similar to Hot DOGs (e.g. \citealt{Goulding18}). Our sources are less massive than these two populations, and thus are less luminous, but are still highly obscured. Compared with other DOG samples \citep{Lanzuisi09, Corral16} in which $\lambda_\mathrm{Edd}$ is not taken into consideration, our sample has similar $L_\mathrm{X}$ but is located in the upper part of the $N_\mathrm{H}$ distribution. This is because their DOGs are probably a heterogeneous population with SF-dominated sources and sources with AGN contributions at different $\lambda_\mathrm{Edd}$ levels that are caught at different evolutionary stages. Our sources, more like Hot DOGs, are at the peak of their evolution. Indeed, AGNs have more contribution in our sample than theirs. The mean fractional contribution of AGN at $12~\mathrm{\mu m}$ for the DOG sample in \citet{Corral16} is 15\%, and many of their sources even do not present AGN activity. Such a value is much less than ours (42\% at $12~\mathrm{\mu m}$ and 53\% between $8-1000~\mathrm{\mu m}$). We also note that our sources are somewhat less obscured than Hot DOGs in Fig.~\ref{NH_LX_map_fig}. That may be because the $\lambda_\mathrm{Edd}$ values of our sources are generally smaller than those of Hot DOGs \citep{Wu18, Jun20}.\par
The outflow and host-galaxy properties provide further evidence for the scenario mentioned in the previous paragraph. As shown in Section~\ref{sec: lEddNH}, in the $\lambda_\mathrm{Edd}-N_\mathrm{H}$ plane (Fig.~\ref{lambdaEdd_NH_fig}), high-$\lambda_\mathrm{Edd}$ DOGs lie just beside the boundary between the forbidden region and the allowed region, which is also seen in Hot DOG samples. In contrast, reddened quasars are further away from the boundary, and thus present stronger outflows. This is because our sources and Hot DOGs are still entering the blow-out phase, and the AGN feedback is not as strong as that in reddened quasars. Based on SED fitting, we find that the host galaxies of our high-$\lambda_\mathrm{Edd}$ DOGs are undergoing intense starbursts, and the deviation from the MS is much more significant than for other less-extreme DOGs (Section~\ref{sec: host_MS}). This contrast is also seen between Hot DOGs and reddened quasars. Previous works found that Hot DOGs were also undergoing strong starbursts, with SFRs reaching up to $10^2-10^4~M_\odot~\mathrm{yr^{-1}}$ (e.g. \citealt{Eisenhardt12, Fan16, Fan18, Diaz-Santos18, Assef19}). In contrast, powerful reddened quasars, which have similar luminosity to Hot DOGs, show significantly lower host SFRs than Hot DOGs ($\lesssim\mathrm{a~few}~10^{2}~M_\odot~\mathrm{yr^{-1}}$; e.g. \citealt{AlaghbandZadeh16}). According to simulations (e.g. \citealt{Narayanan10}), the peaks of star formation and SMBH accretion are close in time, and thus the difference in host-galaxy star formation can be explained under this co-evolution framework.\par
However, even at the peak of SMBH accretion, our sources still do not significantly outshine their host galaxies in the IR band. The distribution of $f_\mathrm{IR}$ spans over a wide range ($32\%-91\%$) with a median value at 53\%. Therefore, the contribution from galaxies is still an important factor even in high-$\lambda_\mathrm{Edd}$ DOGs. Hot DOGs, on the other hand, are dominated by AGNs in the IR band. Also, the IR luminosities of DOGs are generally smaller than those of Hot DOGs, and thus DOGs' dust is cooler because the temperature and IR luminosity are positively correlated (e.g. \citealt{Liang19, Toba19}). Due to the strict selection requirements of Hot DOGs, even the most extreme DOGs may still be unable to be regarded as Hot DOGs (e.g. \citealt{Toba18}). Therefore, even though the evolutionary phases of high-$\lambda_\mathrm{Edd}$ DOGs and Hot DOGs are somewhat similar, their observational characteristics are different.
\subsection{Origins of the broad \iona{M\lowercase{g}}{ii} and H$\beta$ lines}
\label{sec: origin_broad}
So far, we have generally assumed that the broad \iona{Mg}{ii} and H$\beta$ lines are broadened by virial motions of BLRs. In this subsection, we focus on assessing the validity of this virial assumption.\par
\begin{figure*}
\resizebox{\hsize}{!}{
\includegraphics{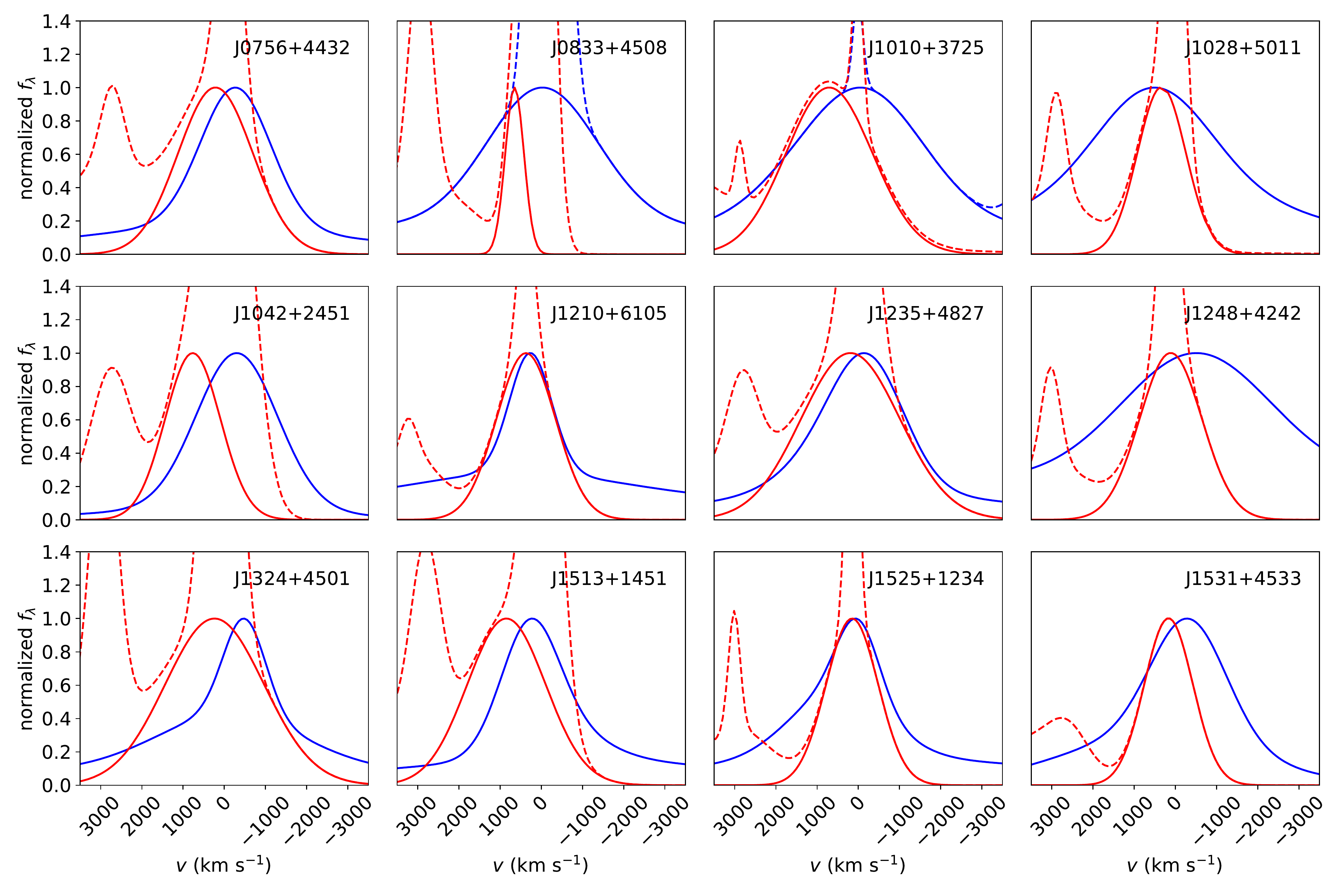}
}
\caption{Comparisons between the broad \iona{Mg}{ii} or H$\beta$ line profiles and the broad [\iona{O}{iii}]~$\lambda5007$ profiles in velocity space, where the velocity is blueshifted velocity compared to the rest wavelength of the corresponding line. Red represents [\iona{O}{iii}], and blue represents \iona{Mg}{ii} for sources other than J1010+3725 or H$\beta$ for J1010+3725. The dashed and solid lines are total (broad + narrow) and broad emission line profiles, respectively. All the spectra are normalized so that the maximums of the solid lines are unity. Note that the red peaks at $\sim3000~\mathrm{km~s^{-1}}$ are [\iona{O}{iii}]~$\lambda4959$.}
\label{complineFig}
\end{figure*}

Following \citet{Jun20}, we compare the broad \iona{Mg}{ii} and H$\beta$ profiles with the outflow-broadened [\iona{O}{iii}] profiles for our sources. Based on the fitting in Section~\ref{sec: estimate_MBH} and Appendix~\ref{append: optspec}, we plot the broad lines (\iona{Mg}{ii} or H$\beta$) used to measure $M_\mathrm{BH}$ in Section~\ref{sec: estimate_MBH}, together with [\iona{O}{iii}] in Fig.~\ref{complineFig}. As the figure shows, in most cases, the broad \iona{Mg}{ii} or H$\beta$ components are broader than the broad [\iona{O}{iii}]~$\lambda5007$ components, but the extent of the difference varies significantly from source to source. Most sources (e.g. J1028+5011) show drastic differences in the widths, but \iona{Mg}{ii} of some sources (e.g. J1235+4827) is comparable with the broad [\iona{O}{iii}] component in width. This phenomenon is different from the finding in \citet{Jun20} for Hot DOGs. They found that all of their broad lines (H$\alpha$, H$\beta$, \iona{Mg}{ii}, and \iona{C}{iv}) had comparable widths with corresponding [\iona{O}{iii}] lines, and thus argued the broad lines might be broadened by outflows. Therefore, the origins of our broad lines may be more complicated, and outflows may only drive the line broadening for a small part of our sample if we assume that outflow-broadened \iona{Mg}{ii} or H$\beta$ lines must have comparable widths with [\iona{O}{iii}] lines.\par
On the other hand, it is possible that the different line widths are just due to a gradient of the outflow velocity in the outflow region. The \iona{Mg}{ii} outflow, if exiting, may not happen at the same radius as the [\iona{O}{iii}] outflow, because radiating \iona{Mg}{ii} requires a much higher density. In the literature, the outflow velocity is often assumed to be constant at different radii (e.g. \citealt{Kakkad16, Perna17}). If this is correct, the difference in the line widths would disfavor the outflow mechanism. However, this assumption itself has not been well tested, and researchers often have to accept it because the outflow geometry is poorly understood (e.g. \citealt{Carniani15}). Therefore, more detailed understanding about the outflow geometry is needed to understand fully the contributions of outflows to the \iona{Mg}{ii} or H$\beta$ broadening and how they influence $M_\mathrm{BH}$ measurements. Nevertheless, even if there is a radial gradient of the outflow velocity, this can hardly explain the significant variation of the difference in the line widths among different sources, and thus is perhaps disfavored. Given all these uncertainties, we do not discriminate which sources have outflow-driven broad \iona{Mg}{ii} or H$\beta$ lines based on their differences in the line widths here, since there is not any reasonable theoretical guideline for judgment.\par

We also found that the \iona{Mg}{ii} FWHMs of our sources are generally lower than those of normal quasars. We select a luminosity-matched quasar sample from \citet{Paris17}, and its median \iona{Mg}{ii} FWHM is $\sim3600~\mathrm{km~s^{-1}}$, while the median value of our sample is only $\sim2500~\mathrm{km~s^{-1}}$. This can be explained if the outflows dominate the line broadening because this is independent of the virial broadening. However, it is also possible that the virial broadening of DOGs is intrinsically lower than normal quasars, as it is expected that SMBHs in obscured DOG phases are less massive than those of luminous type 1 quasars.\par
Overall, the aforementioned facts cannot fully rule out the virial assumption nor the outflow-broadening mechanism, but Fig.~\ref{complineFig} somewhat favors the virial assumption. Besides, we do have some indirect pieces of evidence supporting the $\lambda_\mathrm{Edd}$ measurements and hence the virial assumption, as summarized in the last paragraph of Section~\ref{sec: reliab_MBH}. Thus though the virial assumption is not conservative, it is still reasonable.

\section{Summary and future prospects}
\label{sec: sum_future}
In this work, we analyze the properties of 12 DOGs at $0.3\lesssim z\lesssim1.0$ with broad \iona{Mg}{ii} or H$\beta$ lines based on \textit{Chandra} snapshot observations as well as other multi-wavelength data, assuming that the broad \iona{Mg}{ii} and H$\beta$ lines are from virial motions of BLRs. The main conclusions are the following:
\begin{enumerate}
\item
Our sources are generally moderately luminous ($L_\mathrm{X}\lesssim10^{45}~\mathrm{erg~s^{-1}}$) and highly obscured ($N_\mathrm{H}\gtrsim10^{23}~\mathrm{cm^{-2}}$). In the $N_\mathrm{H}-L_\mathrm{X}$ plane, our sources lie at the upper tip of other DOGs in terms of $N_\mathrm{H}$. See Section~\ref{sec: XSrcProp} and Section~\ref{sec: NHLX}.
\item
The $L_\mathrm{X}-L_\mathrm{6~\mu m}$, $k_\mathrm{bol}-L_\mathrm{bol}$, and $k_\mathrm{bol}-\lambda_\mathrm{Edd}$ relations are all consistent with the well-established ones in the literature, indicating that the SED decompositions (Section~\ref{sec: SED}) and $\lambda_\mathrm{Edd}$ measurements (Section~\ref{sec: MBH}) for these DOGs, though uncertain, are generally reliable. (Section~\ref{sec: luminrelation})
\item
Similar to Hot DOGs, our sources lie just beside the boundary of the forbidden region in the $\lambda_\mathrm{Edd}-N_\mathrm{H}$ plane, and thus our sources are expected to present only moderate outflows. (Section~\ref{sec: lEddnH_plane})
\item
We do not find a significant association between $L_\mathrm{X}$ and outflow strength for our sources, but this may be undermined by our small sample size and the narrow scope of our  parameter space. (Section~\ref{sec: outflow_strength})
\item
Our sources present strong starbursts and deviate more from the MS than other less-extreme DOGs. (Section~\ref{sec: host_MS})
\item
We carefully examined the validity of the virial assumption, and found that some of the \iona{Mg}{ii} lines are broader than the broad [\iona{O}{iii}] lines, while some \iona{Mg}{ii} lines are comparable with [\iona{O}{iii}] in width. Generally, we cannot confidently accept or rule out this assumption, but still somewhat favor it. (Section~\ref{sec: origin_broad})
\end{enumerate}
We argue that our findings can be explained under the co-evolution framework: high-$\lambda_\mathrm{Edd}$ DOGs are at similar evolutionary stages with Hot DOGs such that both SMBH accretion and host star formation are reaching the highest level and outflows have not blown out the gas reservoir, while other DOGs may be from a heterogeneous population that is composed of sources dominated by star formation or AGNs with different accretion rates.\par
Our work suggests that $\lambda_\mathrm{Edd}$ may be one of the key factors for discriminating different kinds of DOGs. Larger samples of DOGs with $\lambda_\mathrm{Edd}$ measurements (including both low-$\lambda_\mathrm{Edd}$ DOGs and, if possible, even higher-$\lambda_\mathrm{Edd}$ DOGs) would be beneficial for further understanding DOGs. Future work should also try to constrain the outflow geometry to offer better guidance for examining the virial assumption. More broadly, it is essential to examine the validity of the application of the single-epoch method to measure $M_\mathrm{BH}$ for DOGs. In terms of the X-ray data, though we have successfully probed the basic \mbox{X-ray} properties of our sample through economical snapshot observations, we would need at least several hundred \mbox{X-ray} photons to perform more reliable X-ray spectral analyses. Stacking (e.g. \citealt{Vito18}) can serve as one remedy for the problem of faintness, but it is difficult to probe more detailed spectral features with these approaches. For example, stacking may induce some biases and smear possible features like iron lines. Therefore, it is necessary to obtain sufficient photons for individual sources to probe more detailed \mbox{X-ray} spectra. The mean net count rate of our sources is \mbox{$\sim1~\mathrm{count~ks^{-1}}$}. Therefore, expensive \textit{Chandra} observations (hundreds of $\mathrm{ks}$ per source) would be required to obtain enough X-ray photons. Fortunately, the next-generation X-ray observatories, including \textit{Athena} \citep{Nandra13} and \textit{Lynx} \citep{Gaskin19}, have high throughput and will be able to shorten the required observing time by a factor of several tens. These future missions would certainly provide excellent opportunities to unveil the mysteries of DOGs.

\section*{Acknowledgements}
We thank the anonymous referee for their useful comments. We thank John Timlin for his help with \mbox{\texttt{PyQSOFit}}. FZ, WNB, and FV acknowledge support from CXC grant GO8-19076X, the V.M. Willaman Endowment, and the Penn State ACIS Instrument Team Contract SV4-74018  (issued by the Chandra X-ray Center, which is operated by the Smithsonian Astrophysical Observatory for and on behalf of NASA under contract NAS8-03060). FV acknowledges financial support from CONICYT and CASSACA through the Fourth call for tenders of the CAS-CONICYT Fund, and financial contribution from CONICYT grants Basal-CATA AFB-170002. The work of DS was carried out at the Jet Propulsion Laboratory, California Institute of Technology, under a contract with NASA. The Chandra ACIS team Guaranteed Time Observations (GTO) utilized were selected by the ACIS Instrument Principal Investigator, Gordon P. Garmire, currently of the Huntingdon Institute for X-ray Astronomy, LLC, which is under contract to the Smithsonian Astrophysical Observatory via Contract SV2-82024.

\section*{Data availability}
The data underlying this article will be shared upon reasonable request to the corresponding author. The relevant \textit{Chandra} data and SDSS spectra are publicly available from the \textit{Chandra} Data Archive and the SDSS Science Archive Server, respectively.

\bibliographystyle{mnras}
\bibliography{Citation_DOG}

\begin{thebibliography}{}
\makeatletter
\relax
\def\mn@urlcharsother{\let\do\@makeother \do\$\do\&\do\#\do\^\do\_\do\%\do\~}
\def\mn@doi{\begingroup\mn@urlcharsother \@ifnextchar [ {\mn@doi@}
  {\mn@doi@[]}}
\def\mn@doi@[#1]#2{\def\@tempa{#1}\ifx\@tempa\@empty \href
  {http://dx.doi.org/#2} {doi:#2}\else \href {http://dx.doi.org/#2} {#1}\fi
  \endgroup}
\def\mn@eprint#1#2{\mn@eprint@#1:#2::\@nil}
\def\mn@eprint@arXiv#1{\href {http://arxiv.org/abs/#1} {{\tt arXiv:#1}}}
\def\mn@eprint@dblp#1{\href {http://dblp.uni-trier.de/rec/bibtex/#1.xml}
  {dblp:#1}}
\def\mn@eprint@#1:#2:#3:#4\@nil{\def\@tempa {#1}\def\@tempb {#2}\def\@tempc
  {#3}\ifx \@tempc \@empty \let \@tempc \@tempb \let \@tempb \@tempa \fi \ifx
  \@tempb \@empty \def\@tempb {arXiv}\fi \@ifundefined
  {mn@eprint@\@tempb}{\@tempb:\@tempc}{\expandafter \expandafter \csname
  mn@eprint@\@tempb\endcsname \expandafter{\@tempc}}}

\bibitem[\protect\citeauthoryear{{Alaghband-Zadeh}, {Banerji}, {Hewett}  \&
  {McMahon}}{{Alaghband-Zadeh} et~al.}{2016}]{AlaghbandZadeh16}
{Alaghband-Zadeh} S.,  {Banerji} M.,  {Hewett} P.~C.,   {McMahon} R.~G.,  2016,
  \mn@doi [\mnras] {10.1093/mnras/stw682}, \href
  {https://ui.adsabs.harvard.edu/abs/2016MNRAS.459..999A} {459, 999}

\bibitem[\protect\citeauthoryear{{Alam} et~al.,}{{Alam} et~al.}{2015}]{Alam15}
{Alam} S.,  et~al., 2015, \mn@doi [\apjs] {10.1088/0067-0049/219/1/12}, \href
  {https://ui.adsabs.harvard.edu/abs/2015ApJS..219...12A} {219, 12}

\bibitem[\protect\citeauthoryear{{Alexander} \& {Hickox}}{{Alexander} \&
  {Hickox}}{2012}]{Alexander12}
{Alexander} D.~M.,  {Hickox} R.~C.,  2012, \mn@doi [\nar]
  {10.1016/j.newar.2011.11.003}, \href
  {https://ui.adsabs.harvard.edu/abs/2012NewAR..56...93A} {56, 93}

\bibitem[\protect\citeauthoryear{{Arnaud}}{{Arnaud}}{1996}]{Arnaud96}
{Arnaud} K.~A.,  1996, in {Jacoby} G.~H.,  {Barnes} J.,  eds,  Astronomical
  Society of the Pacific Conference Series Vol. 101, Astronomical Data Analysis
  Software and Systems V. p.~17

\bibitem[\protect\citeauthoryear{{Assef} et~al.,}{{Assef}
  et~al.}{2010}]{Assef10}
{Assef} R.~J.,  et~al., 2010, \mn@doi [\apj] {10.1088/0004-637X/713/2/970},
  \href {https://ui.adsabs.harvard.edu/abs/2010ApJ...713..970A} {713, 970}

\bibitem[\protect\citeauthoryear{{Assef} et~al.,}{{Assef}
  et~al.}{2015}]{Assef15}
{Assef} R.~J.,  et~al., 2015, \mn@doi [\apj] {10.1088/0004-637X/804/1/27},
  \href {https://ui.adsabs.harvard.edu/abs/2015ApJ...804...27A} {804, 27}

\bibitem[\protect\citeauthoryear{{Assef} et~al.,}{{Assef}
  et~al.}{2016}]{Assef16}
{Assef} R.~J.,  et~al., 2016, \mn@doi [\apj] {10.3847/0004-637X/819/2/111},
  \href {https://ui.adsabs.harvard.edu/abs/2016ApJ...819..111A} {819, 111}

\bibitem[\protect\citeauthoryear{{Assef} et~al.,}{{Assef}
  et~al.}{2019}]{Assef19}
{Assef} R.~J.,  et~al., 2019, arXiv e-prints, \href
  {https://ui.adsabs.harvard.edu/abs/2019arXiv190504320A} {p. arXiv:1905.04320}

\bibitem[\protect\citeauthoryear{{Bahk}, {Woo}  \& {Park}}{{Bahk}
  et~al.}{2019}]{Bahk19}
{Bahk} H.,  {Woo} J.-H.,   {Park} D.,  2019, \mn@doi [\apj]
  {10.3847/1538-4357/ab100d}, \href
  {https://ui.adsabs.harvard.edu/abs/2019ApJ...875...50B} {875, 50}

\bibitem[\protect\citeauthoryear{{Bohlin}, {Savage}  \& {Drake}}{{Bohlin}
  et~al.}{1978}]{Bohlin78}
{Bohlin} R.~C.,  {Savage} B.~D.,   {Drake} J.~F.,  1978, \mn@doi [\apj]
  {10.1086/156357}, \href
  {https://ui.adsabs.harvard.edu/abs/1978ApJ...224..132B} {224, 132}

\bibitem[\protect\citeauthoryear{{Bonzini}, {Padovani}, {Mainieri},
  {Kellermann}, {Miller}, {Rosati}, {Tozzi}  \& {Vattakunnel}}{{Bonzini}
  et~al.}{2013}]{Bonzini13}
{Bonzini} M.,  {Padovani} P.,  {Mainieri} V.,  {Kellermann} K.~I.,  {Miller}
  N.,  {Rosati} P.,  {Tozzi} P.,   {Vattakunnel} S.,  2013, \mn@doi [\mnras]
  {10.1093/mnras/stt1879}, \href
  {https://ui.adsabs.harvard.edu/abs/2013MNRAS.436.3759B} {436, 3759}

\bibitem[\protect\citeauthoryear{{Boquien}, {Burgarella}, {Roehlly}, {Buat},
  {Ciesla}, {Corre}, {Inoue}  \& {Salas}}{{Boquien} et~al.}{2019}]{Boquien19}
{Boquien} M.,  {Burgarella} D.,  {Roehlly} Y.,  {Buat} V.,  {Ciesla} L.,
  {Corre} D.,  {Inoue} A.~K.,   {Salas} H.,  2019, \mn@doi [\aap]
  {10.1051/0004-6361/201834156}, \href
  {https://ui.adsabs.harvard.edu/abs/2019A&A...622A.103B} {622, A103}

\bibitem[\protect\citeauthoryear{{Brandt} \& {Alexander}}{{Brandt} \&
  {Alexander}}{2015}]{Brandt15}
{Brandt} W.~N.,  {Alexander} D.~M.,  2015, \mn@doi [\aapr]
  {10.1007/s00159-014-0081-z}, \href
  {https://ui.adsabs.harvard.edu/abs/2015A&ARv..23....1B} {23, 1}

\bibitem[\protect\citeauthoryear{{Brightman} et~al.,}{{Brightman}
  et~al.}{2013}]{Brightman13}
{Brightman} M.,  et~al., 2013, \mn@doi [\mnras] {10.1093/mnras/stt920}, \href
  {https://ui.adsabs.harvard.edu/abs/2013MNRAS.433.2485B} {433, 2485}

\bibitem[\protect\citeauthoryear{{Broos}, {Feigelson}, {Townsley}, {Getman},
  {Wang}, {Garmire}, {Jiang}  \& {Tsuboi}}{{Broos} et~al.}{2007}]{Broos07}
{Broos} P.~S.,  {Feigelson} E.~D.,  {Townsley} L.~K.,  {Getman} K.~V.,  {Wang}
  J.,  {Garmire} G.~P.,  {Jiang} Z.,   {Tsuboi} Y.,  2007, \mn@doi [\apjs]
  {10.1086/512068}, \href
  {https://ui.adsabs.harvard.edu/abs/2007ApJS..169..353B} {169, 353}

\bibitem[\protect\citeauthoryear{{Brusa} et~al.,}{{Brusa}
  et~al.}{2015}]{Brusa15}
{Brusa} M.,  et~al., 2015, \mn@doi [\mnras] {10.1093/mnras/stu2117}, \href
  {https://ui.adsabs.harvard.edu/abs/2015MNRAS.446.2394B} {446, 2394}

\bibitem[\protect\citeauthoryear{{Bruzual} \& {Charlot}}{{Bruzual} \&
  {Charlot}}{2003}]{Bruzual03}
{Bruzual} G.,  {Charlot} S.,  2003, \mn@doi [\mnras]
  {10.1046/j.1365-8711.2003.06897.x}, \href
  {https://ui.adsabs.harvard.edu/abs/2003MNRAS.344.1000B} {344, 1000}

\bibitem[\protect\citeauthoryear{{Bussmann} et~al.,}{{Bussmann}
  et~al.}{2009}]{Bussmann09}
{Bussmann} R.~S.,  et~al., 2009, \mn@doi [\apj] {10.1088/0004-637X/705/1/184},
  \href {https://ui.adsabs.harvard.edu/abs/2009ApJ...705..184B} {705, 184}

\bibitem[\protect\citeauthoryear{{Calzetti}, {Armus}, {Bohlin}, {Kinney},
  {Koornneef}  \& {Storchi-Bergmann}}{{Calzetti} et~al.}{2000}]{Calzetti00}
{Calzetti} D.,  {Armus} L.,  {Bohlin} R.~C.,  {Kinney} A.~L.,  {Koornneef} J.,
   {Storchi-Bergmann} T.,  2000, \mn@doi [\apj] {10.1086/308692}, \href
  {https://ui.adsabs.harvard.edu/abs/2000ApJ...533..682C} {533, 682}

\bibitem[\protect\citeauthoryear{{Cao}}{{Cao}}{2009}]{Cao09}
{Cao} X.,  2009, \mn@doi [\mnras] {10.1111/j.1365-2966.2008.14347.x}, \href
  {https://ui.adsabs.harvard.edu/abs/2009MNRAS.394..207C} {394, 207}

\bibitem[\protect\citeauthoryear{{Carniani} et~al.,}{{Carniani}
  et~al.}{2015}]{Carniani15}
{Carniani} S.,  et~al., 2015, \mn@doi [\aap] {10.1051/0004-6361/201526557},
  \href {https://ui.adsabs.harvard.edu/abs/2015A&A...580A.102C} {580, A102}

\bibitem[\protect\citeauthoryear{{Cash}}{{Cash}}{1979}]{Cash79}
{Cash} W.,  1979, \mn@doi [\apj] {10.1086/156922}, \href
  {https://ui.adsabs.harvard.edu/abs/1979ApJ...228..939C} {228, 939}

\bibitem[\protect\citeauthoryear{{Chabrier}}{{Chabrier}}{2003}]{Chabrier03}
{Chabrier} G.,  2003, \mn@doi [\apjl] {10.1086/374879}, \href
  {https://ui.adsabs.harvard.edu/abs/2003ApJ...586L.133C} {586, L133}

\bibitem[\protect\citeauthoryear{{Chen} et~al.,}{{Chen} et~al.}{2017}]{Chen17}
{Chen} C.-T.~J.,  et~al., 2017, \mn@doi [\apj] {10.3847/1538-4357/837/2/145},
  \href {https://ui.adsabs.harvard.edu/abs/2017ApJ...837..145C} {837, 145}

\bibitem[\protect\citeauthoryear{{Chen} et~al.,}{{Chen} et~al.}{2019}]{Chen19}
{Chen} X.-Y.,  et~al., 2019, arXiv e-prints, \href
  {https://ui.adsabs.harvard.edu/abs/2019arXiv191104095C} {p. arXiv:1911.04095}

\bibitem[\protect\citeauthoryear{{Ciesla} et~al.,}{{Ciesla}
  et~al.}{2015}]{Ciesla15}
{Ciesla} L.,  et~al., 2015, \mn@doi [\aap] {10.1051/0004-6361/201425252}, \href
  {https://ui.adsabs.harvard.edu/abs/2015A&A...576A..10C} {576, A10}

\bibitem[\protect\citeauthoryear{{Corral} et~al.,}{{Corral}
  et~al.}{2016}]{Corral16}
{Corral} A.,  et~al., 2016, \mn@doi [\aap] {10.1051/0004-6361/201527624}, \href
  {https://ui.adsabs.harvard.edu/abs/2016A&A...592A.109C} {592, A109}

\bibitem[\protect\citeauthoryear{{Cutri}, {}  \& {et al.}}{{Cutri}
  et~al.}{2014}]{Cutri14}
{Cutri} R.~M.,  {}  {et al.} 2014, VizieR Online Data Catalog, \href
  {https://ui.adsabs.harvard.edu/abs/2014yCat.2328....0C} {p. II/328}

\bibitem[\protect\citeauthoryear{{Dale}, {Helou}, {Magdis}, {Armus},
  {D{\'\i}az-Santos}  \& {Shi}}{{Dale} et~al.}{2014}]{Dale14}
{Dale} D.~A.,  {Helou} G.,  {Magdis} G.~E.,  {Armus} L.,  {D{\'\i}az-Santos}
  T.,   {Shi} Y.,  2014, \mn@doi [\apj] {10.1088/0004-637X/784/1/83}, \href
  {https://ui.adsabs.harvard.edu/abs/2014ApJ...784...83D} {784, 83}

\bibitem[\protect\citeauthoryear{{Dermer} \& {Giebels}}{{Dermer} \&
  {Giebels}}{2016}]{Dermer16}
{Dermer} C.~D.,  {Giebels} B.,  2016, \mn@doi [Comptes Rendus Physique]
  {10.1016/j.crhy.2016.04.004}, \href
  {https://ui.adsabs.harvard.edu/abs/2016CRPhy..17..594D} {17, 594}

\bibitem[\protect\citeauthoryear{{Dey} et~al.,}{{Dey} et~al.}{2008}]{Dey08}
{Dey} A.,  et~al., 2008, \mn@doi [\apj] {10.1086/529516}, \href
  {https://ui.adsabs.harvard.edu/abs/2008ApJ...677..943D} {677, 943}

\bibitem[\protect\citeauthoryear{{D{\'\i}az-Santos} et~al.,}{{D{\'\i}az-Santos}
  et~al.}{2018}]{Diaz-Santos18}
{D{\'\i}az-Santos} T.,  et~al., 2018, \mn@doi [Science]
  {10.1126/science.aap7605}, \href
  {https://ui.adsabs.harvard.edu/abs/2018Sci...362.1034D} {362, 1034}

\bibitem[\protect\citeauthoryear{{Eisenhardt} et~al.,}{{Eisenhardt}
  et~al.}{2012}]{Eisenhardt12}
{Eisenhardt} P. R.~M.,  et~al., 2012, \mn@doi [\apj]
  {10.1088/0004-637X/755/2/173}, \href
  {https://ui.adsabs.harvard.edu/abs/2012ApJ...755..173E} {755, 173}

\bibitem[\protect\citeauthoryear{{Fabian}, {Vasudevan}  \& {Gandhi}}{{Fabian}
  et~al.}{2008}]{Fabian08}
{Fabian} A.~C.,  {Vasudevan} R.~V.,   {Gandhi} P.,  2008, \mn@doi [\mnras]
  {10.1111/j.1745-3933.2008.00430.x}, \href
  {https://ui.adsabs.harvard.edu/abs/2008MNRAS.385L..43F} {385, L43}

\bibitem[\protect\citeauthoryear{{Fabian}, {Vasudevan}, {Mushotzky}, {Winter}
  \& {Reynolds}}{{Fabian} et~al.}{2009}]{Fabian09}
{Fabian} A.~C.,  {Vasudevan} R.~V.,  {Mushotzky} R.~F.,  {Winter} L.~M.,
  {Reynolds} C.~S.,  2009, \mn@doi [\mnras] {10.1111/j.1745-3933.2009.00617.x},
  \href {https://ui.adsabs.harvard.edu/abs/2009MNRAS.394L..89F} {394, L89}

\bibitem[\protect\citeauthoryear{{Fan}, {Han}, {Nikutta}, {Drouart}  \&
  {Knudsen}}{{Fan} et~al.}{2016}]{Fan16}
{Fan} L.,  {Han} Y.,  {Nikutta} R.,  {Drouart} G.,   {Knudsen} K.~K.,  2016,
  \mn@doi [\apj] {10.3847/0004-637X/823/2/107}, \href
  {https://ui.adsabs.harvard.edu/abs/2016ApJ...823..107F} {823, 107}

\bibitem[\protect\citeauthoryear{{Fan}, {Gao}, {Knudsen}  \& {Shu}}{{Fan}
  et~al.}{2018}]{Fan18}
{Fan} L.,  {Gao} Y.,  {Knudsen} K.~K.,   {Shu} X.,  2018, \mn@doi [\apj]
  {10.3847/1538-4357/aaaaae}, \href
  {https://ui.adsabs.harvard.edu/abs/2018ApJ...854..157F} {854, 157}

\bibitem[\protect\citeauthoryear{{Fan}, {Knudsen}, {Han}  \& {Tan}}{{Fan}
  et~al.}{2019}]{Fan19}
{Fan} L.,  {Knudsen} K.~K.,  {Han} Y.,   {Tan} Q.-h.,  2019, \mn@doi [\apj]
  {10.3847/1538-4357/ab5059}, \href
  {https://ui.adsabs.harvard.edu/abs/2019ApJ...887...74F} {887, 74}

\bibitem[\protect\citeauthoryear{{Fiore} et~al.,}{{Fiore}
  et~al.}{2009}]{Fiore09}
{Fiore} F.,  et~al., 2009, \mn@doi [\apj] {10.1088/0004-637X/693/1/447}, \href
  {https://ui.adsabs.harvard.edu/abs/2009ApJ...693..447F} {693, 447}

\bibitem[\protect\citeauthoryear{{Fiore} et~al.,}{{Fiore}
  et~al.}{2017}]{Fiore17}
{Fiore} F.,  et~al., 2017, \mn@doi [\aap] {10.1051/0004-6361/201629478}, \href
  {https://ui.adsabs.harvard.edu/abs/2017A&A...601A.143F} {601, A143}

\bibitem[\protect\citeauthoryear{{Franceschini} et~al.,}{{Franceschini}
  et~al.}{2003}]{Franceschini03}
{Franceschini} A.,  et~al., 2003, \mn@doi [\mnras]
  {10.1046/j.1365-8711.2003.06744.x}, \href
  {https://ui.adsabs.harvard.edu/abs/2003MNRAS.343.1181F} {343, 1181}

\bibitem[\protect\citeauthoryear{{Gandhi}, {Horst}, {Smette}, {H{\"o}nig},
  {Comastri}, {Gilli}, {Vignali}  \& {Duschl}}{{Gandhi}
  et~al.}{2009}]{Gandhi09}
{Gandhi} P.,  {Horst} H.,  {Smette} A.,  {H{\"o}nig} S.,  {Comastri} A.,
  {Gilli} R.,  {Vignali} C.,   {Duschl} W.,  2009, \mn@doi [\aap]
  {10.1051/0004-6361/200811368}, \href
  {https://ui.adsabs.harvard.edu/abs/2009A&A...502..457G} {502, 457}

\bibitem[\protect\citeauthoryear{{Gaskin} et~al.,}{{Gaskin}
  et~al.}{2019}]{Gaskin19}
{Gaskin} J.~A.,  et~al., 2019, \mn@doi [Journal of Astronomical Telescopes,
  Instruments, and Systems] {10.1117/1.JATIS.5.2.021001}, \href
  {https://ui.adsabs.harvard.edu/abs/2019JATIS...5b1001G} {5, 021001}

\bibitem[\protect\citeauthoryear{{Glidden}, {Rose}, {Elvis}  \&
  {McDowell}}{{Glidden} et~al.}{2016}]{Glidden16}
{Glidden} A.,  {Rose} M.,  {Elvis} M.,   {McDowell} J.,  2016, \mn@doi [\apj]
  {10.3847/0004-637X/824/1/34}, \href
  {https://ui.adsabs.harvard.edu/abs/2016ApJ...824...34G} {824, 34}

\bibitem[\protect\citeauthoryear{{Glikman}, {LaMassa}, {Piconcelli}, {Urry}  \&
  {Lacy}}{{Glikman} et~al.}{2017}]{Glikman17}
{Glikman} E.,  {LaMassa} S.,  {Piconcelli} E.,  {Urry} M.,   {Lacy} M.,  2017,
  \mn@doi [\apj] {10.3847/1538-4357/aa88ac}, \href
  {https://ui.adsabs.harvard.edu/abs/2017ApJ...847..116G} {847, 116}

\bibitem[\protect\citeauthoryear{{Goulding} et~al.,}{{Goulding}
  et~al.}{2018}]{Goulding18}
{Goulding} A.~D.,  et~al., 2018, \mn@doi [\apj] {10.3847/1538-4357/aab040},
  \href {https://ui.adsabs.harvard.edu/abs/2018ApJ...856....4G} {856, 4}

\bibitem[\protect\citeauthoryear{{Guo}, {Shen}  \& {Wang}}{{Guo}
  et~al.}{2018}]{Guo18}
{Guo} H.,  {Shen} Y.,   {Wang} S.,  2018, {PyQSOFit: Python code to fit the
  spectrum of quasars} (\mn@eprint {ascl} {1809.008})

\bibitem[\protect\citeauthoryear{{HI4PI Collaboration} et~al.,}{{HI4PI
  Collaboration} et~al.}{2016}]{HI4PI16}
{HI4PI Collaboration} et~al., 2016, \mn@doi [\aap]
  {10.1051/0004-6361/201629178}, \href
  {https://ui.adsabs.harvard.edu/abs/2016A&A...594A.116H} {594, A116}

\bibitem[\protect\citeauthoryear{{Hao} et~al.,}{{Hao} et~al.}{2014}]{Hao14}
{Hao} H.,  et~al., 2014, arXiv e-prints, \href
  {https://ui.adsabs.harvard.edu/abs/2014arXiv1408.1090H} {p. arXiv:1408.1090}

\bibitem[\protect\citeauthoryear{{Harrison} et~al.,}{{Harrison}
  et~al.}{2016}]{Harrison16}
{Harrison} C.~M.,  et~al., 2016, \mn@doi [\mnras] {10.1093/mnras/stv2727},
  \href {https://ui.adsabs.harvard.edu/abs/2016MNRAS.456.1195H} {456, 1195}

\bibitem[\protect\citeauthoryear{{Hopkins}, {Hernquist}, {Cox}, {Di Matteo},
  {Robertson}  \& {Springel}}{{Hopkins} et~al.}{2006}]{Hopkins06}
{Hopkins} P.~F.,  {Hernquist} L.,  {Cox} T.~J.,  {Di Matteo} T.,  {Robertson}
  B.,   {Springel} V.,  2006, \mn@doi [\apjs] {10.1086/499298}, \href
  {https://ui.adsabs.harvard.edu/abs/2006ApJS..163....1H} {163, 1}

\bibitem[\protect\citeauthoryear{{Hopkins}, {Richards}  \&
  {Hernquist}}{{Hopkins} et~al.}{2007}]{Hopkins07}
{Hopkins} P.~F.,  {Richards} G.~T.,   {Hernquist} L.,  2007, \mn@doi [\apj]
  {10.1086/509629}, \href
  {https://ui.adsabs.harvard.edu/abs/2007ApJ...654..731H} {654, 731}

\bibitem[\protect\citeauthoryear{{Hopkins}, {Hernquist}, {Cox}  \&
  {Kere{\v{s}}}}{{Hopkins} et~al.}{2008}]{Hopkins08}
{Hopkins} P.~F.,  {Hernquist} L.,  {Cox} T.~J.,   {Kere{\v{s}}} D.,  2008,
  \mn@doi [\apjs] {10.1086/524362}, \href
  {https://ui.adsabs.harvard.edu/abs/2008ApJS..175..356H} {175, 356}

\bibitem[\protect\citeauthoryear{{Ishibashi}, {Fabian}, {Ricci}  \&
  {Celotti}}{{Ishibashi} et~al.}{2018}]{Ishibashi18}
{Ishibashi} W.,  {Fabian} A.~C.,  {Ricci} C.,   {Celotti} A.,  2018, \mn@doi
  [\mnras] {10.1093/mnras/sty1620}, \href
  {https://ui.adsabs.harvard.edu/abs/2018MNRAS.479.3335I} {479, 3335}

\bibitem[\protect\citeauthoryear{{Ishihara} et~al.,}{{Ishihara}
  et~al.}{2010}]{Ishihara10}
{Ishihara} D.,  et~al., 2010, \mn@doi [\aap] {10.1051/0004-6361/200913811},
  \href {https://ui.adsabs.harvard.edu/abs/2010A&A...514A...1I} {514, A1}

\bibitem[\protect\citeauthoryear{{Jones} et~al.,}{{Jones}
  et~al.}{2014}]{Jones14}
{Jones} S.~F.,  et~al., 2014, \mn@doi [\mnras] {10.1093/mnras/stu1157}, \href
  {https://ui.adsabs.harvard.edu/abs/2014MNRAS.443..146J} {443, 146}

\bibitem[\protect\citeauthoryear{{Jun} et~al.,}{{Jun} et~al.}{2020}]{Jun20}
{Jun} H.~D.,  et~al., 2020, \mn@doi [\apj] {10.3847/1538-4357/ab5e7b}, \href
  {https://ui.adsabs.harvard.edu/abs/2020ApJ...888..110J} {888, 110}

\bibitem[\protect\citeauthoryear{{Kaastra}}{{Kaastra}}{2017}]{Kaastra17}
{Kaastra} J.~S.,  2017, \mn@doi [\aap] {10.1051/0004-6361/201629319}, \href
  {https://ui.adsabs.harvard.edu/abs/2017A&A...605A..51K} {605, A51}

\bibitem[\protect\citeauthoryear{{Kakkad} et~al.,}{{Kakkad}
  et~al.}{2016}]{Kakkad16}
{Kakkad} D.,  et~al., 2016, \mn@doi [\aap] {10.1051/0004-6361/201527968}, \href
  {https://ui.adsabs.harvard.edu/abs/2016A&A...592A.148K} {592, A148}

\bibitem[\protect\citeauthoryear{{Kawada} et~al.,}{{Kawada}
  et~al.}{2007}]{Kawada07}
{Kawada} M.,  et~al., 2007, \mn@doi [\pasj] {10.1093/pasj/59.sp2.S389}, \href
  {https://ui.adsabs.harvard.edu/abs/2007PASJ...59S.389K} {59, S389}

\bibitem[\protect\citeauthoryear{{Kormendy} \& {Ho}}{{Kormendy} \&
  {Ho}}{2013}]{Kormendy13}
{Kormendy} J.,  {Ho} L.~C.,  2013, \mn@doi [\araa]
  {10.1146/annurev-astro-082708-101811}, \href
  {https://ui.adsabs.harvard.edu/abs/2013ARA&A..51..511K} {51, 511}

\bibitem[\protect\citeauthoryear{{Koz{\l}owski}}{{Koz{\l}owski}}{2017}]{Kozlowski17}
{Koz{\l}owski} S.,  2017, \mn@doi [\apjs] {10.3847/1538-4365/228/1/9}, \href
  {https://ui.adsabs.harvard.edu/abs/2017ApJS..228....9K} {228, 9}

\bibitem[\protect\citeauthoryear{{LaMassa} et~al.,}{{LaMassa}
  et~al.}{2016}]{LaMassa16}
{LaMassa} S.~M.,  et~al., 2016, \mn@doi [\apj] {10.3847/0004-637X/820/1/70},
  \href {https://ui.adsabs.harvard.edu/abs/2016ApJ...820...70L} {820, 70}

\bibitem[\protect\citeauthoryear{{LaMassa} et~al.,}{{LaMassa}
  et~al.}{2017}]{LaMassa17}
{LaMassa} S.~M.,  et~al., 2017, \mn@doi [\apj] {10.3847/1538-4357/aa87b5},
  \href {https://ui.adsabs.harvard.edu/abs/2017ApJ...847..100L} {847, 100}

\bibitem[\protect\citeauthoryear{{Lansbury}, {Banerji}, {Fabian}  \&
  {Temple}}{{Lansbury} et~al.}{2019}]{Lansbury19}
{Lansbury} G.~B.,  {Banerji} M.,  {Fabian} A.~C.,   {Temple} M.~J.,  2019,
  arXiv e-prints, \href {https://ui.adsabs.harvard.edu/abs/2019arXiv191000602L}
  {p. arXiv:1910.00602}

\bibitem[\protect\citeauthoryear{{Lanzuisi}, {Piconcelli}, {Fiore}, {Feruglio},
  {Vignali}, {Salvato}  \& {Gruppioni}}{{Lanzuisi} et~al.}{2009}]{Lanzuisi09}
{Lanzuisi} G.,  {Piconcelli} E.,  {Fiore} F.,  {Feruglio} C.,  {Vignali} C.,
  {Salvato} M.,   {Gruppioni} C.,  2009, \mn@doi [\aap]
  {10.1051/0004-6361/200811282}, \href
  {https://ui.adsabs.harvard.edu/abs/2009A&A...498...67L} {498, 67}

\bibitem[\protect\citeauthoryear{{Le}, {Woo}  \& {Xue}}{{Le}
  et~al.}{2020}]{Le20}
{Le} H. A.~N.,  {Woo} J.-H.,   {Xue} Y.,  2020, arXiv e-prints, \href
  {https://ui.adsabs.harvard.edu/abs/2020arXiv200802990L} {p. arXiv:2008.02990}

\bibitem[\protect\citeauthoryear{{Liang} et~al.,}{{Liang}
  et~al.}{2019}]{Liang19}
{Liang} L.,  et~al., 2019, \mn@doi [\mnras] {10.1093/mnras/stz2134}, \href
  {https://ui.adsabs.harvard.edu/abs/2019MNRAS.489.1397L} {489, 1397}

\bibitem[\protect\citeauthoryear{{Luo} et~al.,}{{Luo} et~al.}{2015}]{Luo15}
{Luo} B.,  et~al., 2015, \mn@doi [\apj] {10.1088/0004-637X/805/2/122}, \href
  {https://ui.adsabs.harvard.edu/abs/2015ApJ...805..122L} {805, 122}

\bibitem[\protect\citeauthoryear{{Lusso} et~al.,}{{Lusso}
  et~al.}{2010}]{Lusso10}
{Lusso} E.,  et~al., 2010, \mn@doi [\aap] {10.1051/0004-6361/200913298}, \href
  {https://ui.adsabs.harvard.edu/abs/2010A&A...512A..34L} {512, A34}

\bibitem[\protect\citeauthoryear{{Lusso} et~al.,}{{Lusso}
  et~al.}{2012}]{Lusso12}
{Lusso} E.,  et~al., 2012, \mn@doi [\mnras] {10.1111/j.1365-2966.2012.21513.x},
  \href {https://ui.adsabs.harvard.edu/abs/2012MNRAS.425..623L} {425, 623}

\bibitem[\protect\citeauthoryear{{Lutz}, {Maiolino}, {Spoon}  \&
  {Moorwood}}{{Lutz} et~al.}{2004}]{Lutz04}
{Lutz} D.,  {Maiolino} R.,  {Spoon} H.~W.~W.,   {Moorwood} A.~F.~M.,  2004,
  \mn@doi [\aap] {10.1051/0004-6361:20035838}, \href
  {https://ui.adsabs.harvard.edu/abs/2004A&A...418..465L} {418, 465}

\bibitem[\protect\citeauthoryear{{Maiolino}, {Marconi}, {Salvati}, {Risaliti},
  {Severgnini}, {Oliva}, {La Franca}  \& {Vanzi}}{{Maiolino}
  et~al.}{2001}]{Maiolino01}
{Maiolino} R.,  {Marconi} A.,  {Salvati} M.,  {Risaliti} G.,  {Severgnini} P.,
  {Oliva} E.,  {La Franca} F.,   {Vanzi} L.,  2001, \mn@doi [\aap]
  {10.1051/0004-6361:20000177}, \href
  {https://ui.adsabs.harvard.edu/abs/2001A&A...365...28M} {365, 28}

\bibitem[\protect\citeauthoryear{{Marconi}, {Risaliti}, {Gilli}, {Hunt},
  {Maiolino}  \& {Salvati}}{{Marconi} et~al.}{2004}]{Marconi04}
{Marconi} A.,  {Risaliti} G.,  {Gilli} R.,  {Hunt} L.~K.,  {Maiolino} R.,
  {Salvati} M.,  2004, \mn@doi [\mnras] {10.1111/j.1365-2966.2004.07765.x},
  \href {https://ui.adsabs.harvard.edu/abs/2004MNRAS.351..169M} {351, 169}

\bibitem[\protect\citeauthoryear{{Martocchia} et~al.,}{{Martocchia}
  et~al.}{2017}]{Martocchia17}
{Martocchia} S.,  et~al., 2017, \mn@doi [\aap] {10.1051/0004-6361/201731314},
  \href {https://ui.adsabs.harvard.edu/abs/2017A&A...608A..51M} {608, A51}

\bibitem[\protect\citeauthoryear{{Matsuoka} et~al.,}{{Matsuoka}
  et~al.}{2018}]{Matsuoka18}
{Matsuoka} K.,  et~al., 2018, \mn@doi [\aap] {10.1051/0004-6361/201833943},
  \href {https://ui.adsabs.harvard.edu/abs/2018A&A...620L...3M} {620, L3}

\bibitem[\protect\citeauthoryear{{Melbourne} et~al.,}{{Melbourne}
  et~al.}{2011}]{Melbourne11}
{Melbourne} J.,  et~al., 2011, \mn@doi [\aj] {10.1088/0004-6256/141/4/141},
  \href {https://ui.adsabs.harvard.edu/abs/2011AJ....141..141M} {141, 141}

\bibitem[\protect\citeauthoryear{{Melbourne} et~al.,}{{Melbourne}
  et~al.}{2012}]{Melbourne12}
{Melbourne} J.,  et~al., 2012, \mn@doi [\aj] {10.1088/0004-6256/143/5/125},
  \href {https://ui.adsabs.harvard.edu/abs/2012AJ....143..125M} {143, 125}

\bibitem[\protect\citeauthoryear{{Mountrichas} et~al.,}{{Mountrichas}
  et~al.}{2017}]{Mountrichas17}
{Mountrichas} G.,  et~al., 2017, \mn@doi [\mnras] {10.1093/mnras/stx572}, \href
  {https://ui.adsabs.harvard.edu/abs/2017MNRAS.468.3042M} {468, 3042}

\bibitem[\protect\citeauthoryear{{Nandra} et~al.,}{{Nandra}
  et~al.}{2013}]{Nandra13}
{Nandra} K.,  et~al., 2013, arXiv e-prints, \href
  {https://ui.adsabs.harvard.edu/abs/2013arXiv1306.2307N} {p. arXiv:1306.2307}

\bibitem[\protect\citeauthoryear{{Narayanan} et~al.,}{{Narayanan}
  et~al.}{2010}]{Narayanan10}
{Narayanan} D.,  et~al., 2010, \mn@doi [\mnras]
  {10.1111/j.1365-2966.2010.16997.x}, \href
  {https://ui.adsabs.harvard.edu/abs/2010MNRAS.407.1701N} {407, 1701}

\bibitem[\protect\citeauthoryear{{P{\^a}ris} et~al.,}{{P{\^a}ris}
  et~al.}{2017}]{Paris17}
{P{\^a}ris} I.,  et~al., 2017, \mn@doi [\aap] {10.1051/0004-6361/201527999},
  \href {https://ui.adsabs.harvard.edu/abs/2017A&A...597A..79P} {597, A79}

\bibitem[\protect\citeauthoryear{{Perna}, {Lanzuisi}, {Brusa}, {Mignoli}  \&
  {Cresci}}{{Perna} et~al.}{2017}]{Perna17}
{Perna} M.,  {Lanzuisi} G.,  {Brusa} M.,  {Mignoli} M.,   {Cresci} G.,  2017,
  \mn@doi [\aap] {10.1051/0004-6361/201630369}, \href
  {https://ui.adsabs.harvard.edu/abs/2017A&A...603A..99P} {603, A99}

\bibitem[\protect\citeauthoryear{{Pons}, {McMahon}, {Simcoe}, {Banerji},
  {Hewett}  \& {Reed}}{{Pons} et~al.}{2019}]{Pons19}
{Pons} E.,  {McMahon} R.~G.,  {Simcoe} R.~A.,  {Banerji} M.,  {Hewett} P.~C.,
  {Reed} S.~L.,  2019, \mn@doi [\mnras] {10.1093/mnras/stz292}, \href
  {https://ui.adsabs.harvard.edu/abs/2019MNRAS.484.5142P} {484, 5142}

\bibitem[\protect\citeauthoryear{{Pope} et~al.,}{{Pope} et~al.}{2008}]{Pope08}
{Pope} A.,  et~al., 2008, \mn@doi [\apj] {10.1086/592739}, \href
  {https://ui.adsabs.harvard.edu/abs/2008ApJ...689..127P} {689, 127}

\bibitem[\protect\citeauthoryear{{Prevot}, {Lequeux}, {Maurice}, {Prevot}  \&
  {Rocca-Volmerange}}{{Prevot} et~al.}{1984}]{Prevot84}
{Prevot} M.~L.,  {Lequeux} J.,  {Maurice} E.,  {Prevot} L.,
  {Rocca-Volmerange} B.,  1984, \aap, \href
  {https://ui.adsabs.harvard.edu/abs/1984A&A...132..389P} {132, 389}

\bibitem[\protect\citeauthoryear{{Reines} \& {Volonteri}}{{Reines} \&
  {Volonteri}}{2015}]{Reines15}
{Reines} A.~E.,  {Volonteri} M.,  2015, \mn@doi [\apj]
  {10.1088/0004-637X/813/2/82}, \href
  {https://ui.adsabs.harvard.edu/abs/2015ApJ...813...82R} {813, 82}

\bibitem[\protect\citeauthoryear{{Ricci} et~al.,}{{Ricci}
  et~al.}{2017}]{Ricci17}
{Ricci} C.,  et~al., 2017, \mn@doi [\apj] {10.3847/1538-4357/835/1/105}, \href
  {https://ui.adsabs.harvard.edu/abs/2017ApJ...835..105R} {835, 105}

\bibitem[\protect\citeauthoryear{{Rodighiero} et~al.,}{{Rodighiero}
  et~al.}{2011}]{Rodighiero11}
{Rodighiero} G.,  et~al., 2011, \mn@doi [\apjl] {10.1088/2041-8205/739/2/L40},
  \href {https://ui.adsabs.harvard.edu/abs/2011ApJ...739L..40R} {739, L40}

\bibitem[\protect\citeauthoryear{{Roig}, {Blanton}  \& {Ross}}{{Roig}
  et~al.}{2014}]{Roig14}
{Roig} B.,  {Blanton} M.~R.,   {Ross} N.~P.,  2014, \mn@doi [\apj]
  {10.1088/0004-637X/781/2/72}, \href
  {https://ui.adsabs.harvard.edu/abs/2014ApJ...781...72R} {781, 72}

\bibitem[\protect\citeauthoryear{{Rose}, {Tadhunter}, {Holt}, {Ramos Almeida}
  \& {Littlefair}}{{Rose} et~al.}{2011}]{Rose11}
{Rose} M.,  {Tadhunter} C.~N.,  {Holt} J.,  {Ramos Almeida} C.,   {Littlefair}
  S.~P.,  2011, \mn@doi [\mnras] {10.1111/j.1365-2966.2011.18639.x}, \href
  {https://ui.adsabs.harvard.edu/abs/2011MNRAS.414.3360R} {414, 3360}

\bibitem[\protect\citeauthoryear{{Rose}, {Elvis}  \& {Tadhunter}}{{Rose}
  et~al.}{2015}]{Rose15}
{Rose} M.,  {Elvis} M.,   {Tadhunter} C.~N.,  2015, \mn@doi [\mnras]
  {10.1093/mnras/stv113}, \href
  {https://ui.adsabs.harvard.edu/abs/2015MNRAS.448.2900R} {448, 2900}

\bibitem[\protect\citeauthoryear{{Sanders}, {Soifer}, {Elias}, {Madore},
  {Matthews}, {Neugebauer}  \& {Scoville}}{{Sanders} et~al.}{1988}]{Sanders88}
{Sanders} D.~B.,  {Soifer} B.~T.,  {Elias} J.~H.,  {Madore} B.~F.,  {Matthews}
  K.,  {Neugebauer} G.,   {Scoville} N.~Z.,  1988, \mn@doi [\apj]
  {10.1086/165983}, \href
  {https://ui.adsabs.harvard.edu/abs/1988ApJ...325...74S} {325, 74}

\bibitem[\protect\citeauthoryear{{Shankar} et~al.,}{{Shankar}
  et~al.}{2016}]{Shankar16}
{Shankar} F.,  et~al., 2016, \mn@doi [\mnras] {10.1093/mnras/stw678}, \href
  {https://ui.adsabs.harvard.edu/abs/2016MNRAS.460.3119S} {460, 3119}

\bibitem[\protect\citeauthoryear{{Shemmer}, {Brandt}, {Netzer}, {Maiolino}  \&
  {Kaspi}}{{Shemmer} et~al.}{2008}]{Shemmer08}
{Shemmer} O.,  {Brandt} W.~N.,  {Netzer} H.,  {Maiolino} R.,   {Kaspi} S.,
  2008, \mn@doi [\apj] {10.1086/588776}, \href
  {https://ui.adsabs.harvard.edu/abs/2008ApJ...682...81S} {682, 81}

\bibitem[\protect\citeauthoryear{{Shen}}{{Shen}}{2013}]{Shen13}
{Shen} Y.,  2013, Bulletin of the Astronomical Society of India, \href
  {https://ui.adsabs.harvard.edu/abs/2013BASI...41...61S} {41, 61}

\bibitem[\protect\citeauthoryear{{Shen} et~al.,}{{Shen} et~al.}{2011}]{Shen11}
{Shen} Y.,  et~al., 2011, \mn@doi [\apjs] {10.1088/0067-0049/194/2/45}, \href
  {https://ui.adsabs.harvard.edu/abs/2011ApJS..194...45S} {194, 45}

\bibitem[\protect\citeauthoryear{{Shen} et~al.,}{{Shen} et~al.}{2016}]{Shen16}
{Shen} Y.,  et~al., 2016, \mn@doi [\apj] {10.3847/0004-637X/831/1/7}, \href
  {https://ui.adsabs.harvard.edu/abs/2016ApJ...831....7S} {831, 7}

\bibitem[\protect\citeauthoryear{{Silva}, {Maiolino}  \& {Granato}}{{Silva}
  et~al.}{2004}]{Silva04}
{Silva} L.,  {Maiolino} R.,   {Granato} G.~L.,  2004, \mn@doi [\mnras]
  {10.1111/j.1365-2966.2004.08380.x}, \href
  {https://ui.adsabs.harvard.edu/abs/2004MNRAS.355..973S} {355, 973}

\bibitem[\protect\citeauthoryear{{Speagle}, {Steinhardt}, {Capak}  \&
  {Silverman}}{{Speagle} et~al.}{2014}]{Speagle14}
{Speagle} J.~S.,  {Steinhardt} C.~L.,  {Capak} P.~L.,   {Silverman} J.~D.,
  2014, \mn@doi [\apjs] {10.1088/0067-0049/214/2/15}, \href
  {https://ui.adsabs.harvard.edu/abs/2014ApJS..214...15S} {214, 15}

\bibitem[\protect\citeauthoryear{{Stalevski}, {Fritz}, {Baes}, {Nakos}  \&
  {Popovi{\'c}}}{{Stalevski} et~al.}{2012}]{Stalevski12}
{Stalevski} M.,  {Fritz} J.,  {Baes} M.,  {Nakos} T.,   {Popovi{\'c}}
  L.~{\v{C}}.,  2012, \mn@doi [\mnras] {10.1111/j.1365-2966.2011.19775.x},
  \href {https://ui.adsabs.harvard.edu/abs/2012MNRAS.420.2756S} {420, 2756}

\bibitem[\protect\citeauthoryear{{Stalevski}, {Ricci}, {Ueda}, {Lira}, {Fritz}
  \& {Baes}}{{Stalevski} et~al.}{2016}]{Stalevski16}
{Stalevski} M.,  {Ricci} C.,  {Ueda} Y.,  {Lira} P.,  {Fritz} J.,   {Baes} M.,
  2016, \mn@doi [\mnras] {10.1093/mnras/stw444}, \href
  {https://ui.adsabs.harvard.edu/abs/2016MNRAS.458.2288S} {458, 2288}

\bibitem[\protect\citeauthoryear{{Stern}}{{Stern}}{2015}]{Stern15}
{Stern} D.,  2015, \mn@doi [\apj] {10.1088/0004-637X/807/2/129}, \href
  {https://ui.adsabs.harvard.edu/abs/2015ApJ...807..129S} {807, 129}

\bibitem[\protect\citeauthoryear{{Stern} et~al.,}{{Stern}
  et~al.}{2014}]{Stern14}
{Stern} D.,  et~al., 2014, \mn@doi [\apj] {10.1088/0004-637X/794/2/102}, \href
  {https://ui.adsabs.harvard.edu/abs/2014ApJ...794..102S} {794, 102}

\bibitem[\protect\citeauthoryear{{Sun} et~al.,}{{Sun} et~al.}{2015}]{Sun15}
{Sun} M.,  et~al., 2015, \mn@doi [\apj] {10.1088/0004-637X/802/1/14}, \href
  {https://ui.adsabs.harvard.edu/abs/2015ApJ...802...14S} {802, 14}

\bibitem[\protect\citeauthoryear{{Temple}, {Banerji}, {Hewett}, {Coatman},
  {Maddox}  \& {Peroux}}{{Temple} et~al.}{2019}]{Temple19}
{Temple} M.~J.,  {Banerji} M.,  {Hewett} P.~C.,  {Coatman} L.,  {Maddox} N.,
  {Peroux} C.,  2019, \mn@doi [\mnras] {10.1093/mnras/stz1420}, \href
  {https://ui.adsabs.harvard.edu/abs/2019MNRAS.487.2594T} {487, 2594}

\bibitem[\protect\citeauthoryear{{Teng} \& {Veilleux}}{{Teng} \&
  {Veilleux}}{2010}]{Teng10}
{Teng} S.~H.,  {Veilleux} S.,  2010, \mn@doi [\apj]
  {10.1088/0004-637X/725/2/1848}, \href
  {https://ui.adsabs.harvard.edu/abs/2010ApJ...725.1848T} {725, 1848}

\bibitem[\protect\citeauthoryear{{Timlin}, {Brandt}, {Ni}, {Luo}, {Pu},
  {Schneider}, {Vivek}  \& {Yi}}{{Timlin} et~al.}{2020}]{Timlin20}
{Timlin} J.~D.,  {Brandt} W.~N.,  {Ni} Q.,  {Luo} B.,  {Pu} X.,  {Schneider}
  D.~P.,  {Vivek} M.,   {Yi} W.,  2020, \mn@doi [\mnras]
  {10.1093/mnras/stz3433}, \href
  {https://ui.adsabs.harvard.edu/abs/2020MNRAS.492..719T} {492, 719}

\bibitem[\protect\citeauthoryear{{Toba} \& {Nagao}}{{Toba} \&
  {Nagao}}{2016}]{Toba16}
{Toba} Y.,  {Nagao} T.,  2016, \mn@doi [\apj] {10.3847/0004-637X/820/1/46},
  \href {https://ui.adsabs.harvard.edu/abs/2016ApJ...820...46T} {820, 46}

\bibitem[\protect\citeauthoryear{{Toba} et~al.,}{{Toba} et~al.}{2015}]{Toba15}
{Toba} Y.,  et~al., 2015, \mn@doi [\pasj] {10.1093/pasj/psv057}, \href
  {https://ui.adsabs.harvard.edu/abs/2015PASJ...67...86T} {67, 86}

\bibitem[\protect\citeauthoryear{{Toba}, {Bae}, {Nagao}, {Woo}, {Wang},
  {Wagner}, {Sun}  \& {Chang}}{{Toba} et~al.}{2017}]{Toba17}
{Toba} Y.,  {Bae} H.-J.,  {Nagao} T.,  {Woo} J.-H.,  {Wang} W.-H.,  {Wagner}
  A.~Y.,  {Sun} A.-L.,   {Chang} Y.-Y.,  2017, \mn@doi [\apj]
  {10.3847/1538-4357/aa918a}, \href
  {https://ui.adsabs.harvard.edu/abs/2017ApJ...850..140T} {850, 140}

\bibitem[\protect\citeauthoryear{{Toba}, {Ueda}, {Lim}, {Wang}, {Nagao},
  {Chang}, {Saito}  \& {Kawabe}}{{Toba} et~al.}{2018}]{Toba18}
{Toba} Y.,  {Ueda} J.,  {Lim} C.-F.,  {Wang} W.-H.,  {Nagao} T.,  {Chang}
  Y.-Y.,  {Saito} T.,   {Kawabe} R.,  2018, \mn@doi [\apj]
  {10.3847/1538-4357/aab3cf}, \href
  {https://ui.adsabs.harvard.edu/abs/2018ApJ...857...31T} {857, 31}

\bibitem[\protect\citeauthoryear{{Toba} et~al.,}{{Toba} et~al.}{2019}]{Toba19}
{Toba} Y.,  et~al., 2019, arXiv e-prints, \href
  {https://ui.adsabs.harvard.edu/abs/2019arXiv191205813T} {p. arXiv:1912.05813}

\bibitem[\protect\citeauthoryear{{Toba} et~al.,}{{Toba} et~al.}{2020}]{Toba20}
{Toba} Y.,  et~al., 2020, \mn@doi [\apj] {10.3847/1538-4357/ab5718}, \href
  {https://ui.adsabs.harvard.edu/abs/2020ApJ...888....8T} {888, 8}

\bibitem[\protect\citeauthoryear{{Trakhtenbrot} et~al.,}{{Trakhtenbrot}
  et~al.}{2015}]{Trakhtenbrot15}
{Trakhtenbrot} B.,  et~al., 2015, \mn@doi [Science] {10.1126/science.aaa4506},
  \href {https://ui.adsabs.harvard.edu/abs/2015Sci...349..168T} {349, 168}

\bibitem[\protect\citeauthoryear{{Tsai} et~al.,}{{Tsai} et~al.}{2015}]{Tsai15}
{Tsai} C.-W.,  et~al., 2015, \mn@doi [\apj] {10.1088/0004-637X/805/2/90}, \href
  {https://ui.adsabs.harvard.edu/abs/2015ApJ...805...90T} {805, 90}

\bibitem[\protect\citeauthoryear{{Urrutia}, {Lacy}, {Gregg}  \&
  {Becker}}{{Urrutia} et~al.}{2005}]{Urrutia05}
{Urrutia} T.,  {Lacy} M.,  {Gregg} M.~D.,   {Becker} R.~H.,  2005, \mn@doi
  [\apj] {10.1086/430165}, \href
  {https://ui.adsabs.harvard.edu/abs/2005ApJ...627...75U} {627, 75}

\bibitem[\protect\citeauthoryear{{Vasudevan} \& {Fabian}}{{Vasudevan} \&
  {Fabian}}{2009}]{Vasudevan09}
{Vasudevan} R.~V.,  {Fabian} A.~C.,  2009, \mn@doi [\mnras]
  {10.1111/j.1365-2966.2008.14108.x}, \href
  {https://ui.adsabs.harvard.edu/abs/2009MNRAS.392.1124V} {392, 1124}

\bibitem[\protect\citeauthoryear{{Vito} et~al.,}{{Vito} et~al.}{2018}]{Vito18}
{Vito} F.,  et~al., 2018, \mn@doi [\mnras] {10.1093/mnras/stx3120}, \href
  {https://ui.adsabs.harvard.edu/abs/2018MNRAS.474.4528V} {474, 4528}

\bibitem[\protect\citeauthoryear{{Vito} et~al.,}{{Vito} et~al.}{2019}]{Vito19}
{Vito} F.,  et~al., 2019, \mn@doi [\aap] {10.1051/0004-6361/201936217}, \href
  {https://ui.adsabs.harvard.edu/abs/2019A&A...630A.118V} {630, A118}

\bibitem[\protect\citeauthoryear{{Wang} et~al.,}{{Wang} et~al.}{2019}]{Wang19}
{Wang} S.,  et~al., 2019, \mn@doi [\apj] {10.3847/1538-4357/ab322b}, \href
  {https://ui.adsabs.harvard.edu/abs/2019ApJ...882....4W} {882, 4}

\bibitem[\protect\citeauthoryear{{Weisskopf}, {Wu}, {Trimble}, {O'Dell},
  {Elsner}, {Zavlin}  \& {Kouveliotou}}{{Weisskopf} et~al.}{2007}]{Weisskopf07}
{Weisskopf} M.~C.,  {Wu} K.,  {Trimble} V.,  {O'Dell} S.~L.,  {Elsner} R.~F.,
  {Zavlin} V.~E.,   {Kouveliotou} C.,  2007, \mn@doi [\apj] {10.1086/510776},
  \href {https://ui.adsabs.harvard.edu/abs/2007ApJ...657.1026W} {657, 1026}

\bibitem[\protect\citeauthoryear{{Whitaker}, {van Dokkum}, {Brammer}  \&
  {Franx}}{{Whitaker} et~al.}{2012}]{Whitaker12}
{Whitaker} K.~E.,  {van Dokkum} P.~G.,  {Brammer} G.,   {Franx} M.,  2012,
  \mn@doi [\apjl] {10.1088/2041-8205/754/2/L29}, \href
  {https://ui.adsabs.harvard.edu/abs/2012ApJ...754L..29W} {754, L29}

\bibitem[\protect\citeauthoryear{{Wright} et~al.,}{{Wright}
  et~al.}{2010}]{Wright10}
{Wright} E.~L.,  et~al., 2010, \mn@doi [\aj] {10.1088/0004-6256/140/6/1868},
  \href {https://ui.adsabs.harvard.edu/abs/2010AJ....140.1868W} {140, 1868}

\bibitem[\protect\citeauthoryear{{Wu} et~al.,}{{Wu} et~al.}{2012}]{Wu12}
{Wu} J.,  et~al., 2012, \mn@doi [\apj] {10.1088/0004-637X/756/1/96}, \href
  {https://ui.adsabs.harvard.edu/abs/2012ApJ...756...96W} {756, 96}

\bibitem[\protect\citeauthoryear{{Wu} et~al.,}{{Wu} et~al.}{2018}]{Wu18}
{Wu} J.,  et~al., 2018, \mn@doi [\apj] {10.3847/1538-4357/aa9ff3}, \href
  {https://ui.adsabs.harvard.edu/abs/2018ApJ...852...96W} {852, 96}

\bibitem[\protect\citeauthoryear{{Yamamura}, {Makiuti}, {Ikeda}, {Fukuda},
  {Oyabu}, {Koga}  \& {White}}{{Yamamura} et~al.}{2010}]{Yamamura10}
{Yamamura} I.,  {Makiuti} S.,  {Ikeda} N.,  {Fukuda} Y.,  {Oyabu} S.,  {Koga}
  T.,   {White} G.~J.,  2010, VizieR Online Data Catalog, \href
  {https://ui.adsabs.harvard.edu/abs/2010yCat.2298....0Y} {p. II/298}

\bibitem[\protect\citeauthoryear{{Yang} et~al.,}{{Yang}
  et~al.}{2019a}]{YangG19}
{Yang} G.,  et~al., 2019a, \mn@doi [\mnras] {10.1093/mnras/stz3001}, \href
  {https://ui.adsabs.harvard.edu/abs/2019MNRAS.tmp.2604Y} {p.~2604}

\bibitem[\protect\citeauthoryear{{Yang} et~al.,}{{Yang}
  et~al.}{2019b}]{YangQ19}
{Yang} Q.,  et~al., 2019b, arXiv e-prints, \href
  {https://ui.adsabs.harvard.edu/abs/2019arXiv190410912Y} {p. arXiv:1904.10912}

\bibitem[\protect\citeauthoryear{{Zappacosta} et~al.,}{{Zappacosta}
  et~al.}{2018}]{Zappacosta18}
{Zappacosta} L.,  et~al., 2018, \mn@doi [\aap] {10.1051/0004-6361/201732557},
  \href {https://ui.adsabs.harvard.edu/abs/2018A&A...618A..28Z} {618, A28}

\bibitem[\protect\citeauthoryear{{Zou}, {Yang}, {Brandt}  \& {Xue}}{{Zou}
  et~al.}{2019}]{Zou19}
{Zou} F.,  {Yang} G.,  {Brandt} W.~N.,   {Xue} Y.,  2019, \mn@doi [\apj]
  {10.3847/1538-4357/ab1eb1}, \href
  {https://ui.adsabs.harvard.edu/abs/2019ApJ...878...11Z} {878, 11}

\makeatother
\end{thebibliography}

\appendix
\section{SDSS spectra}
\label{append: optspec}

\begin{figure}
\resizebox{\hsize}{!}{\includegraphics{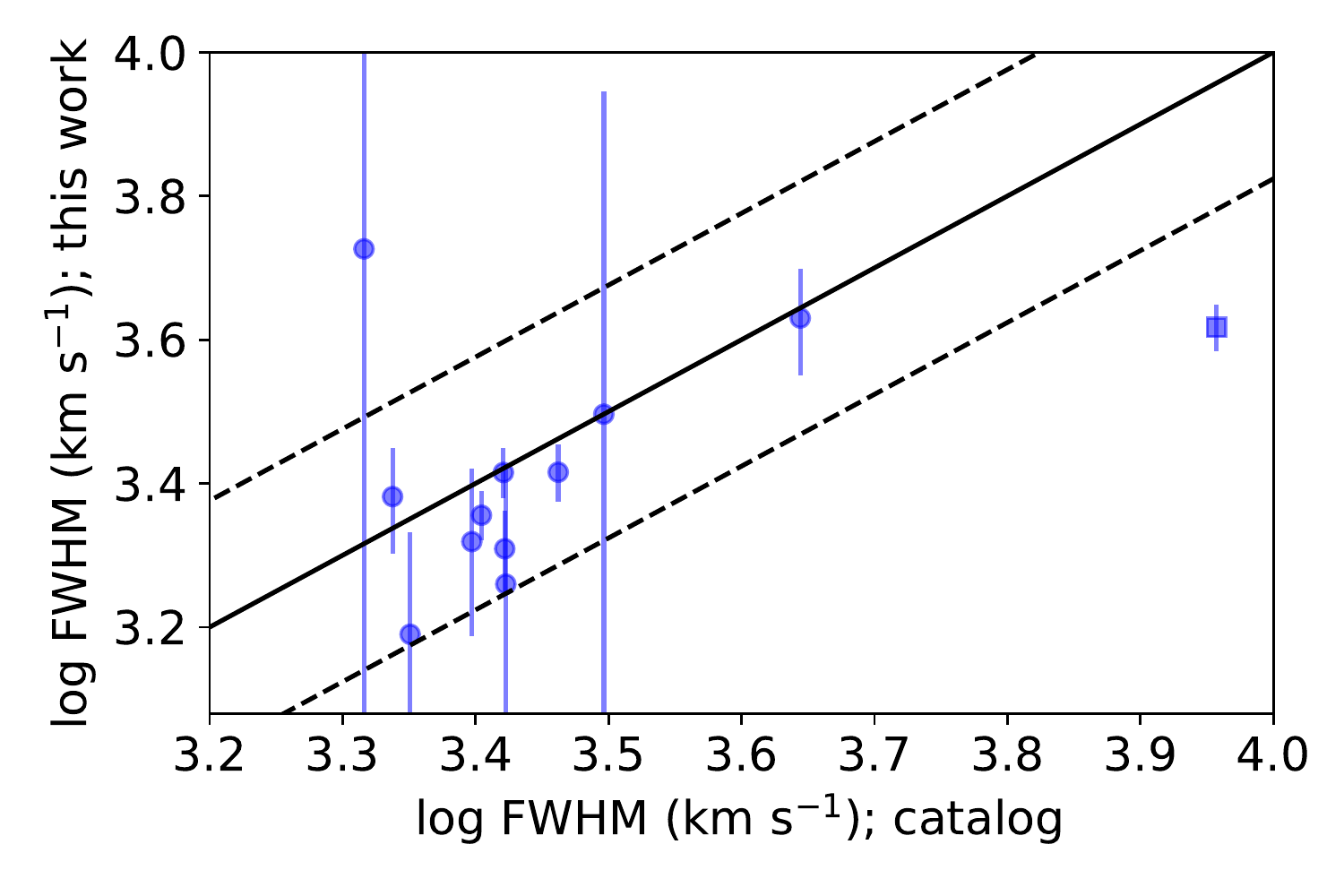}}
\caption{Comparison between our FWHM measurements and the FWHM values in \citet{Shen11} (H$\beta$ for J1010+3725; the square) and \citet{Paris17} (\iona{Mg}{ii}; circular points). The solid black line indicates a one-to-one relationship. The dashed black lines indicate an $1\sigma$ dispersion (0.18 dex) on the differences between the FWHM measurements in \citet{Shen11} and \citet{Paris17}. All of our \iona{Mg}{ii} measurements are generally consistent with the cataloged values within $1\sigma$, but the H$\beta$ square seems to be an outlier.}
\label{compfwhmFig}
\end{figure}

Fig.~\ref{compfwhmFig} compares our FWHM values and those in \citet{Shen11} and \citet{Paris17}. The FWHMs are generally consistent except for J1010+3725. Our fitting for this source is visually convincing (Fig.~\ref{HbOIIISpecs}) and is also visually consistent with the fitting in \citet{Toba17} (see their Fig. 14), and thus we suspect that the fitting in \citet{Shen11} for this source might inappropriately attribute some parts of the [\iona{O}{iii}] outflow to the broad H$\beta$ component. Indeed, if we do not add broad Gaussian lines to model its [\iona{O}{iii}] outflow, we would obtain $\mathrm{FWHM=8260~km~s^{-1}}$, similar to the cataloged value ($\mathrm{9062~km~s^{-1}}$).

\begin{figure*}
\raggedleft
\resizebox{\hsize}{!}{
\includegraphics{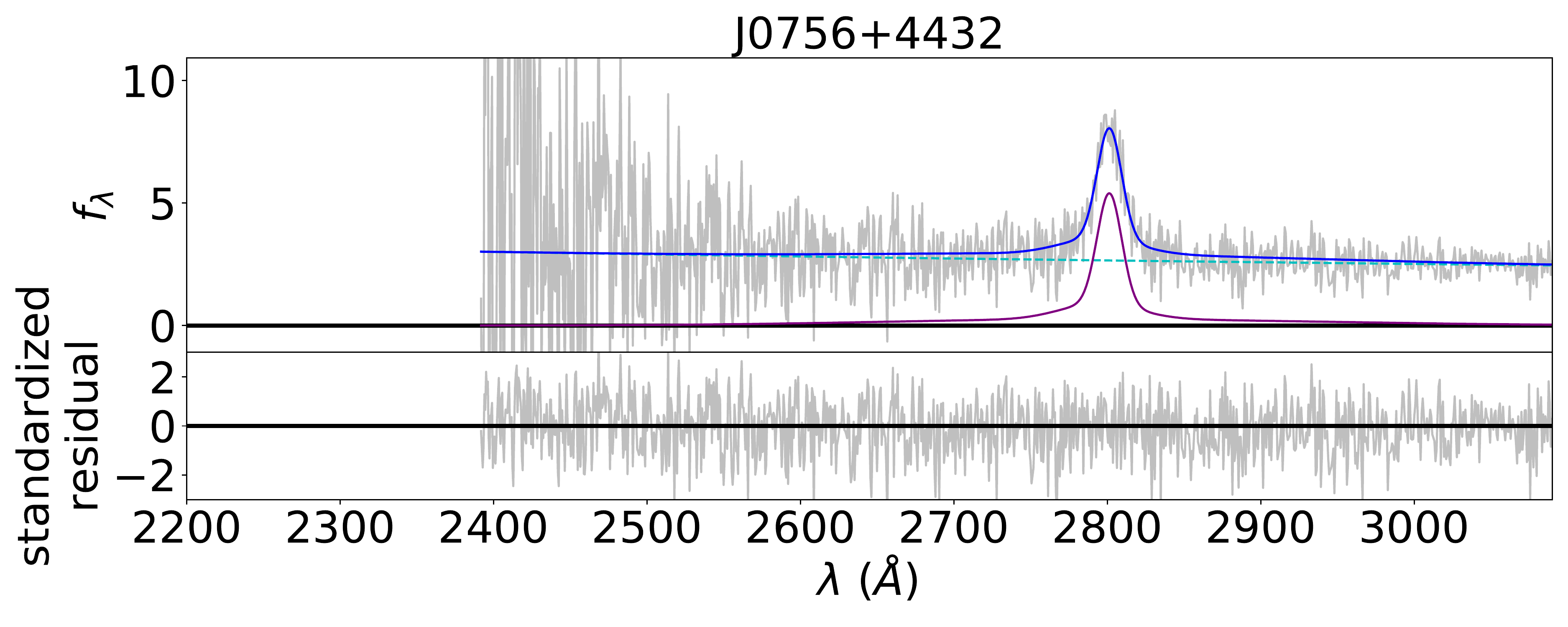}
\includegraphics{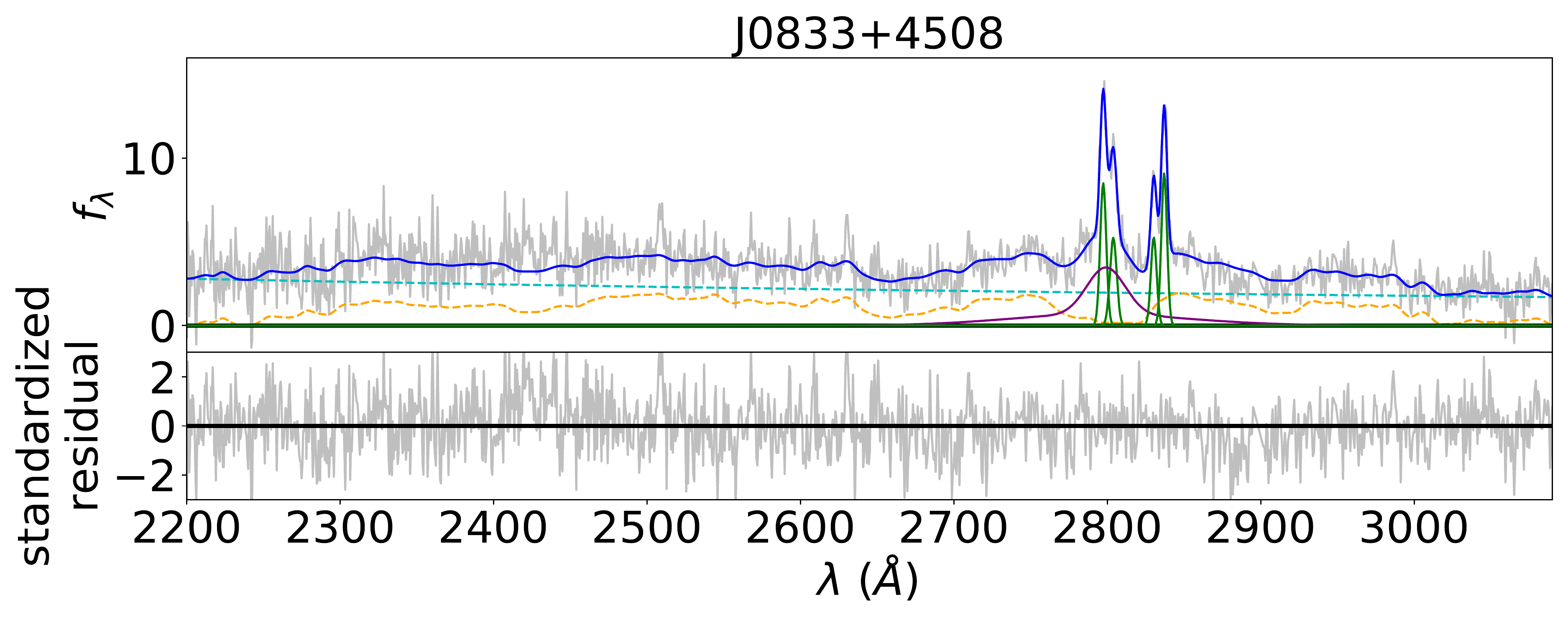}
}
\resizebox{0.5\hsize}{!}{
\includegraphics{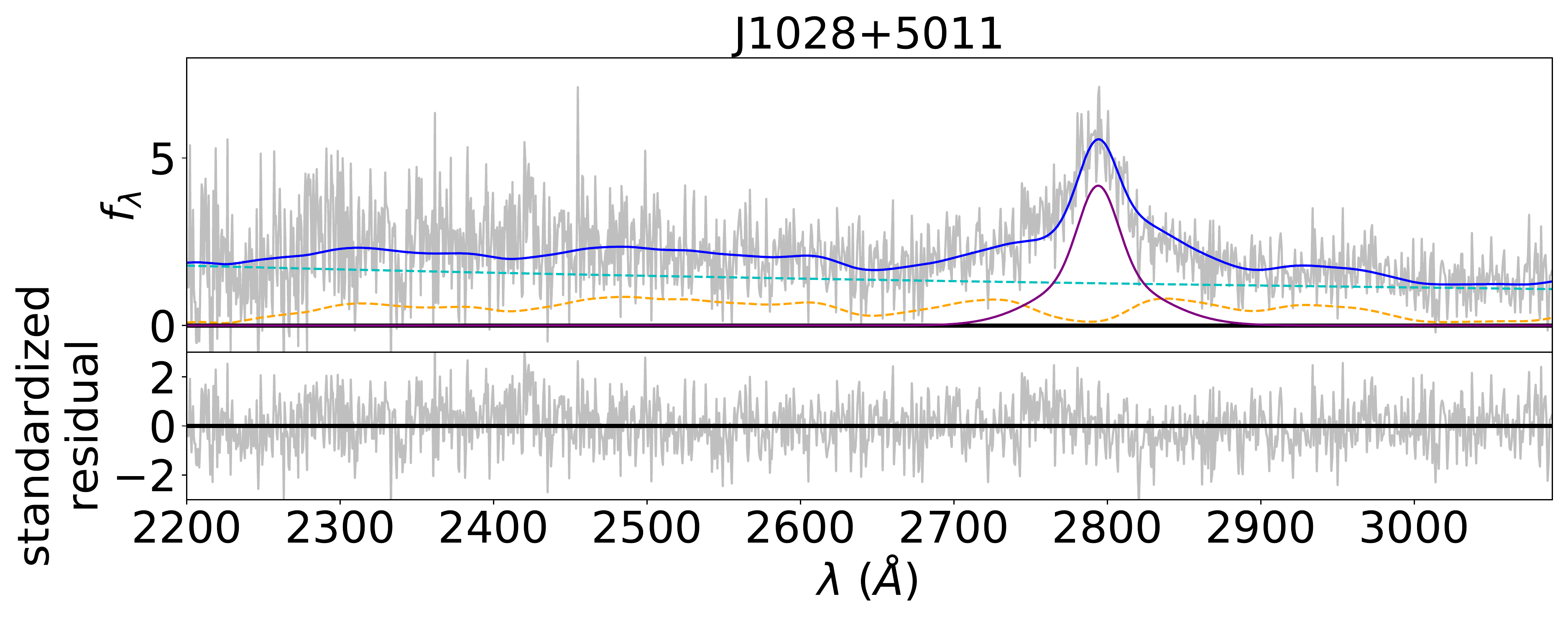}
}
\resizebox{\hsize}{!}{
\includegraphics{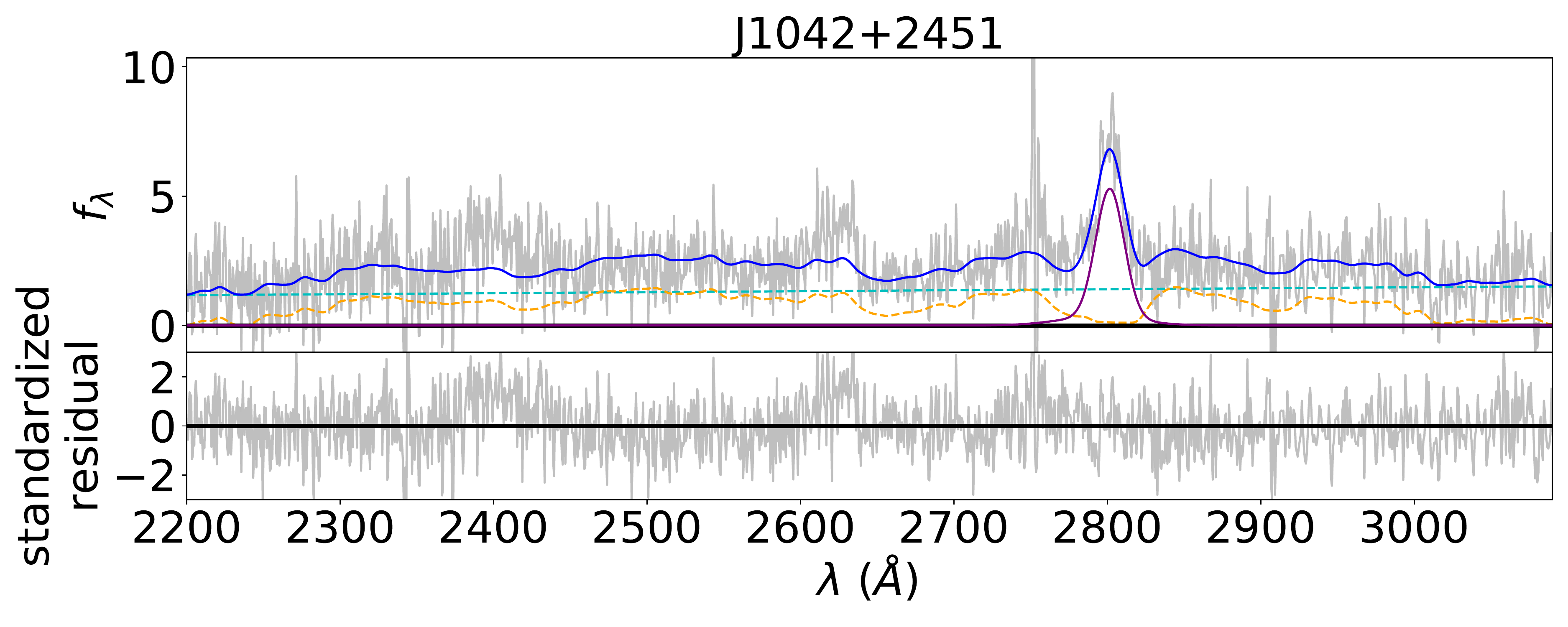}
\includegraphics{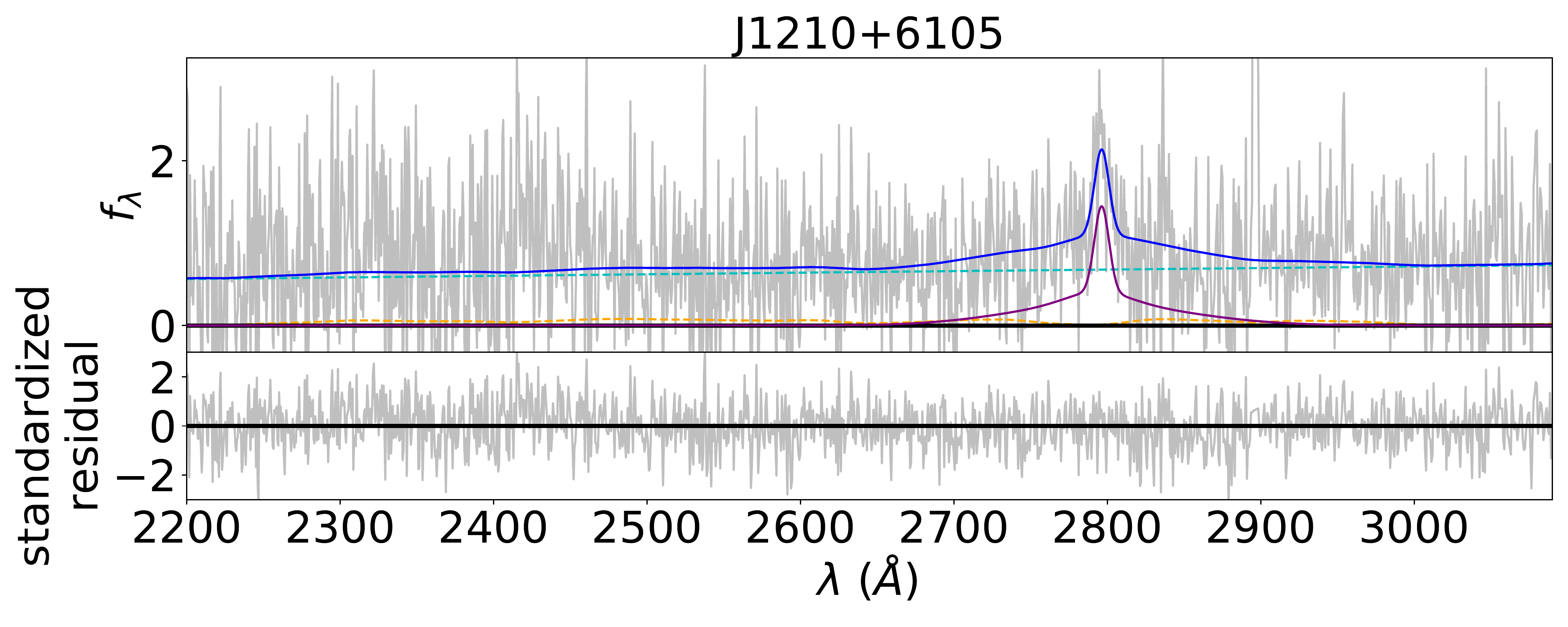}
}
\resizebox{\hsize}{!}{
\includegraphics{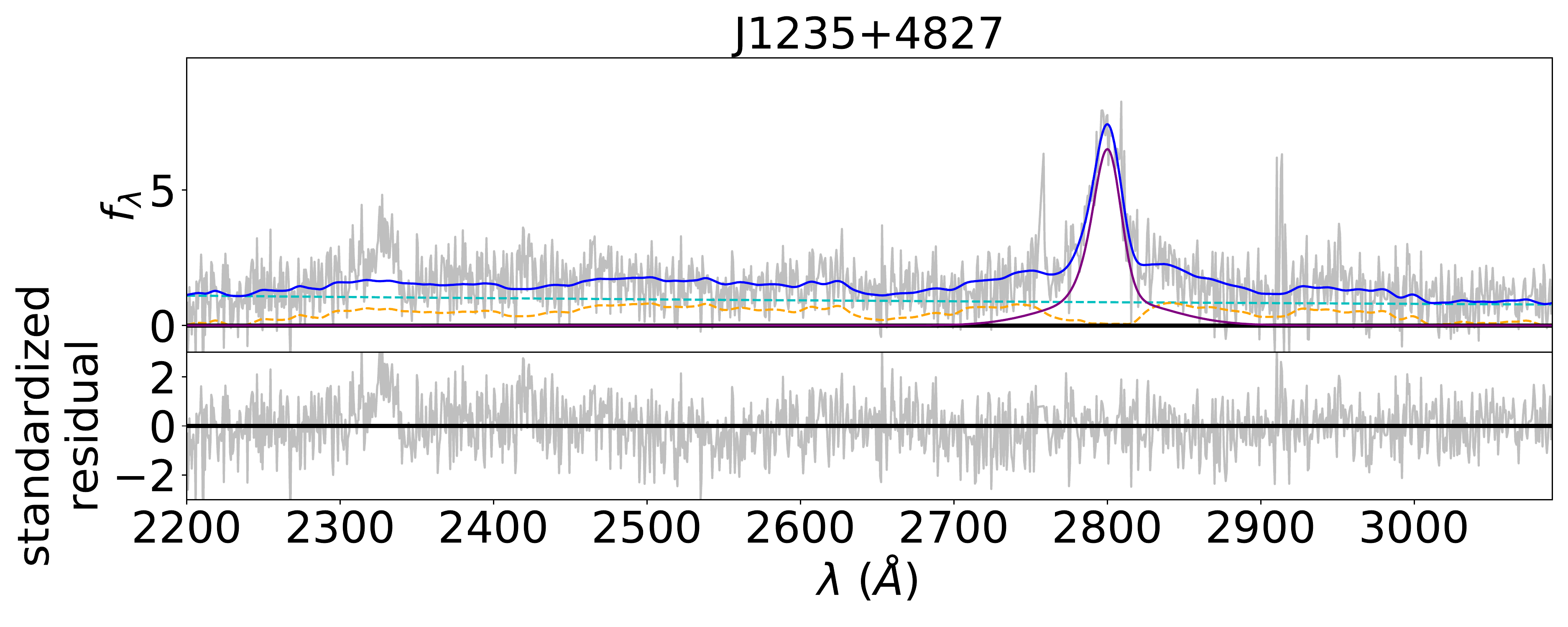}
\includegraphics{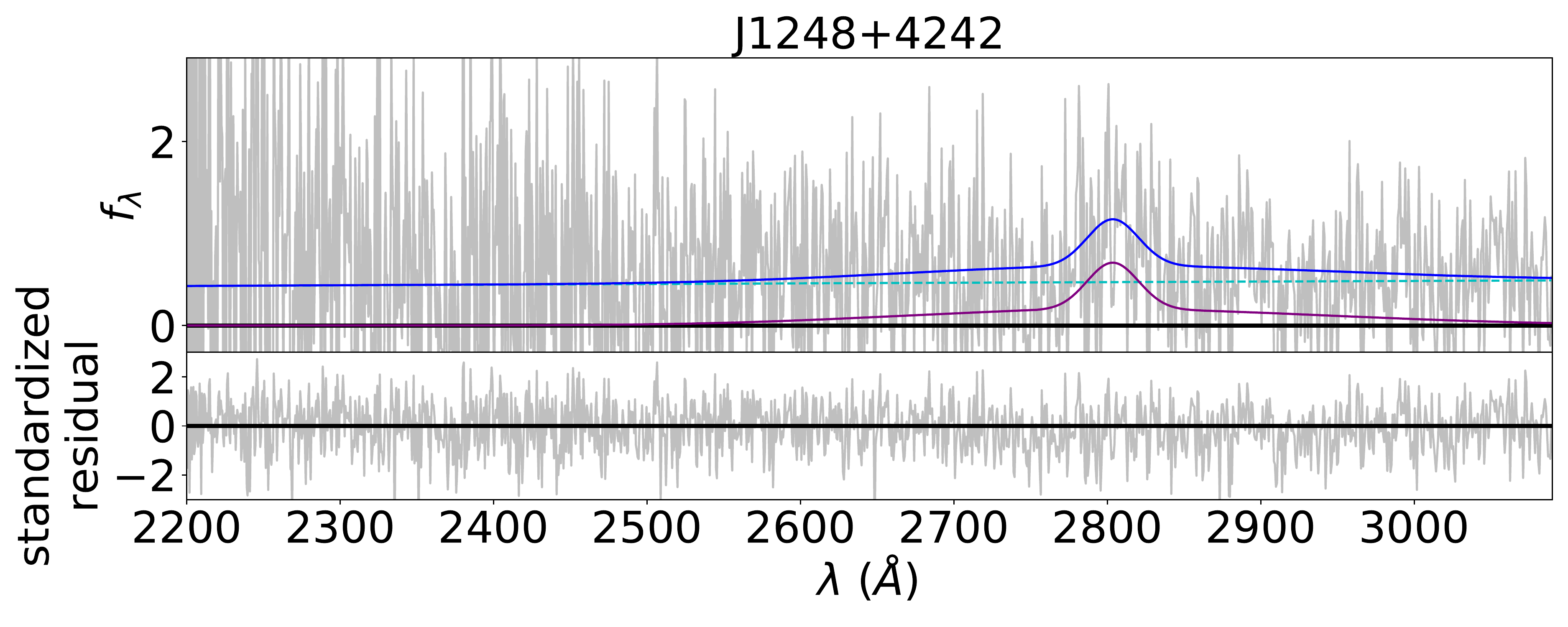}
}
\resizebox{\hsize}{!}{
\includegraphics{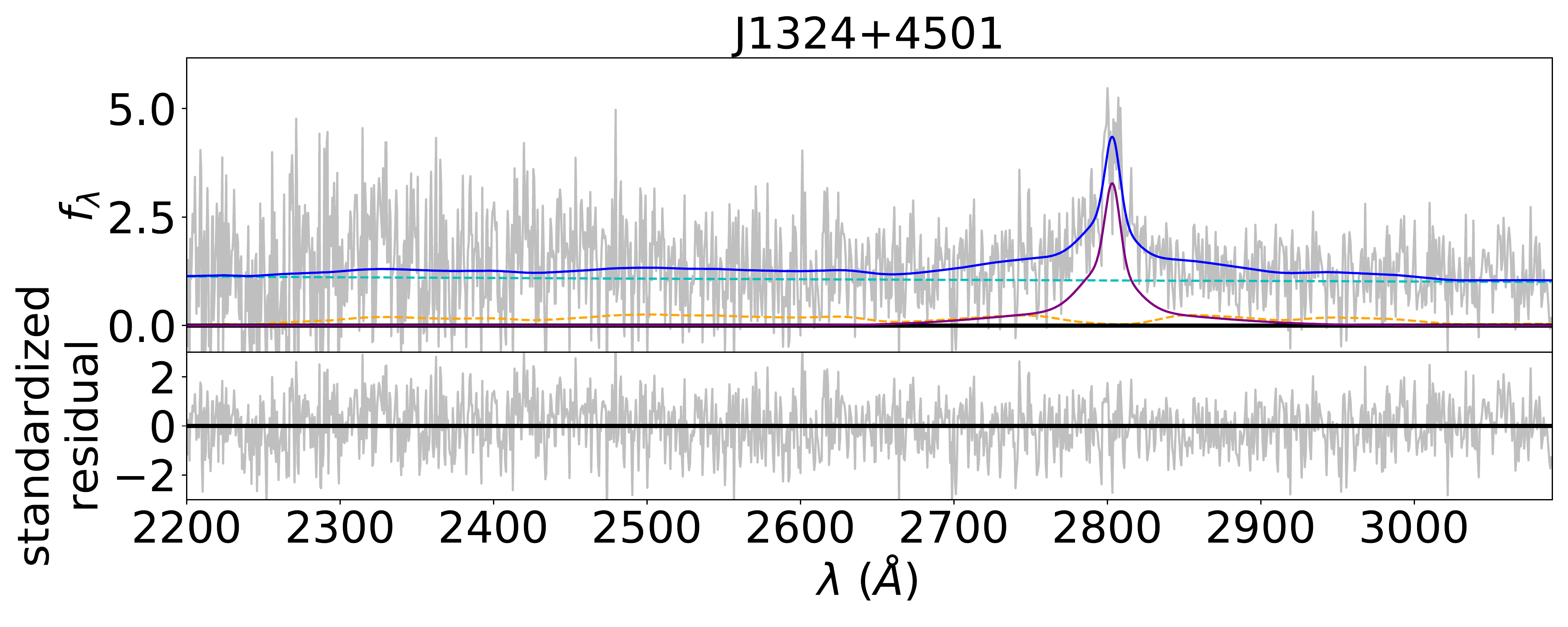}
\includegraphics{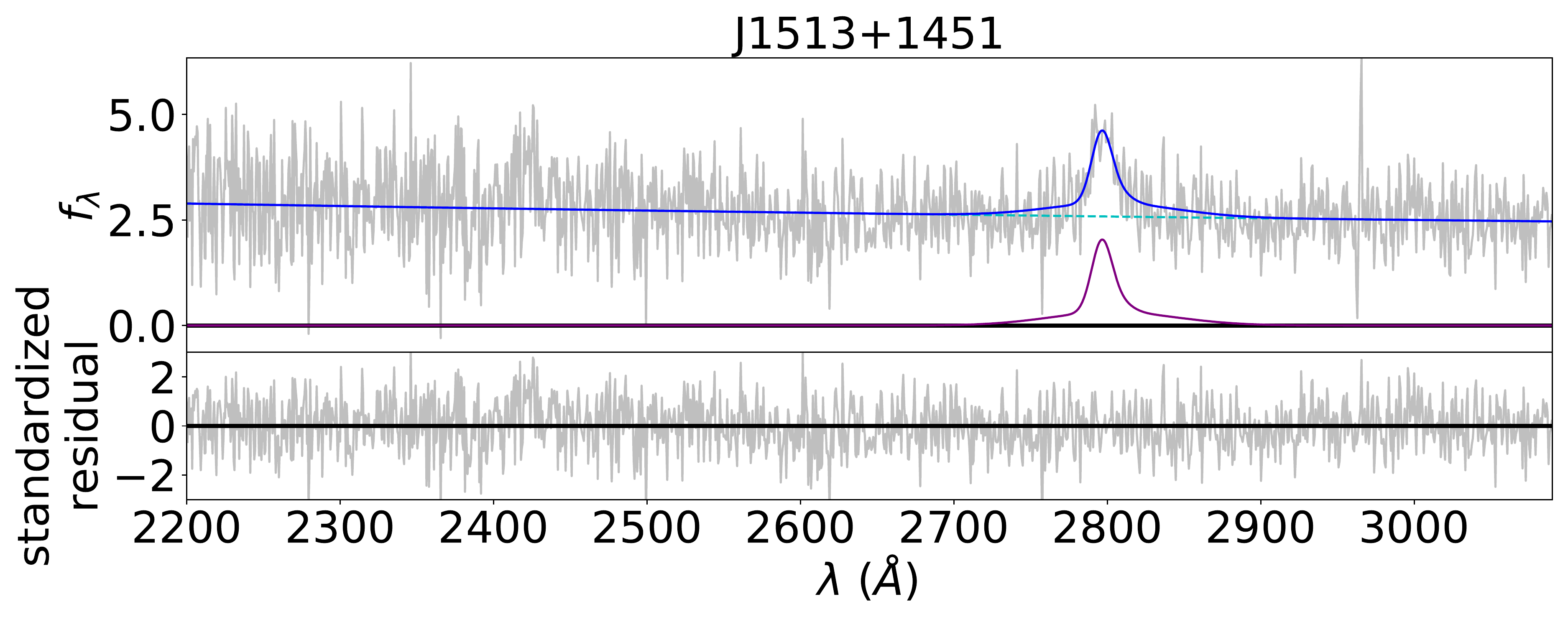}
}
\resizebox{\hsize}{!}{
\includegraphics{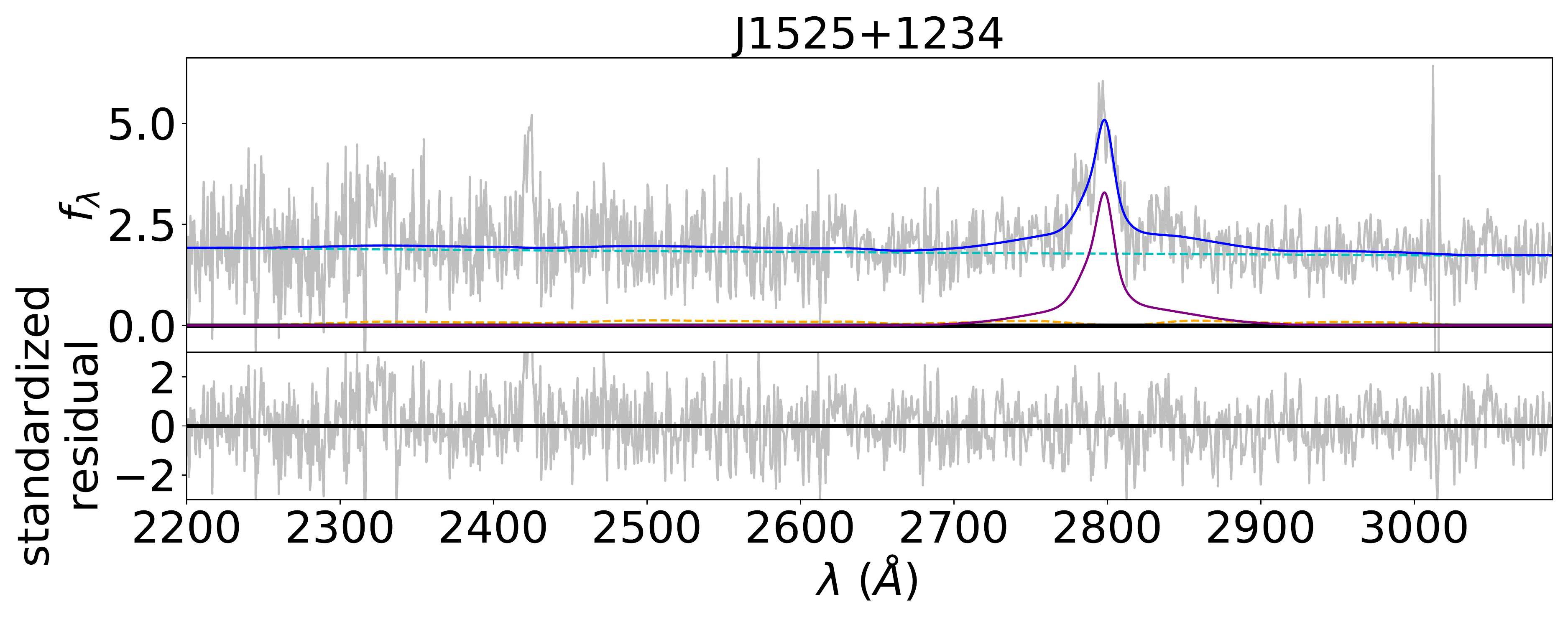}
\includegraphics{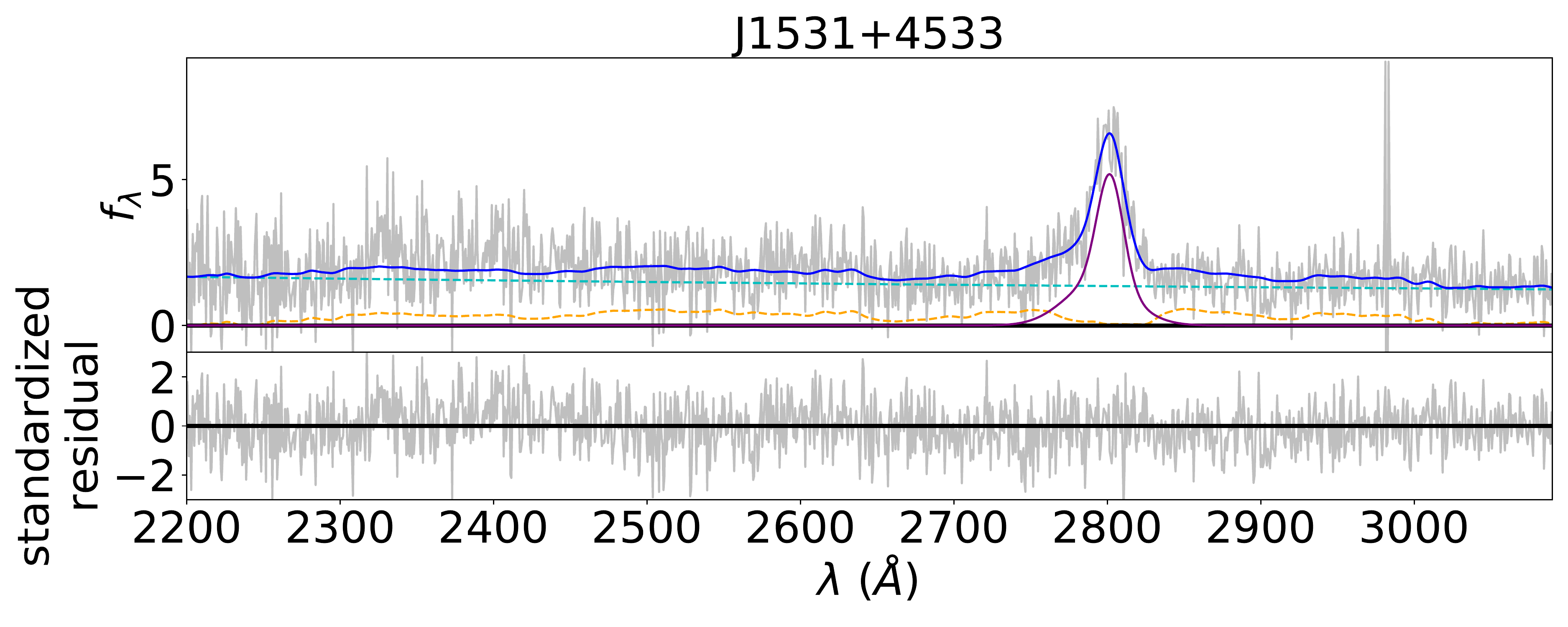}
}\caption{\textit{Top}: rest-frame \iona{Mg}{ii} spectra, where $f_\lambda$ is in units of $\mathrm{10^{-17}~erg~s^{-1}~cm^{-2}~\textup{\AA}^{-1}}$. The solid blue lines show our best fits, which are composed of a power-law (cyan dashed lines), \iona{Fe}{ii} emission (orange dashed lines), and \iona{Mg}{ii} emission lines (solid purple lines). The solid green lines of J0833+4508 indicate narrow \iona{Mg}{ii} cores and [\iona{Fe}{iv}] emission lines. \textit{Bottom}: standardized residuals of the fits, i.e. $\mathrm{(data-model)/noise}$. This figure is arranged in the same manner as Fig.~\ref{HbOIIISpecs}, and the ``hole'' in the second row indicates that J1010+3725 has no \iona{Mg}{ii} coverage.}
\label{MgIISpecs}
\end{figure*}

\begin{figure*}
\resizebox{\hsize}{!}{
\includegraphics{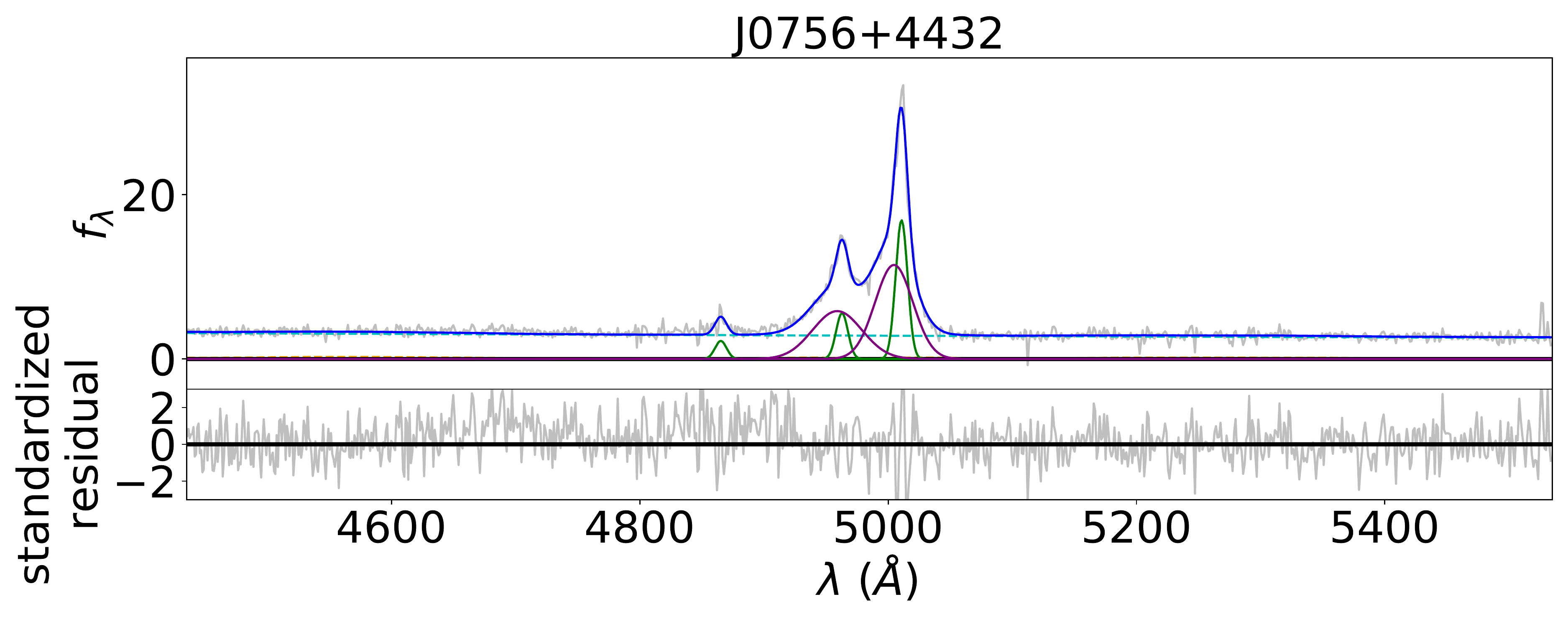}
\includegraphics{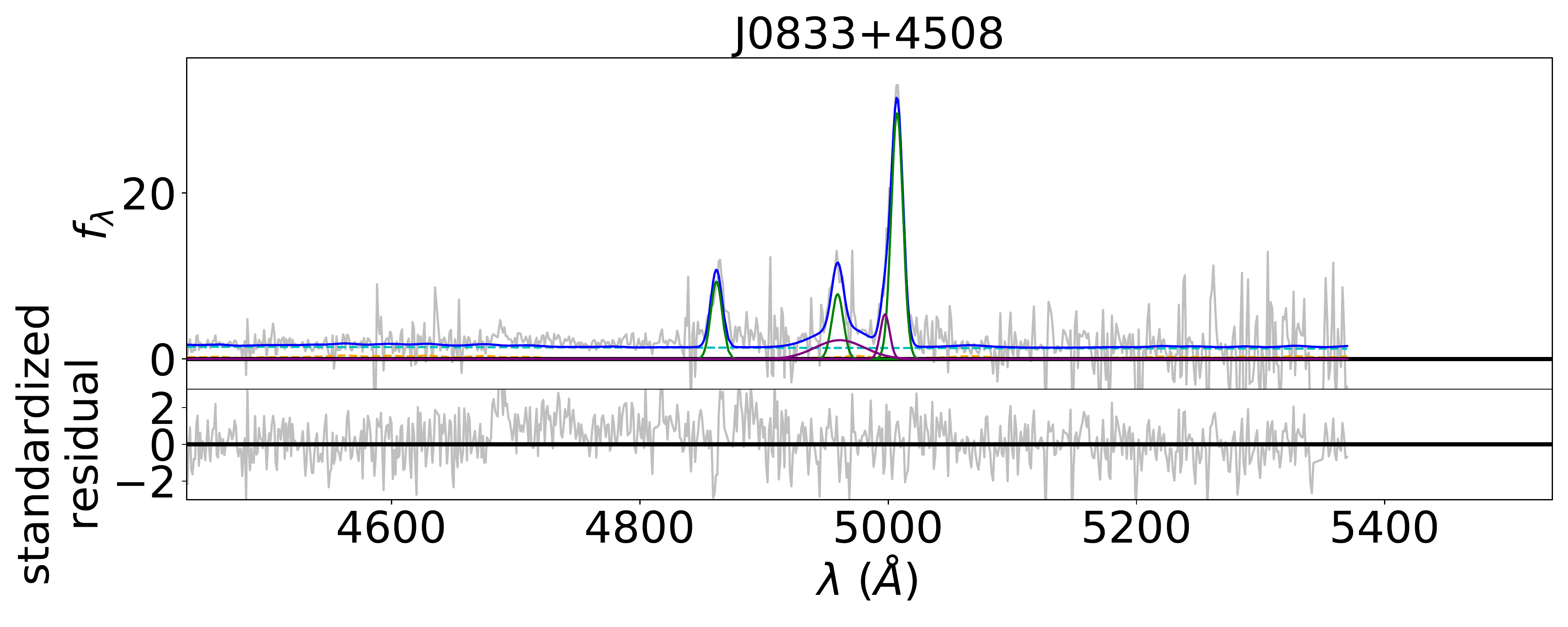}
}
\resizebox{\hsize}{!}{
\includegraphics{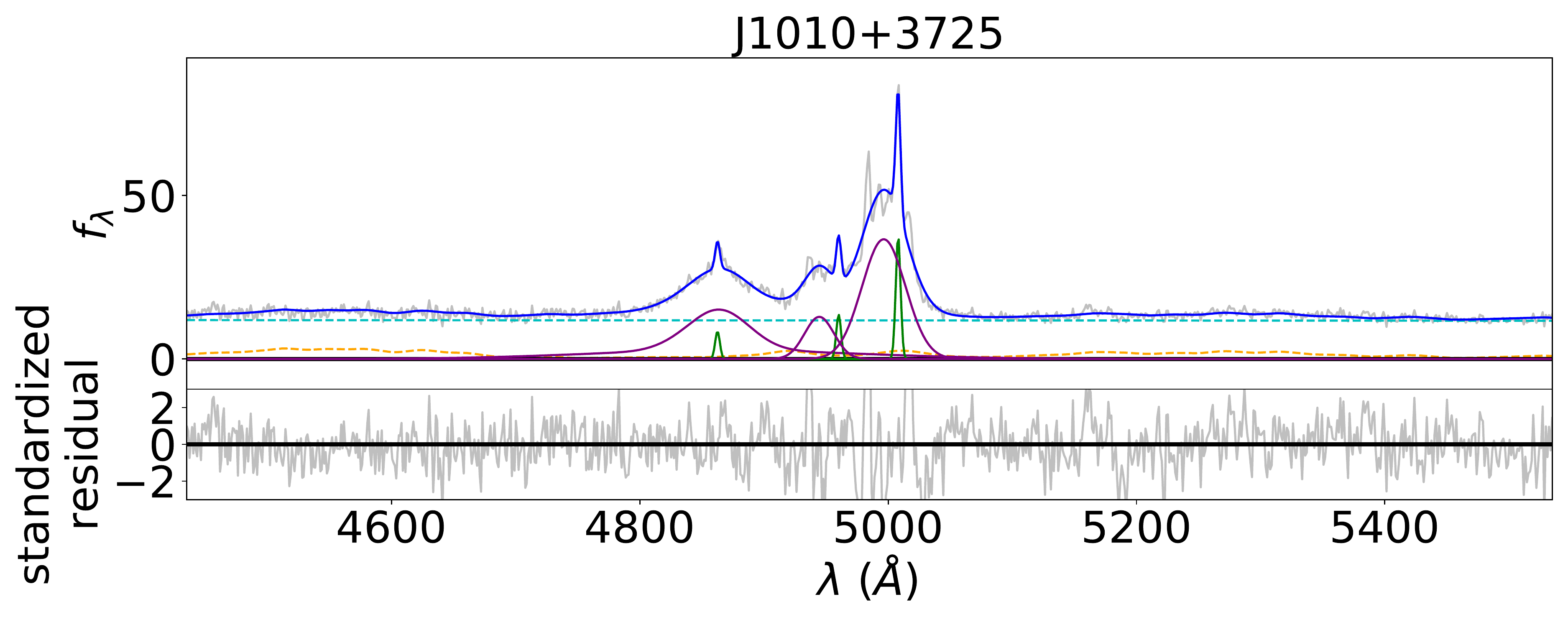}
\includegraphics{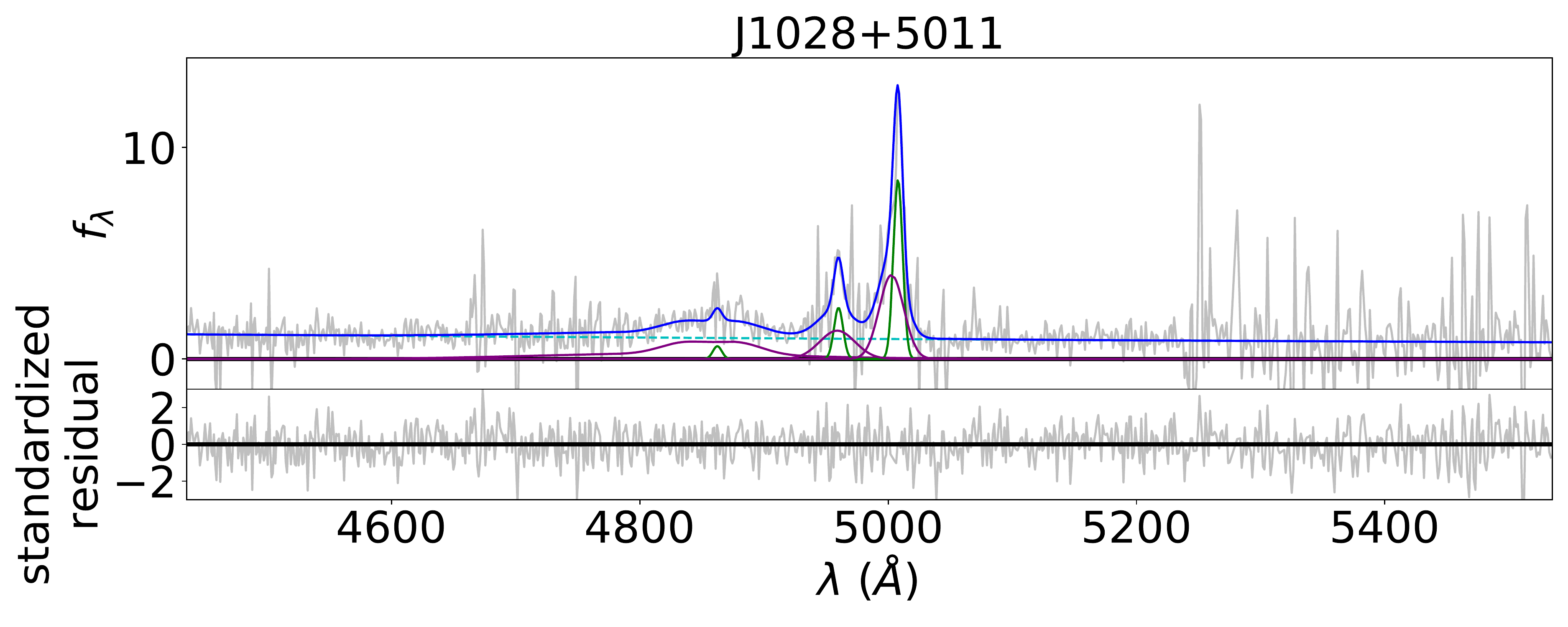}
}
\resizebox{\hsize}{!}{
\includegraphics{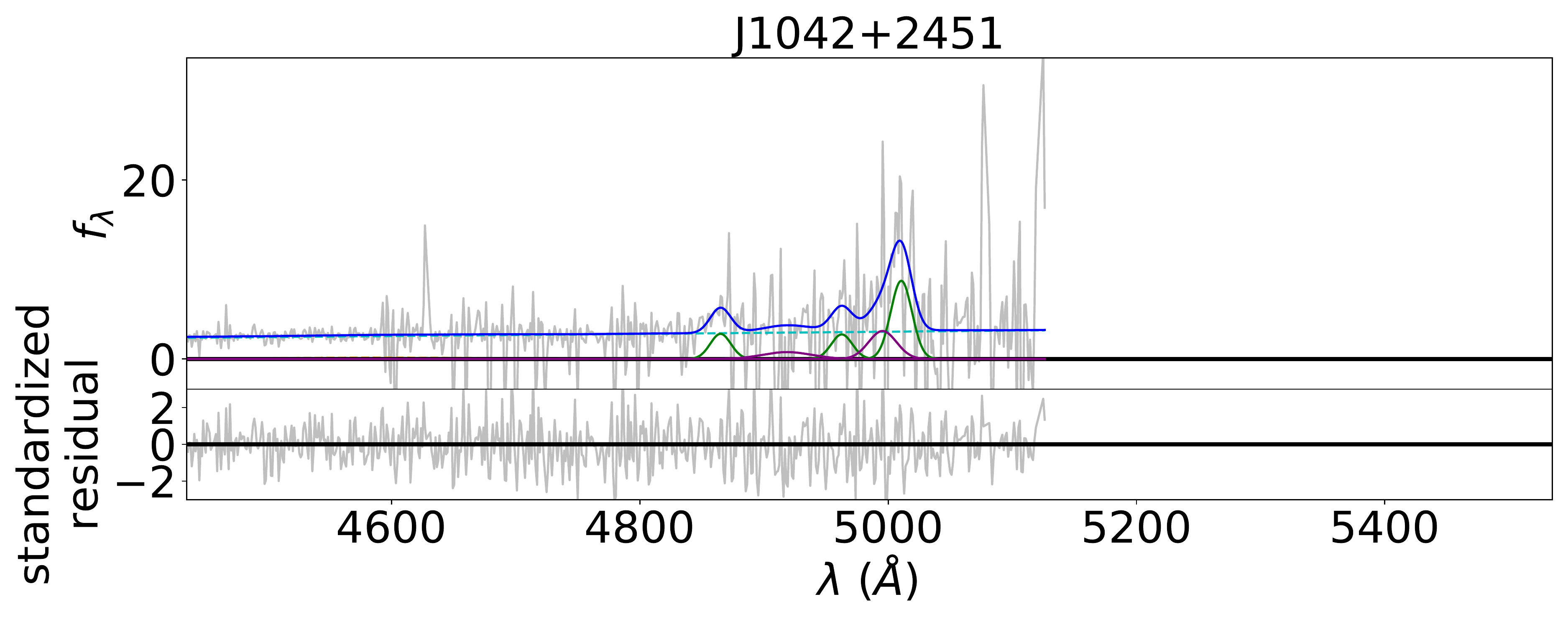}
\includegraphics{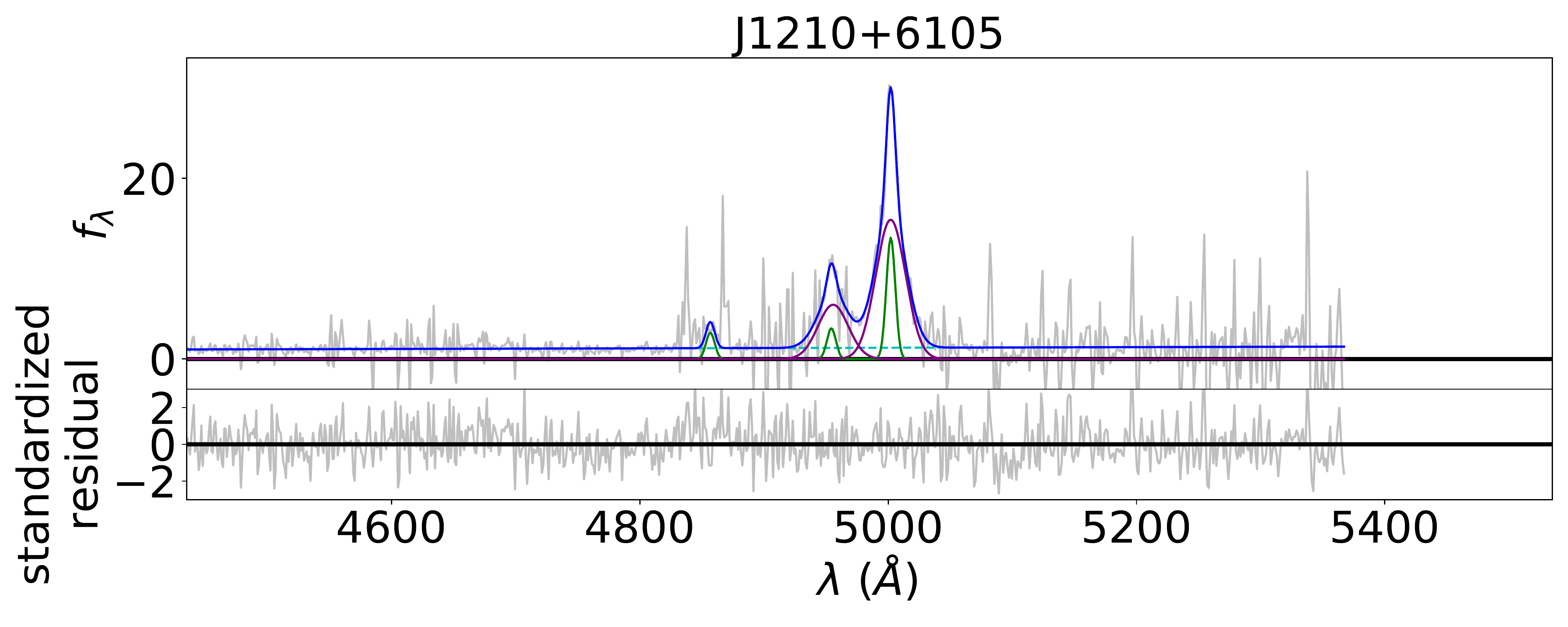}
}
\resizebox{\hsize}{!}{
\includegraphics{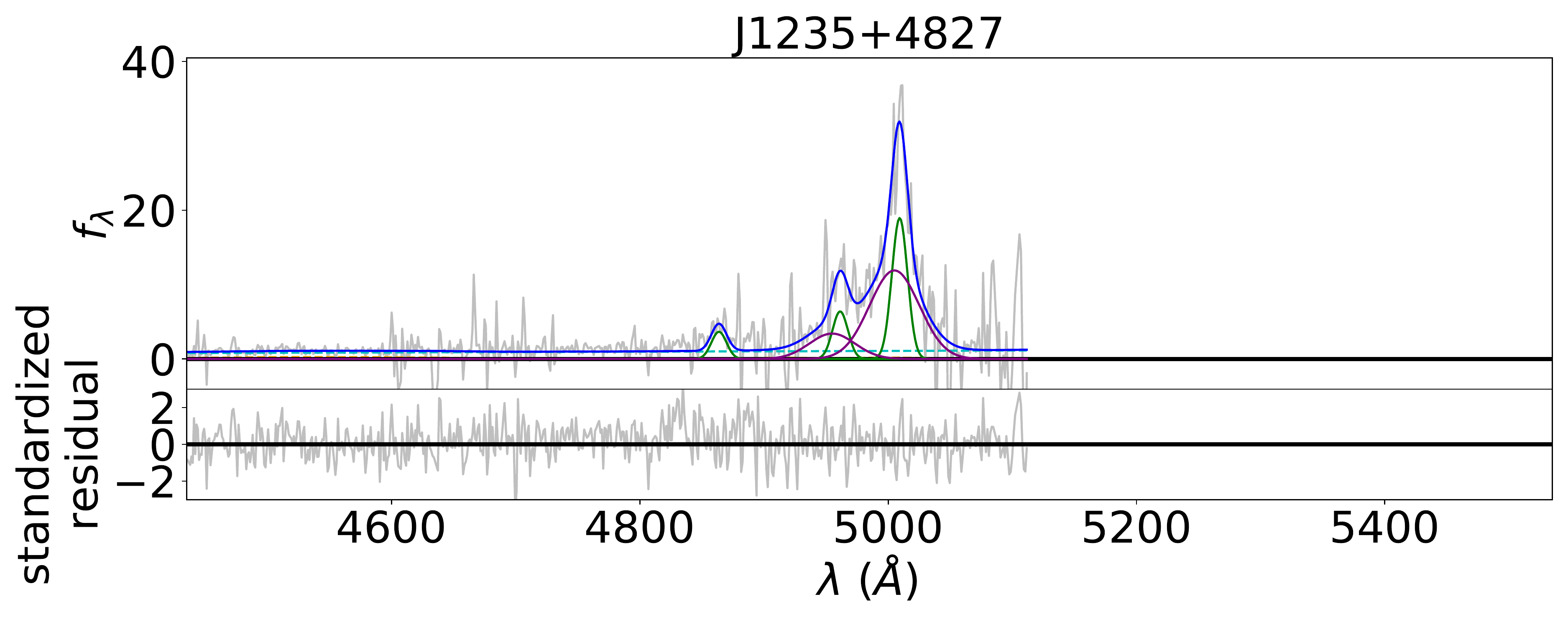}
\includegraphics{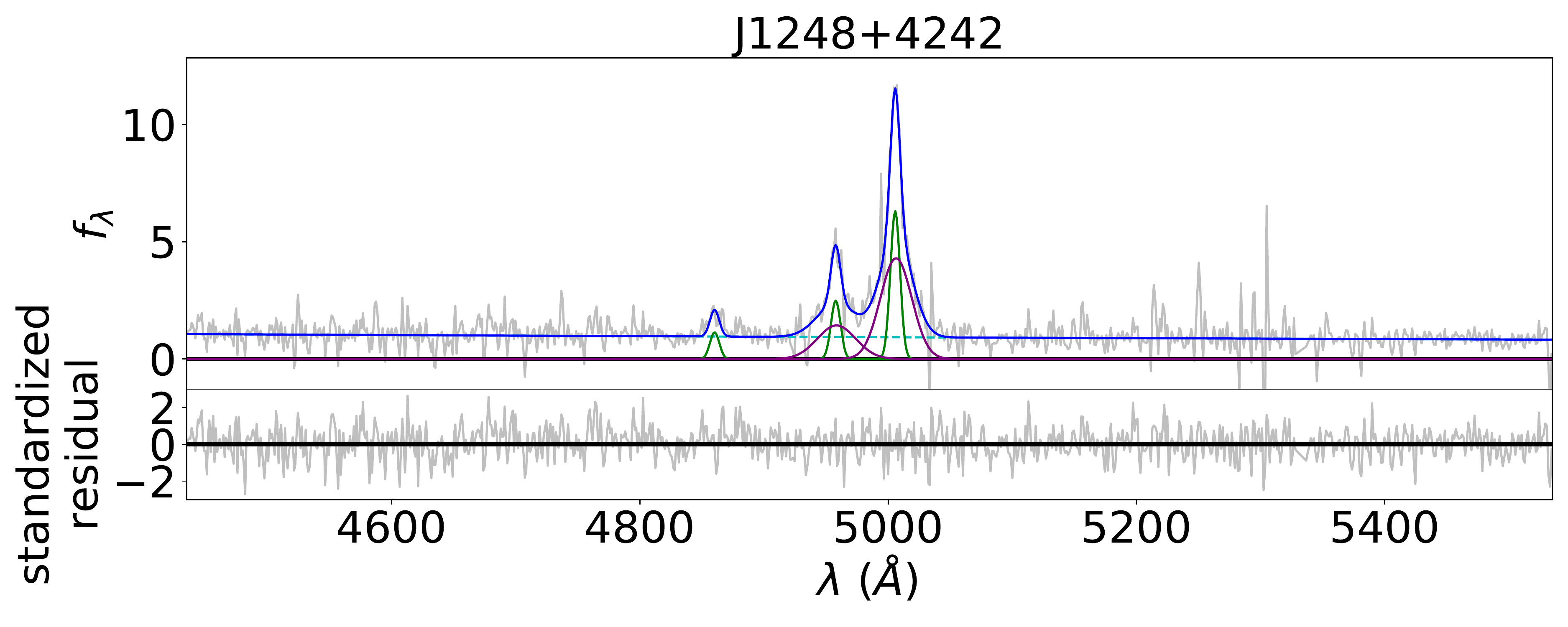}
}
\resizebox{\hsize}{!}{
\includegraphics{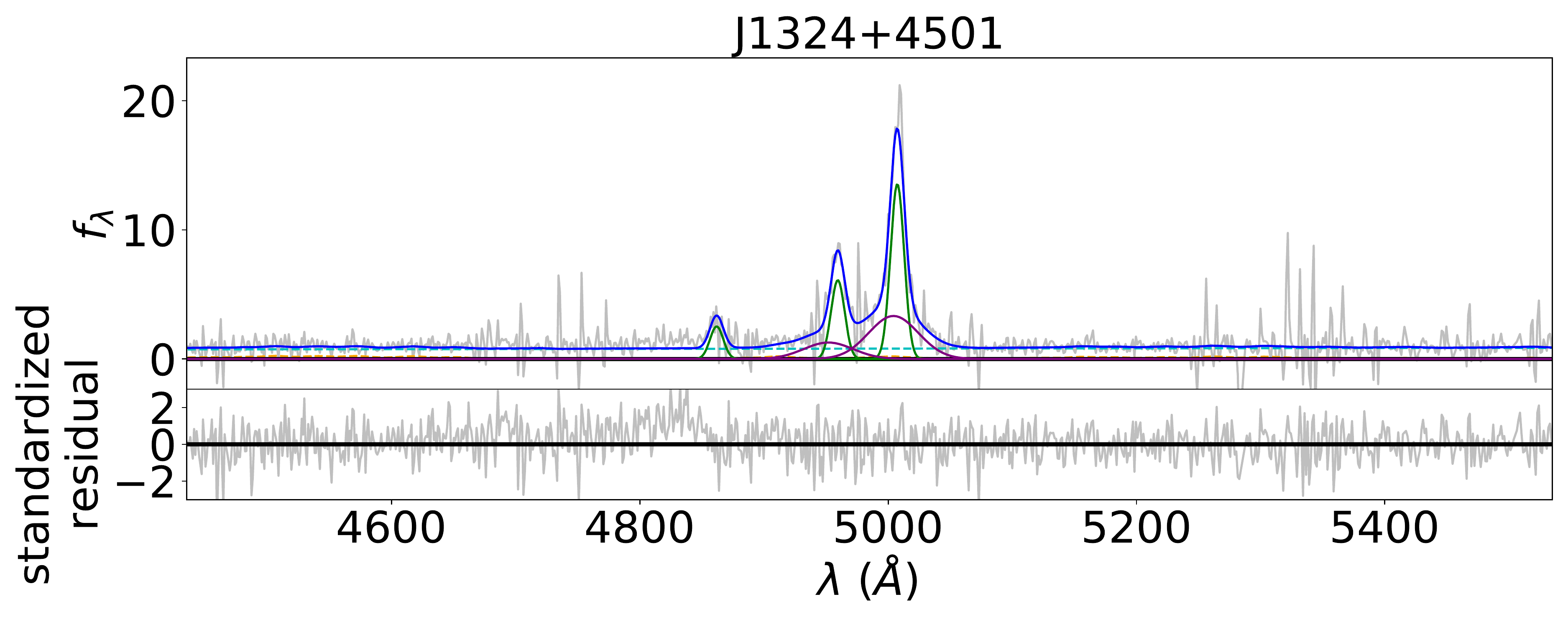}
\includegraphics{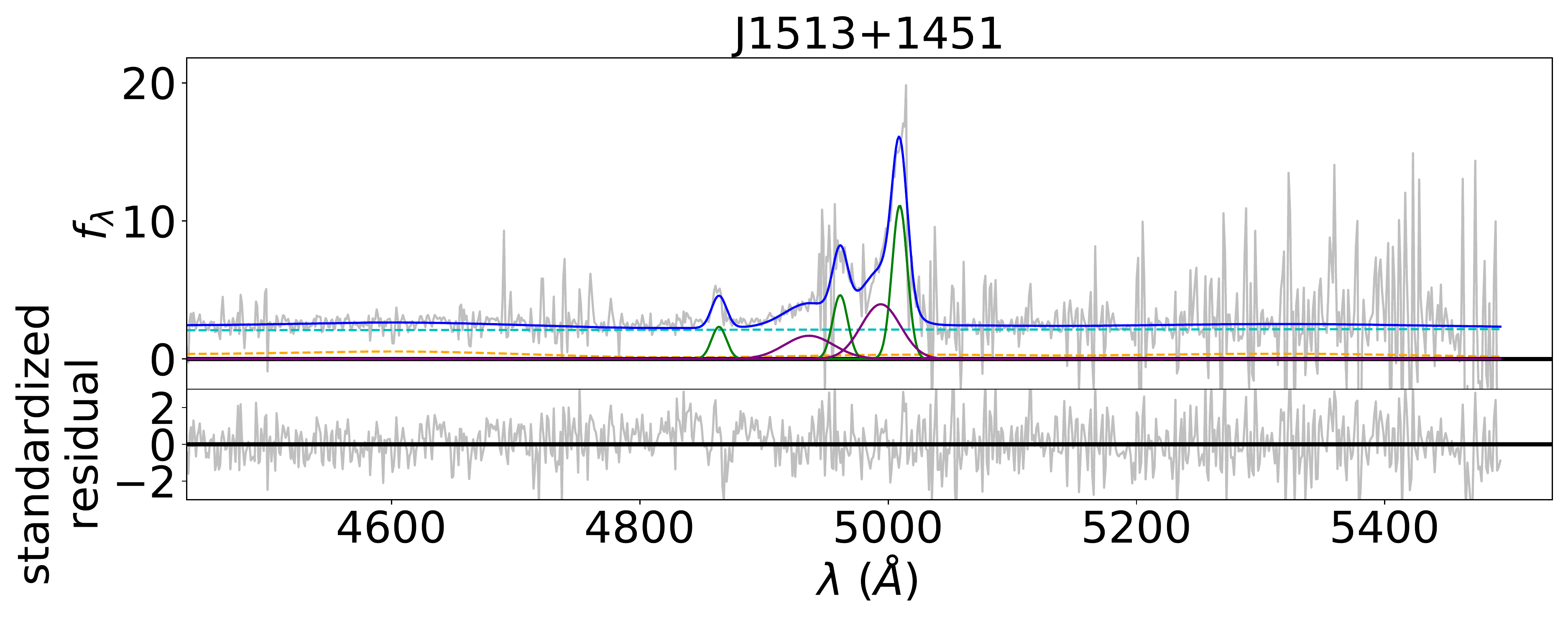}
}
\resizebox{\hsize}{!}{
\includegraphics{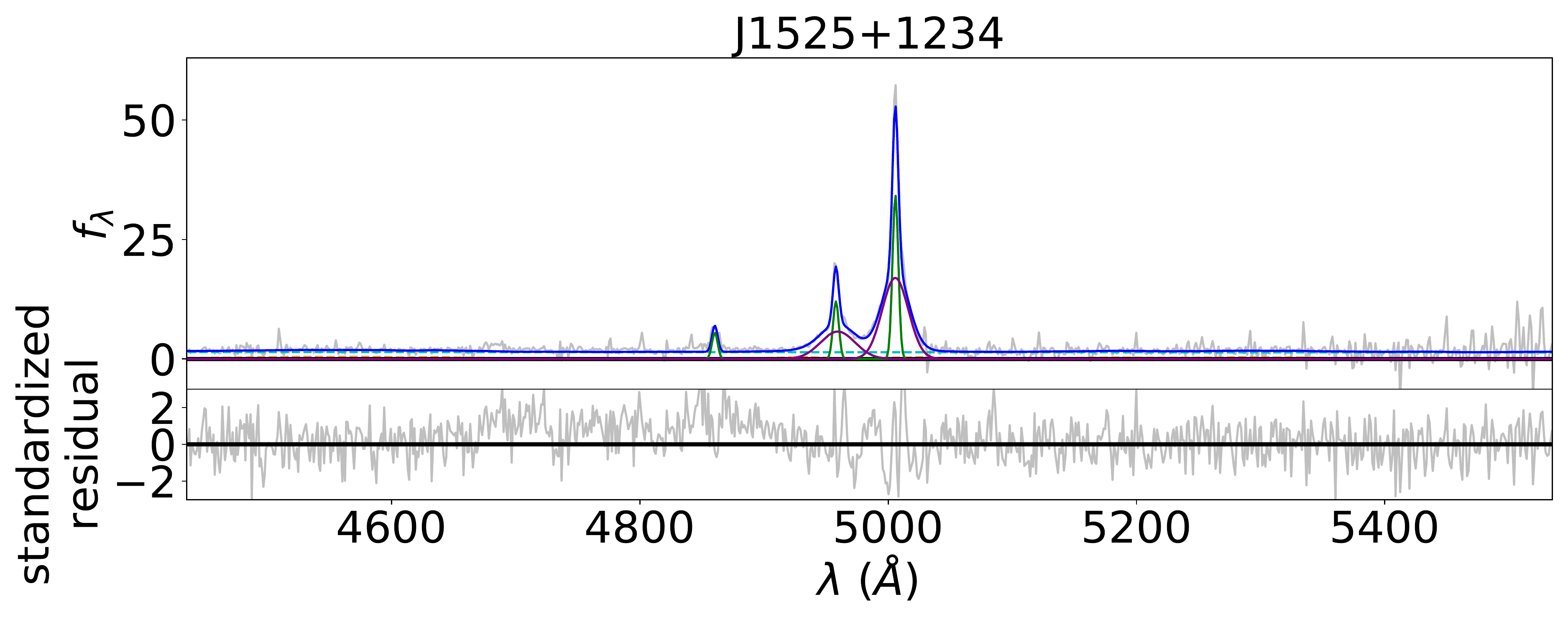}
\includegraphics{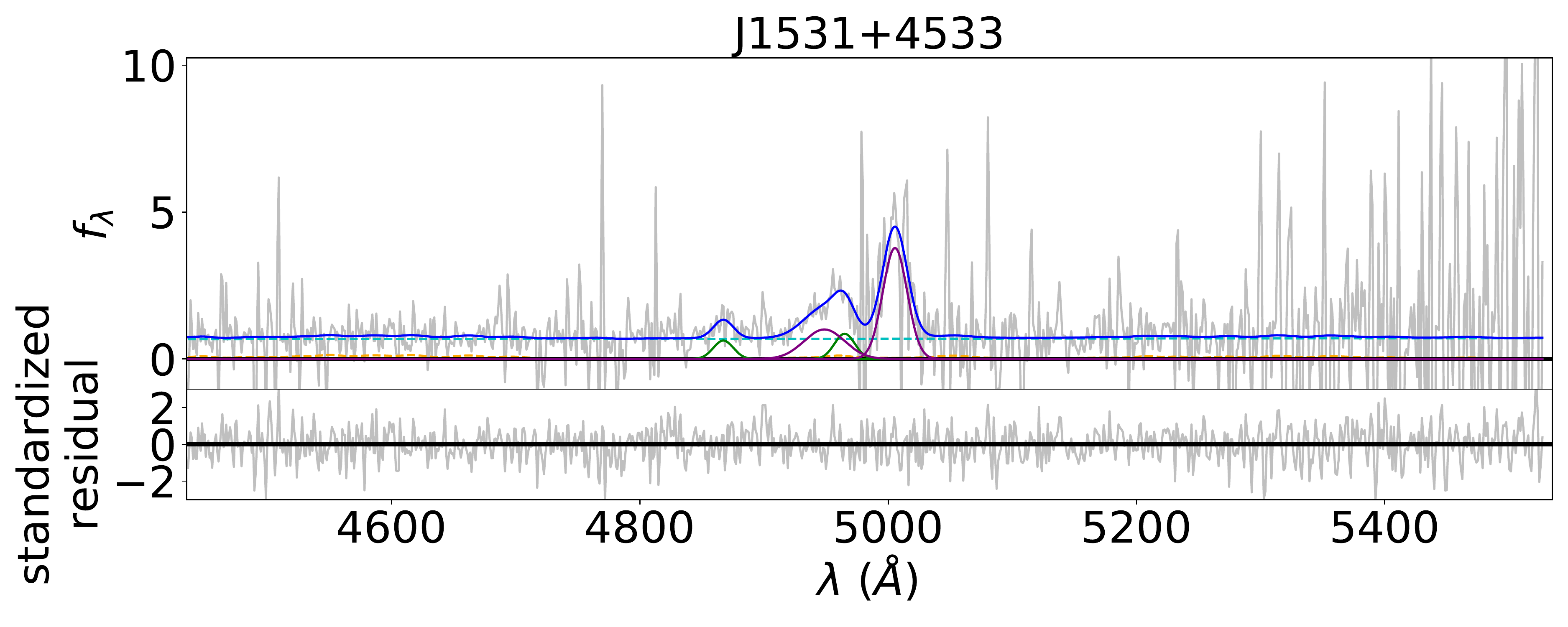}
}
\caption{Similar to Fig.~\ref{MgIISpecs}, but for H$\beta$ and [\iona{O}{iii}]. Purple and green lines represent broad and narrow components, respectively.}
\label{HbOIIISpecs}
\end{figure*}

Fig.~\ref{MgIISpecs} shows best-fit \iona{Mg}{ii} spectra for the 11 sources with \iona{Mg}{ii} coverage. Fig.~\ref{HbOIIISpecs} shows best-fit H$\beta$-[\iona{O}{iii}] spectra for all of our 12 sources.\par

\bsp	
\label{lastpage}
\end{document}